\documentclass[pra,twocolumn,superscriptaddress]{revtex4}
\newcommand{\FigDirectory}{.}

\usepackage{amsmath}                          
\usepackage{amsfonts}                         
\usepackage{amssymb}                          
\usepackage{amscd}                            
\usepackage{amsthm}                           

\usepackage{dsfont}                           
\usepackage[usenames]{color}                  

\usepackage{graphicx}                         
\usepackage{epstopdf}                         
\usepackage{subfigure}			              
\usepackage{comment}

\usepackage[colorlinks=true,
            urlcolor=blue,
            citecolor=blue,
            linkcolor=black,
            pdfstartview=FitH,
            pdfpagemode=None]{hyperref}

\usepackage{datetime}                         
\usepackage{longtable}
\usepackage{multirow}  
\usepackage{algorithm}
\usepackage{algorithmic}
\usepackage{enumerate}

\newcommand{\RR}{{\mathds{R}}}

\newcommand{\BB}{{\mathds{B}}}

\newcommand{\poly}{\mathop{\mathrm{poly}}}

\renewcommand{\>}{\rangle}

\newcommand{\lsim}{\mathrel{\raise.3ex\hbox{$<$\kern-.75em\lower1ex\hbox{$\sim$}}}}
\newcommand{\gsim}{\mathrel{\raise.3ex\hbox{$>$\kern-.75em\lower1ex\hbox{$\sim$}}}}

\def\QECCnk[[#1,#2]]{[\![#1, #2]\!]}

\def\QECCnkq[[#1,#2,#3]]{[\![#1, #2]\!]_{#3}^{\vphantom{T}}}

\def\QECCnkd[[#1,#2,#3]]{[\![#1, #2, #3]\!]}

\def\QECCnkdq[[#1,#2,#3,#4]]{[\![#1, #2, #3]\!]_{#4}^{\vphantom{T}}}

\def\QECCnkgd[[#1,#2,#3,#4]]{[\![#1, #2, #3, #4]\!]}

\def\QECCnkgdq[[#1,#2,#3,#4,#5]]{[\![#1, #2, #3, #4]\!]_{#5}^{\vphantom{T}}}

\def\QECCnkdc[[#1,#2,#3,#4]]{[\![#1, #2, #3; #4]\!]}

\def\QECCnkdcq[[#1,#2,#3,#4,#5]]{[\![#1, #2, #3; #4]\!]_{#5}^{\vphantom{T}}}

\def\QECCnkgdcq[[#1,#2,#3,#4,#5,#6]]{%
  [\![#1, #2, #3, #4; #5]\!]_{#6}^{\vphantom{T}}}

\newcommand{\bigO}{{\cal O}}

\newcommand\CNOT{\ensuremath{\textit{CNOT\/}}}

\def\openone{\leavevmode\hbox{\small1\normalsize\kern-.33em1}}

\newcommand{\etal}{\textit{et~al.}}
\newcommand{\viz}{\textit{viz.}}
\newcommand{\ie}{\textit{i.e.}}
\newcommand{\etc}{\textit{etc.}}

\newcommand{\eg}{\textit{e.g.}}

\long\def\symbolfootnote[#1]#2{\begingroup%
\def\thefootnote{\fnsymbol{footnote}}\footnote[#1]{#2}\endgroup}

\input{Qcircuit.tex}

\theoremstyle{definition}  

\begin{document}


%
\title{Fault-tolerant quantum computing with color codes}

\author{Andrew J. \surname{Landahl}}
\email[]{alandahl@sandia.gov}
\affiliation{Advanced Device Technologies,
             Sandia National Laboratories,
             Albuquerque, NM, 87185, USA}
\affiliation{Center for Quantum Information and Control,
             University of New Mexico,
             Albuquerque, NM, 87131, USA}
\author{Jonas T. \surname{Anderson}}
\email[]{jander10@unm.edu}
\affiliation{Center for Quantum Information and Control,
             University of New Mexico,
             Albuquerque, NM, 87131, USA}
\author{Patrick R. \surname{Rice}}
\email[]{rekniht81@gmail.com}
\affiliation{Center for Quantum Information and Control,
             University of New Mexico,
             Albuquerque, NM, 87131, USA}
\affiliation{Quantum Institute,
             Los Alamos National Laboratories,
             Los Alamos, NM, 87545, USA}



\begin{abstract}

We present and analyze protocols for fault-tolerant quantum computing using
color codes.   To process these codes, no qubit movement is necessary;
nearest-neighbor gates in two spatial dimensions suffices.  Our focus is on
the color codes defined by the 4.8.8 semiregular lattice, as they provide
the best error protection per physical qubit among color codes.  We present
circuit-level schemes for extracting the error syndrome of these codes
fault-tolerantly.  We further present an integer-program-based decoding
algorithm for identifying the most likely error given the (possibly faulty)
syndrome.  We simulated our syndrome extraction and decoding algorithms
against three physically-motivated noise models using Monte Carlo methods,
and used the simulations to estimate the corresponding accuracy thresholds
for fault-tolerant quantum error correction.  We also used a self-avoiding
walk analysis to lower-bound the accuracy threshold for two of these noise
models.  We present two methods for fault-tolerantly computing with these
codes.  In the first, many of the operations are transversal and therefore
spatially local if two-dimensional arrays of qubits are stacked atop each
other.  In the second, code deformation techniques are used so that all
quantum processing is spatially local in just two dimensions.  In both
cases, the accuracy threshold for computation is comparable to that for
error correction.  Our analysis demonstrates that color codes perform
slightly better than Kitaev's surface codes when circuit details are
ignored.  When these details are considered, we estimate that color codes
achieve a threshold of 0.082(3)\%, which is higher than the threshold of
$1.3 \times 10^{-5}$ achieved by concatenated coding schemes restricted to
nearest-neighbor gates in two dimensions [Spedalieri and Roychowdhury,
Quant.\ Inf.\ Comp.\  \textbf{9}, 666 (2009)] but lower than the threshold
of $0.75\%$ to $1.1\%$ reported for the Kitaev codes subject to the same
restrictions [Raussendorf and Harrington, Phys.\ Rev.\ Lett.\ \textbf{98},
190504 (2007); Wang \etal, Phys. Rev. A \textbf{83}, 020302(R) (2011)].
Finally, because the behavior of our decoder's performance for two of the
noise models we consider maps onto an order-disorder phase transition in the
three-body random-bond Ising model in 2D and the corresponding
random-plaquette gauge model in 3D, our results also answer the Nishimori
conjecture for these models in the negative: the statistical-mechanical
classical spin systems associated to the 4.8.8 color codes are
counterintuitively more ordered at positive temperature than at zero
temperature.

\end{abstract}

%
\maketitle


%
\section{Introduction}

The promise of fault-tolerant quantum computing is a crowning achievement of
quantum information science \cite{Shor:1996a, Aharonov:1997a,
Aharonov:1999a, Kitaev:1997b, Steane:1997a, Knill:1998a, Preskill:1998a,
Preskill:1998c}.  Under a specific set of noise and control assumptions, the
promise is that any ideal quantum circuit of size $L$ can be simulated to
any desired precision $\varepsilon$ by a faulty quantum circuit whose size
is at most $\bigO(\varepsilon^{-1} L \log^a L)$ for some (small) constant
$a$.  Fault-tolerant quantum computing protocols are judged by the resources
they employ in the course of a simulation.  Examples of such resources include
the constant $a$, the hidden constant in the big-$\bigO$ notation, and the
requirements imposed by the noise and control assumptions.  Often protocols
are compared by a requirement encapsulated in a single number, the
\emph{accuracy threshold}, which is an upper bound on the error probability
per elementary operation that a faulty circuit must satisfy for the
protocol to work.  A variety of fault-tolerant quantum computing protocols
have been developed, with threshold estimates ranging from as low as
$10^{-6}$ \cite{Stephens:2008a} to as high as $3\%$ \cite{Knill:2004a,
Knill:2004b, Knill:2004c}, depending on the protocol and the noise and
control assumptions.

An important control constraint relevant for several quantum computing
technologies is that the only multi-qubit gates that are possible are those
between nearest-neighbor qubits, where the qubits are laid out in some 2D
geometry in which each qubit neighbors a constant number of other qubits.
Fault-tolerant quantum computing protocols based on concatenated quantum
error-correcting codes have a fractal structure that is not 
commensurate with such a geometry.  Indeed, forcing such codes into a
semiregular 2D geometry requires that one introduce a substantial number of
additional qubit-movement operations that expose the protocol to more
errors, thereby diminishing its accuracy threshold.  The largest accuracy
threshold of which we are aware for a concatenated-coding protocol in a
semiregular 2D geometry is $1.3 \times 10^{-5}$ \cite{Spedalieri:2009a};
that protocol is based on the concatenated nine-qubit Bacon-Shor code
\cite{Bacon:2006a} embedded in the 2D square lattice.

Cognizant of the constraints imposed by 2D geometry, Kitaev introduced a
family of quantum error-correcting codes called \emph{surface codes} that
require only local quantum processing, where locality is defined by a graph
embedded in a surface \cite{Kitaev:1996a}.  Several fault-tolerant quantum
computing protocols have been developed around surface codes
\cite{Dennis:2002a, Raussendorf:2007a, Fowler:2008a}, and these protocols
have significantly higher accuracy thresholds than their concatenated-coding
counterparts.  Numerical threshold estimates for surface-code protocols
range from 0.75\% to 1.1\% \cite{Raussendorf:2007a, Fowler:2008a,
Wang:2011a}; an analytic proof in Ref.~\cite{Dennis:2002a} guarantees that
it is no less than $1.7 \times 10^{-4}$.

Recently Bombin and Martin-Delgado proposed a new family of quantum
error-correcting codes they call \emph{color codes} which are also defined
to be local relative to a graph embedded in a surface \cite{Bombin:2006b}.
Specifically, they are defined by face-three-colorable trivalent graphs in
the following way: on each vertex of the graph lies a qubit, and for each
face $f$ of the graph, one defines two ``stabilizer generators'' or
``checks,'' $X_f$ and $Z_f$.  $X_f$ is the tensor product of Pauli $X$
operators on each qubit incident on face $f$, while $Z_f$ is the tensor
product of Pauli $Z$ operators on each qubit incident on face $f$.  The
color code's codespace is defined as the simultaneous $+1$ eigenspace of
each of the check operators.

\begin{figure}[!ht]

\center{
  \subfigure[\ 4.8.8]{\includegraphics[width=0.25\columnwidth]{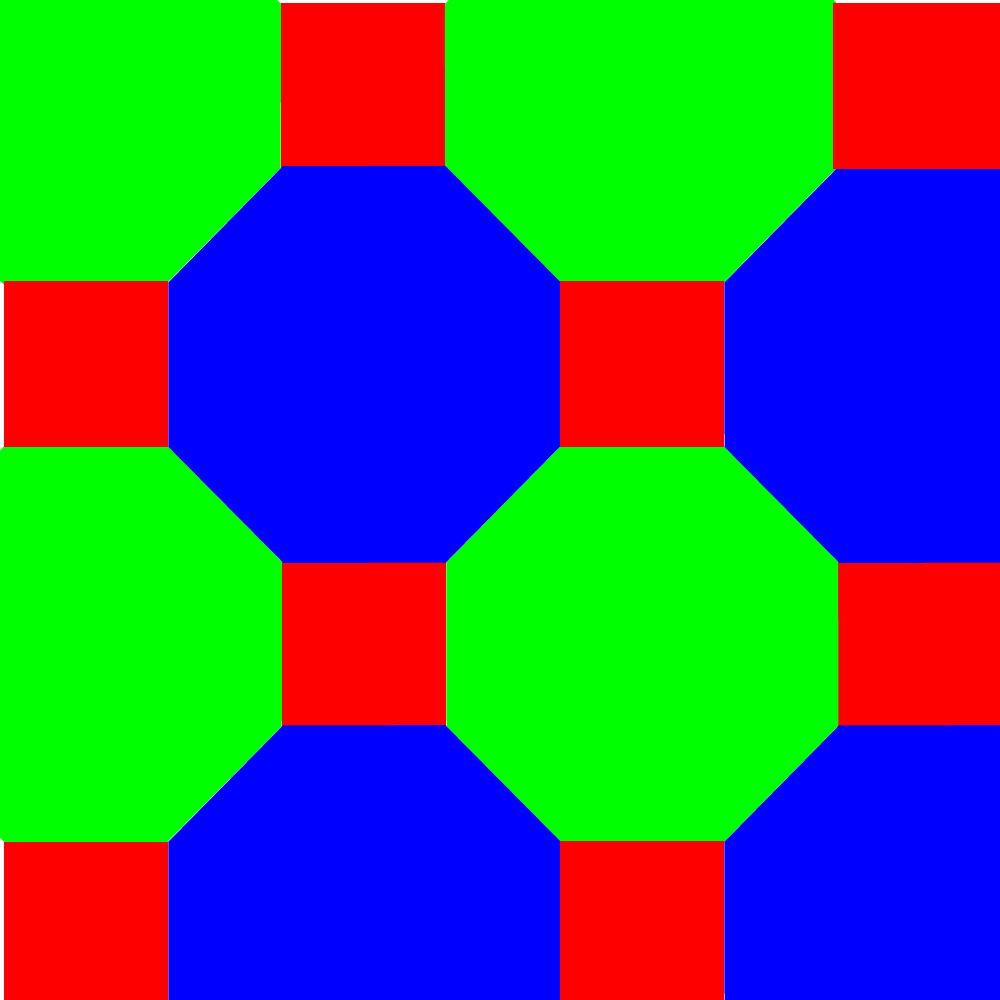}}\hspace{1em}
  \subfigure[\ 6.6.6]{\includegraphics[width=0.25\columnwidth]{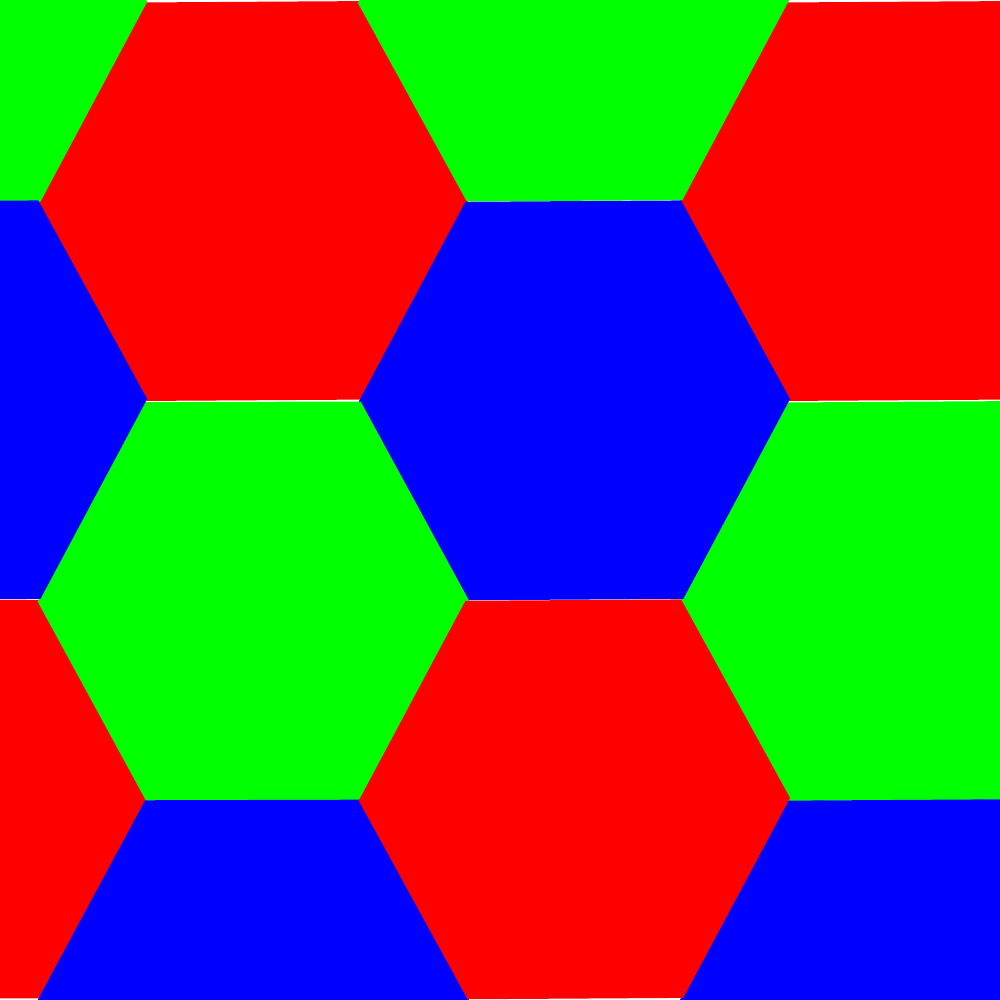}}\hspace{1em}
  \subfigure[\ 4.6.12]{\includegraphics[width=0.25\columnwidth]{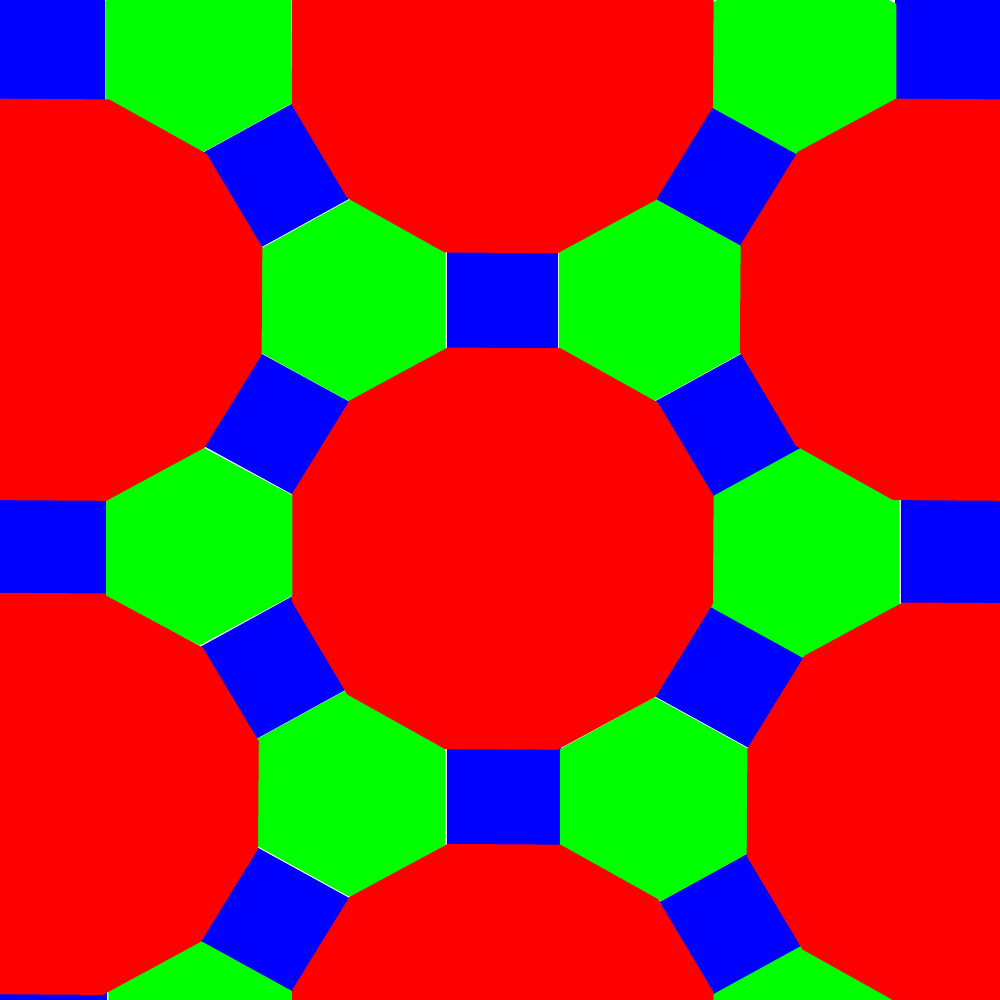}}
}
\caption{\label{fig:trivalent-convex-uniform-tilings}The three possible face-three-colorable
trivalent uniform tilings of the plane.}
\end{figure}

A fault-tolerant quantum computing protocol based on color codes requires an
infinite family of color codes of increasing size in order to be able to
simulate arbitrarily large ideal quantum circuits to increasing precision.
A natural source for an infinite color-code family is a uniform tiling of
the plane by a trivalent face-three-colorable lattice.  Such a lattice can
be embedded in any orientable surface, although later we will restrict
attention to embeddings in planar discs.  These ``semiregular'' or
``Archimedean'' lattices are described in \emph{vertex notation} as $r.s.t$,
where each vertex is locally surrounded by an $r$-gon, an $s$-gon, and a
$t$-gon.  The only possible trivalent face-three-colorable tilings of the
plane are the $4.8.8$ lattice, the $6.6.6$ (hex) lattice, and the $4.6.12$
lattice, depicted in Fig.~\ref{fig:trivalent-convex-uniform-tilings}
\cite{Wiki-Uniform-Tilings:2011a}.

Accuracy thresholds for fault-tolerant quantum computing have been estimated
for color codes in several highly idealized noise models numerically.  The
values of these thresholds are summarized in Table
\ref{tab:code-thresholds}, along with analogous estimates for a well-studied
surface code and two recently-proposed topological subsystem codes.  This
table contains numerous gaps, some of which we fill in with the results of
this Article---the entries containing our results are highlighted in bold.
The most significant gap, which we fill, is an estimate of the accuracy
threshold for noise that afflicts the individual quantum circuit elements
used in a fault-tolerant color-code-based quantum computing protocol.  The
accuracy threshold for noise afflicting the circuit model is perhaps the
most instructive of all table entries.  This is because this threshold
establishes the target error rate per elementary operation that a quantum
technology must meet to admit fault-tolerant quantum computation using these
codes.  It also allows for a fair ``apples-to-apples'' comparison to the
high thresholds estimated for Kitaev's surface codes in the circuit model.

\begin{table*}[ht]
\centering
\begin{tabular}{c|r@{.}l|r@{.}l|r@{.}l|r@{.}l|r@{.}l|c|l} \hline \hline
 & \multicolumn{6}{c|}{Code Capacity} & \multicolumn{4}{c|}{Phenomenological} & \multicolumn{2}{c}{Circuit-based} \\ \hline
Code & \multicolumn{2}{c|}{Other} & \multicolumn{2}{c|}{MLE} &
\multicolumn{2}{c|}{Optimal} & \multicolumn{2}{c|}{MLE} &
\multicolumn{2}{c|}{Optimal} & Other & \multicolumn{1}{c}{MLE} \\ \hline
\multirow{2}{*}{4.8.8} & 8&87\,\%\footnote[1]{Reference computes threshold against DP channel,
not BP channel.  For non-circuit-based noise models, the decoder used does
not account for correlations between bit flips and phase flips in DP
channel.  In these models, we reported the result for the equivalent effective BP
channel of strength $\frac{2}{3}p$.}$^,$\footnote[2]{Decoder based on 
hypergraph matching heuristic.} \cite{Wang:2009b} &
\textbf{10}&\textbf{56(1)}\,\% & 10&9(2)\,\%
\cite{Katzgraber:2009a} & \textbf{3}&\textbf{05(4)\,\%}     &
\multicolumn{2}{c|}{}     & ``$\sim
0.1$\,\%''$^{a,b,}$\footnote[3]{{Limited numerics only weakly suggest
this value.}} \cite{Wang:2009b} & \textbf{0.082(3)\,\%}      \\
    & 8&7\,\%\footnote[4]{Decoder based on mapping to two Kitaev codes.}
\cite{Duclos-Cianci:2011a} & \multicolumn{2}{c|}{(Our result)} &
10&925(5)\,\% \cite{Ohzeki:2009b} & \multicolumn{2}{c|}{(Our result)} &
\multicolumn{2}{c|}{} & & \multicolumn{1}{c}{(Our result)}  \\ \hline
\multirow{2}{*}{6.6.6} & \multicolumn{2}{c|}{} & \multicolumn{2}{c|}{} &
10&9(2)\,\% \cite{Katzgraber:2009a} & \multicolumn{2}{c|}{} &
4&5(2)\,\% \cite{Andrist:2010a} & &      \\ 
    & \multicolumn{2}{c|}{} & \multicolumn{2}{c|}{} & 10&97(1)\,\%
\cite{Ohzeki:2009b} & \multicolumn{2}{c|}{} & \multicolumn{2}{c|}{} & &  \\ \hline
4.6.12 & \multicolumn{2}{c|}{} & \multicolumn{2}{c|}{} &
\multicolumn{2}{c|}{} & \multicolumn{2}{c|}{} & \multicolumn{2}{c|}{} & & \\ \hline
\multirow{2}{*}{4.4.4.4 Kitaev} & \multicolumn{2}{c|}{} & 10&31(1)\,\% \cite{Wang:2003a}    &
10&9187\,\%
\cite{Ohzeki:2009a} &  2&93(2)\,\% \cite{Wang:2003a}  &  3&3\,\%
\cite{Ohno:2004a} & & 0.75\,\%$^{a}$ \cite{Raussendorf:2007a}       \\
        & \multicolumn{2}{c|}{} & \multicolumn{2}{c|}{} & 10&939(6)\,\%
\cite{deQueiroz:2009a} & \multicolumn{2}{c|}{} & \multicolumn{2}{c|}{} & & 1.1\,\%$^{a}$ \cite{Wang:2011a}  \\ \hline
3.4.6.4 TSCC & 1&3\,\%$^{a,c}$ \cite{Duclos-Cianci:2011a} & \multicolumn{2}{c|}{} &
\multicolumn{2}{c|}{} & \multicolumn{2}{c|}{}  &  \multicolumn{2}{c|}{}    &
&       \\ \hline 
``SBT'' \cite{Suchara:2010a} & 1&3\,\%$^{a,c}$ \cite{Suchara:2010a} & \multicolumn{2}{c|}{} &
\multicolumn{2}{c|}{} & \multicolumn{2}{c|}{}  &  \multicolumn{2}{c|}{}    &
&       \\ \hline \hline
\end{tabular}

\caption{Numerically-estimated accuracy thresholds for several topological
quantum error-correcting codes, noise models, and decoding algorithms.  The
first three codes (4.8.8, 6.6.6, 4.6.12) are the color codes described in
Fig.~\ref{fig:trivalent-convex-uniform-tilings} and its preceding text.  The
last three codes are the Kitaev surface code on the square lattice
\cite{Kitaev:1996a}, a topological subsystem color code on the 3.4.6.4
lattice \cite{Bombin:2010a}, and a hypergraph-based topological subsystem
code proposed by Suchara, Bravyi, and Terhal \cite{Suchara:2010a}.  The
details of the noise models (code capacity, phenomenological, and
circuit-based) and decoders (MLE, optimal, and other) are discussed in the
text; when possible, results from other references have been translated into one
of these models.  The notation ``$x.y_1\cdots y_k(z)\%$'' means $x.y_1\cdots
y_k\% \pm (z \times 10^{-k})\%$.  When such notation is not used, it means that
the no error analysis was reported in the reference from which the value was
drawn.}
\label{tab:code-thresholds}
\end{table*}

In this Article, we analyze the accuracy threshold of the 4.8.8 color codes
for fault-tolerant quantum computation under several noise and control
models.  We have restricted our analysis to protocols which use the decoder
that identifies the most likely error (MLE) given the error syndrome.  We
formulate the MLE decoder as an integer program (IP), which in general is
NP-hard to solve \cite{Berlekamp:1978a}.  Although the decoder is
inefficient, it establishes a threshold that we expect is close the the
maximum threshold possible for these codes, namely the one obtainable by an
optimal decoder, which identifies the most likely logical operation
given the error syndrome.  For small codes, the MLE IP can be
solved ``offline'' ahead of time to generate a lookup table that can be used
during the course of a ``live'' fault-tolerant quantum computing protocol.
Our results comprise both numerical estimates of the accuracy threshold
achieved via Monte Carlo simulations and a rigorous lower bound on the
accuracy threshold that we prove using combinatorial counting arguments.

The remainder of this Article is organized as follows.  In
Sec.~\ref{sec:noise-and-control-model}, we lay out the control model and the
three noise models we consider.  In
Sec.~\ref{sec:fault-tolerant-error-correction}, we summarize the properties
of the 4.8.8 triangular color codes we study, present two circuit schedules
for extracting the error syndrome in these codes, and formulate MLE decoders
for these codes as integer programs for each of the noise models that we
consider.  In Sec.~\ref{sec:numerical-estimate-of-the-accuracy-threshold},
we report our numerical estimates for the accuracy threshold for
fault-tolerant quantum error correction of these codes for each of the noise
models that we consider.  In Sec.~\ref{sec:analytic-bound}, we use a
self-avoiding-walk analysis to prove rigorous lower bounds for the accuracy
thresholds of fault-tolerant quantum error correction of these codes against
two of the noise models that we consider.  In
Sec.~\ref{sec:ftqc-with-color-codes}, we relate the quantum error correction
accuracy threshold to the quantum computation accuracy threshold for two
scenarios: one in which logical qubits are associated with 2D planes that
are stacked atop one another like pancakes and the other in which logical
qubits are associated with ``defects'' in a single 2D substrate.  In
Sec.~\ref{sec:conclusions} we conclude, summarizing and interpreting our
results both in terms of the accuracy thresholds we report and in terms of
their consequences for ``re-entrant behavior'' of an order-disorder phase
transition in two associated classical statistical-mechanical models.  We
cap off our conclusions with some parting thoughts about future directions
that we believe are worthy of study.

%
\section{Noise and control model}
\label{sec:noise-and-control-model}

The performance of a fault-tolerant quantum computing (FTQC) protocol is
strongly influenced by underlying architectural assumptions, so it is
important to clearly list what they are.  Indeed, when those assumptions are
not borne out in real quantum information technologies, an FTQC protocol may
fail entirely~\cite{Levy:2009a, Levy:2011a}.

Every existing FTQC protocol makes the following architectural
assumptions---assumptions which appear to be necessary:
\begin{enumerate}

\item \textbf{\textsl{Nonincreasing error rate}}.  The asymptotic scaling of
the error rate as a function of the circuit's size is nonincreasing.  This
allows the performance of fault-tolerant circuits to increase
asymptotically.

\item \textbf{\textsl{Parallel operation}}.  The asymptotic
parallel-processing rate is larger than a constant times the asymptotic
error rate.  This allows error correction to keep ahead of the errors.

\item \textbf{\textsl{Reusable memory}}.  The asymptotic rate at which one
can erase or replace qubits is larger than a constant times the asymptotic
error rate.  This allows entropy to be flushed from the computer faster
than it is generated by errors.

\end{enumerate}

Some FTQC protocols also make the following architectural assumptions, which
generally lead to higher accuracy thresholds; we make these assumptions here:
\begin{enumerate}
\setcounter{enumi}{3}
\item \textbf{\textsl{Reliable classical computation}}.  Classical
computations always return the correct result.
\item \textbf{\textsl{Fast classical computation}}.  Classical computations
are instantaneous.
\item \textbf{\textsl{No qubit leakage}}.  Qubits never ``leak'' out of the
computational Hilbert space.
\item \textbf{\textsl{Uncorrelated noise}}.  Each qubit and gate is
afflicted by an independent noise source.
\end{enumerate}

Some additional architectural assumptions, which have a less clear impact on
the accuracy threshold, are frequently made as well; we also make these
assumptions:
\begin{enumerate}
\setcounter{enumi}{7}
\item \textbf{\textsl{Standard gate basis}}. The set of (faulty) quantum
gates (including preparation and measurement) available consists of $|0\>$,
$|+\>$, $I$, $X$, $Z$, $T$, $S$, $\CNOT$, $M_Z$, and $M_X$.  The
definition of what these gates are can be found in standard textbooks, \eg,
in Refs.~\cite{Nielsen:2000a, Preskill:1998b}.
\item \textbf{\textsl{Equal-time gates}}.  Each gate, including preparations
and measurements, takes the same amount of time to complete.
\item \textbf{\textsl{\label{eq:uniformly-faulty-gates}Uniformly faulty
gates}}.  Each $k$-qubit gate, including preparations and measurements, is
as equally as faulty as every other $k$-qubit gate.  
\end{enumerate}

Inspired by the limitations of 2D geometry for some quantum computing
technologies, we also make the following assumptions:
\begin{enumerate}

\setcounter{enumi}{10}

\item \textbf{\textsl{2D layout}}.  Qubits are laid out on a structure
describable by a graph embedded in a two-dimensional surface.

\item \textbf{\textsl{Local quantum processing}}.  Gates can only couple
nearest-neighbor qubits in the graph describing their layout.

\end{enumerate}

Finally, we make the following three variants of a thirteenth assumption
about the noise model afflicting each gate.  Of all the assumptions we make,
we have found that this one is most likely to vary in the literature.
Commonly-studied alternatives for this assumption include stochastic
adversarial noise \cite{Aliferis:2006a, Aliferis:2007a, Aliferis:2008a,
Aliferis:2009a}, purely depolarizing noise \cite{Cross:2009a}, and noise
that has a strong bias, such as having phase flips significantly more
probable than bit flips \cite{Aliferis:2008b}.

\begin{enumerate}[{13(}$a$).]

\item \textbf{\textsl{Circuit-level noise}}.\label{item:circuit-level-noise}
Each faulty single-qubit preparation and faulty coherent single-qubit gate
($|0\>$, $|+\>$, $I$, $X$, $Z$, $H$, $T$, $S$) acts ideally, followed by the
bit-flip channel of strength $p$, which applies bit flips (Pauli $X$
operators) with probability $p$, followed by the phase-flip channel of
strength $p$, which applies phase flips (Pauli $Z$ operators) with
probability $p$.  We call this channel the \emph{BP channel}.  Each faulty
single-qubit measurement ($M_X$, $M_Z$) acts as the BP channel of
probability $p$ followed by a measurement that returns the incorrect result
with probability $p$.  Importantly, this noise model assumes that the state
after a measurement is in an eigenstate of the observable measured, just
perhaps not the eigenstate that the measurement indicates.  Each $\CNOT$
gate acts ideally followed by a channel in which each of the 16 two-factor
Pauli products ($II$, $IX$, $XI$, $XY$, \etc) is applied with probability
$p/16$.  We call this channel the \emph{DP channel}.  This model differs
slightly from a frequently-studied variant in the literature in which each
of the 15 nontrivial two-factor Pauli products is applied with probability
$p/15$ and the identity is applied with probability $1-p$.

\item \textbf{\textsl{Phenomenological noise}}.  This noise model is the
same as the circuit-level noise model
(13($\ref{item:circuit-level-noise}$)), except that the circuit for
syndrome extraction (described later) is modeled ``phenomenologically,'' having a
probability $p$ for returning the wrong syndrome bit value.  In
this model, the propagation of errors between data qubits and between data
and ancilla qubits induced by the syndrome extraction circuit are ignored.
Single-qubit and two-qubit gates on the data qubits in circuits other than
those used for syndrome extraction (\eg, for encoded computation) are still
subject to the BP and DP channels, respectively, as before.

\item \textbf{\textsl{Code capacity noise}}.  This model is the same as the
phenomenological noise model, except that the syndrome-bit error rate is
assumed to be zero.  Because there is no need to repeat syndrome
measurements in this model, and because the accuracy threshold for
``defect-braided'' quantum computation is the same as that for quantum
memory (as argued later), the accuracy threshold for this noise model is the
same as what in quantum information theory is called the single-shot,
single-letter quantum capacity for color codes subject to the BP channel.

\end{enumerate}

%
\section{Fault-tolerant error correction of color codes}
\label{sec:fault-tolerant-error-correction}

%
\subsection{Code family}

We confine our analysis of color codes to the 4.8.8 color codes; our choice
is motivated by two factors.  First, of the three color codes on semiregular
2D lattices, the 4.8.8 code uses the fewest qubits per code distance.
Second, the 4.8.8 code is the only one of the three which can realize
encoded versions of the entire ``Clifford group'' \cite{Nielsen:2000a} of
quantum gates, namely the gates which conjugate Pauli operators to Pauli
operators in the Heisenberg picture, in a transversal fashion
\cite{Bombin:2006b}, \ie, by applying the same operation to every qubit in a
code block or between corresponding qubits in two code blocks.  In
particular, the gates $X$, $Z$, $H$, $S$, and $\CNOT$ have transversal
encoded implementations for these codes.    When encoded gates are
implemented transversally, fault-tolerant quantum computing protocols for
simulating these gates are generally simpler, leading to more favorable
accuracy thresholds.  The Clifford group of gates is an important group of
gates for stabilizer codes such as the color codes, since error correction
can be carried out solely using those gates \cite{Gottesman:1999b}.

\begin{figure}[h!]
\center{
  \subfigure[\ 4.8.8 code]{\includegraphics[width=0.3\columnwidth]{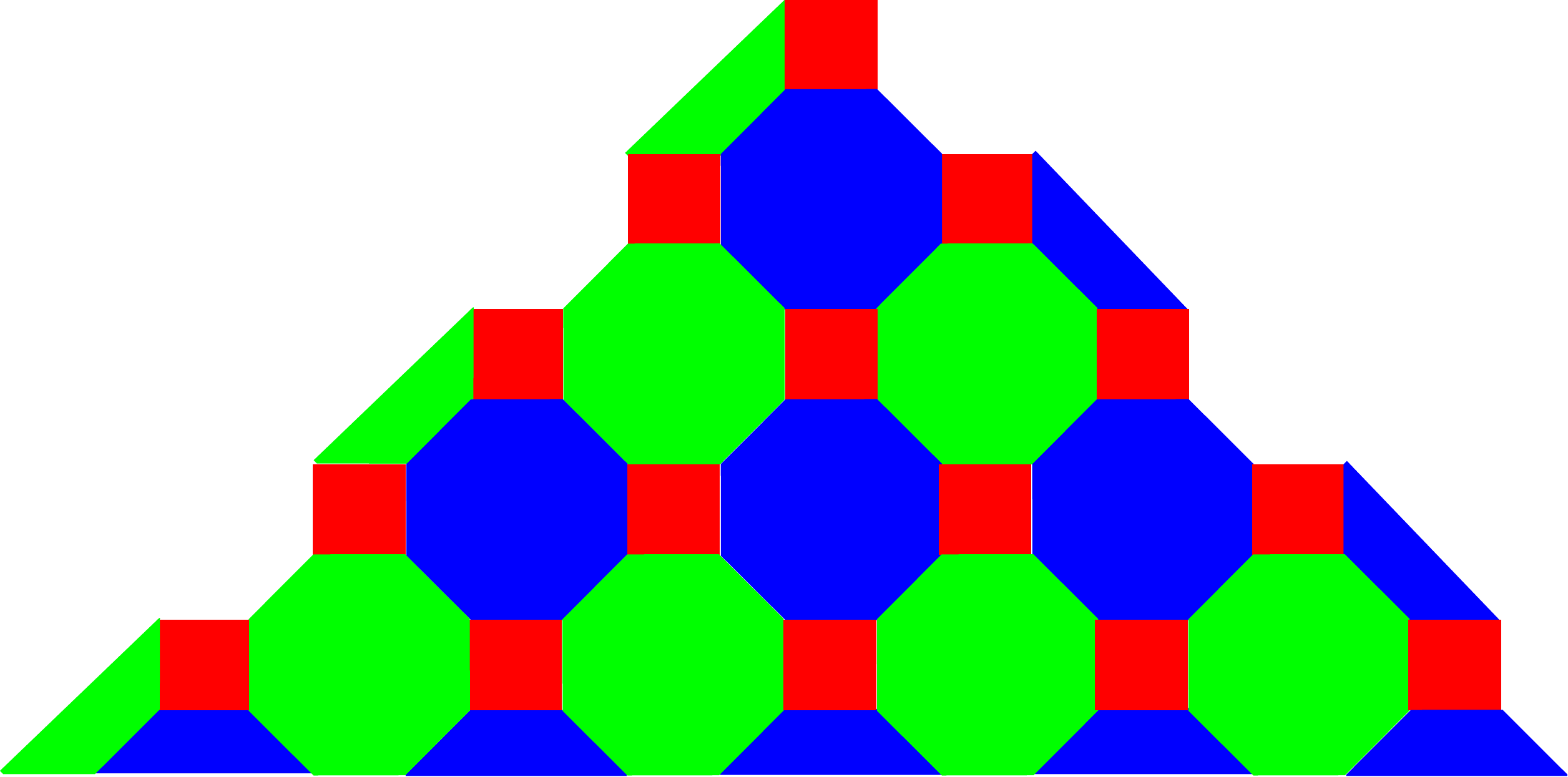}}
  \subfigure[\ 6.6.6 code]{\includegraphics[width=0.3\columnwidth]{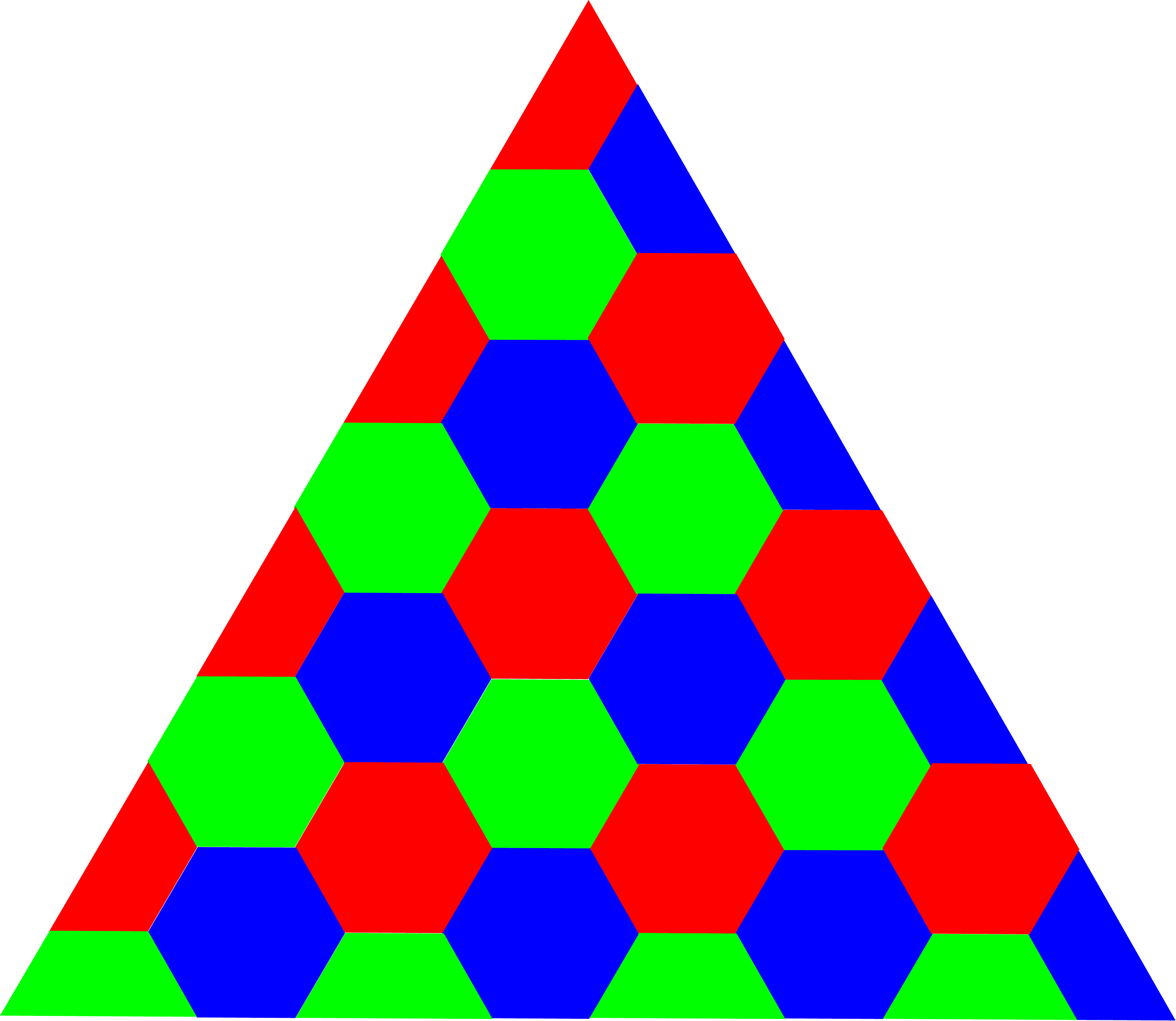}}
  \subfigure[\ 4.6.12 code]{\includegraphics[width=0.3\columnwidth]{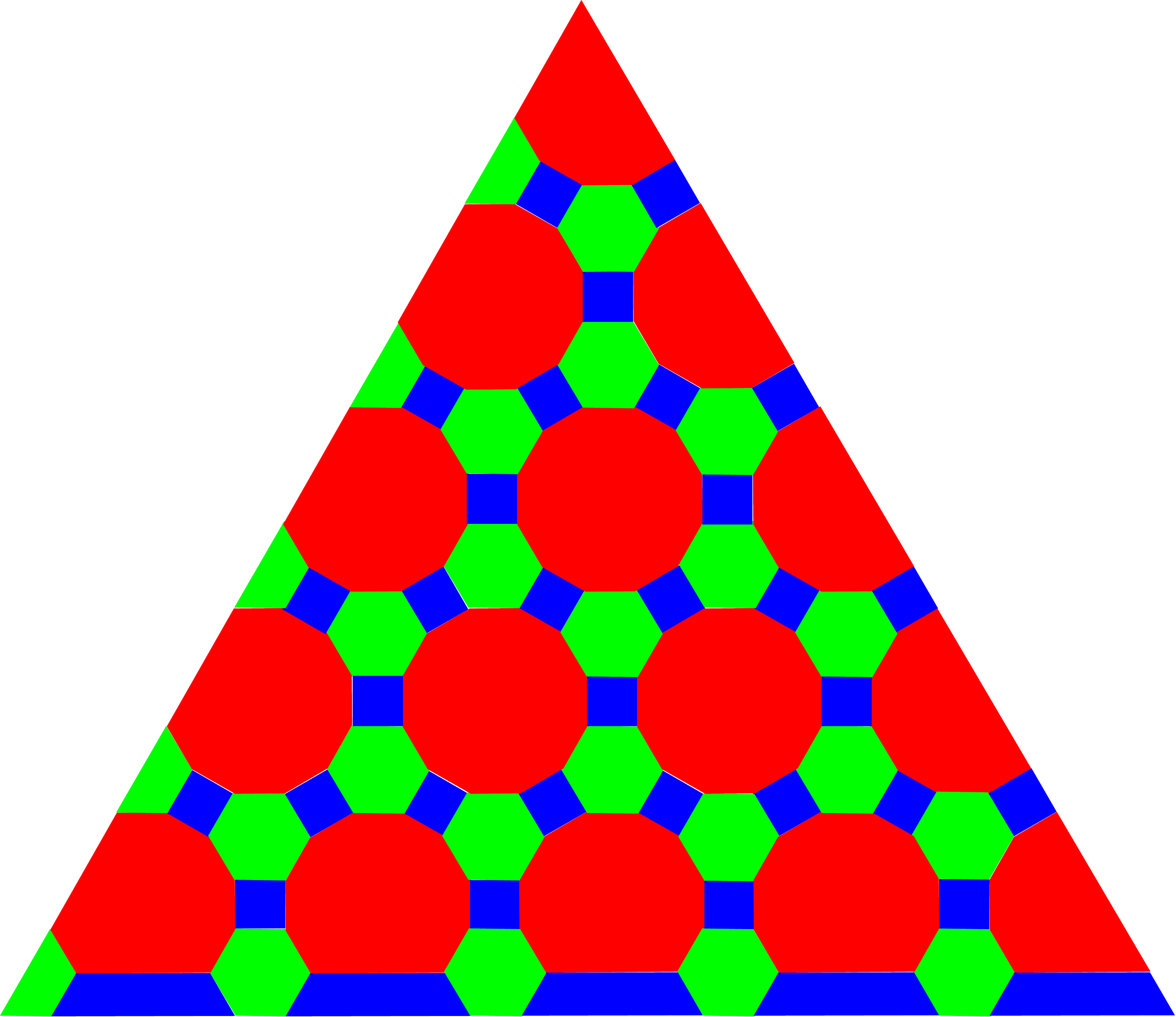}}
}
\caption{\small{\label{fig:triangular-codes}
Three distance $d=11$ triangular codes encoding one qubit, drawn from the $4.8.8$,
$6.6.6$, and $4.6.12$ lattices respectively.  For general $d$, these codes
have length $n$ equal to $\frac{1}{2}d^2 + d - \frac{1}{2}$, $\frac{3}{4}d^2
+ \frac{1}{4}$ and $\frac{3}{2}d^2 - 3d + \frac{5}{2}$ respectively. The
asymptotic ratio of $d^2$ to $n$ is highest for the 4.8.8 codes.}}
\end{figure}

We further restrict our analysis to \emph{planar} color codes, namely those
which are embedded in the disc (a sphere with one puncture).  We do this
because, for all quantum-computing technologies of which we are aware,
arranging qubits on a flat disc is more plausible than arranging them on a
more general surface like a torus.  The graph constraints defining color
codes require that planar color codes have a boundary shaped like a polygon
having $3m$ sides for some positive integer $m$.  A $3m$-sided planar color
code encodes $m$ logical qubits; we restrict attention to the simplest case
in which $m=1$.  In other words, our focus on this paper is on
\emph{triangular} color codes.  Examples of three different triangular color
codes are depicted in Fig.~\ref{fig:triangular-codes}.

\begin{figure}[h!]
\center{
  \subfigure[\ $d=3$]{\includegraphics[width=0.15\columnwidth]{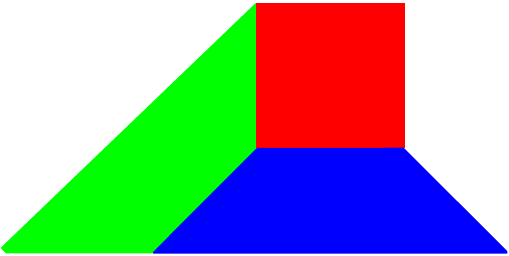}}
  \subfigure[\ $d=5$]{\includegraphics[width=0.3\columnwidth]{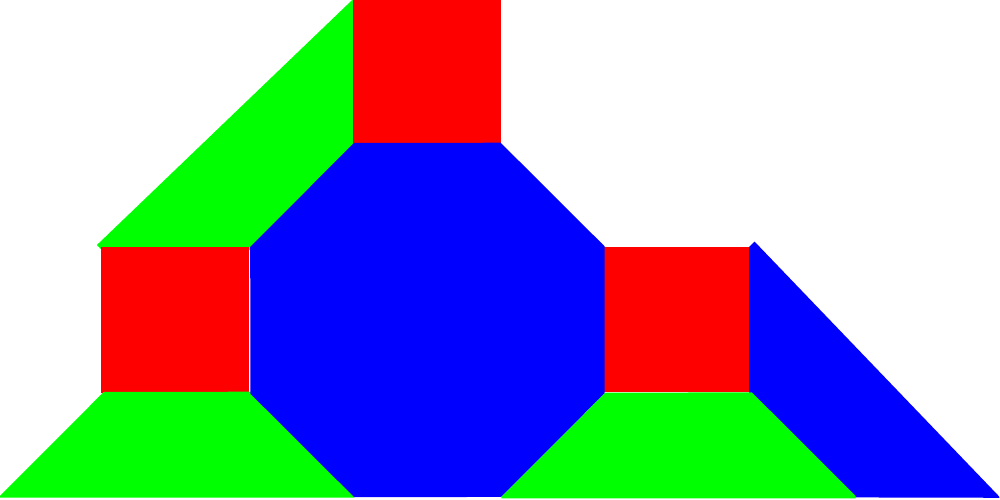}}
  \subfigure[\ $d=7$]{\includegraphics[width=0.4\columnwidth]{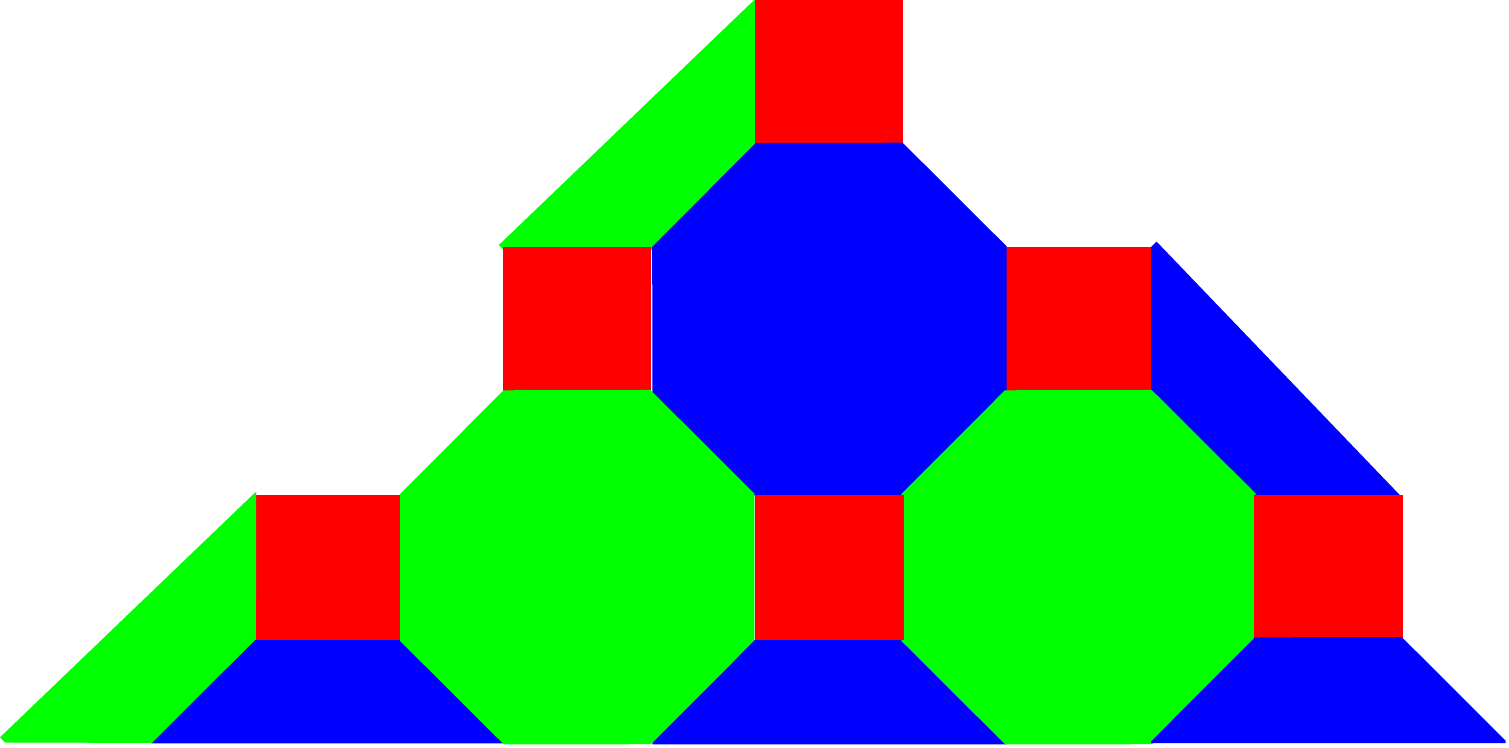}}
}
\caption{\small{\label{fig:4.8.8-triangular-codes}
4.8.8 color codes of sizes 3, 5, and 7.
}}
\end{figure}

The code distance of a triangular color code is equal to its side length,
namely the number of qubits along a side of the defining triangle.  To see
this, notice that the logical $X$ and $Z$ operators for the logical qubit
are transversal because they are encoded Clifford gates.  Thus, when one
multiplies a logical $X$ or $Z$ operator by all checks of the same Pauli
type, except the checks incident on a specified side, one obtains an
equivalent logical operator whose Pauli-weight is equal to the that side's
length.  The family of 4.8.8 triangular codes we study is generated
according to the pattern depicted in Fig.~\ref{fig:4.8.8-triangular-codes}.
Note that the smallest triangular code (for any of three triangular code
families depicted in Fig.~\ref{fig:triangular-codes}) is equivalent to the
well-known Steane $[\![7, 1, 3]\!]$ code \cite{Steane:1996a}; triangular
codes offer a way to generate an infinite code family from the Steane code
by a means other than concatenation \footnote{Similarly, the
three-dimensional color codes \cite{Bombin:2007b} offer a way to generate an
infinite code family from the fifteen-qubit Reed-Muller code by a means
other than concatenation.}.

\begin{figure}[h!]
\center{\includegraphics[width=0.8\columnwidth]{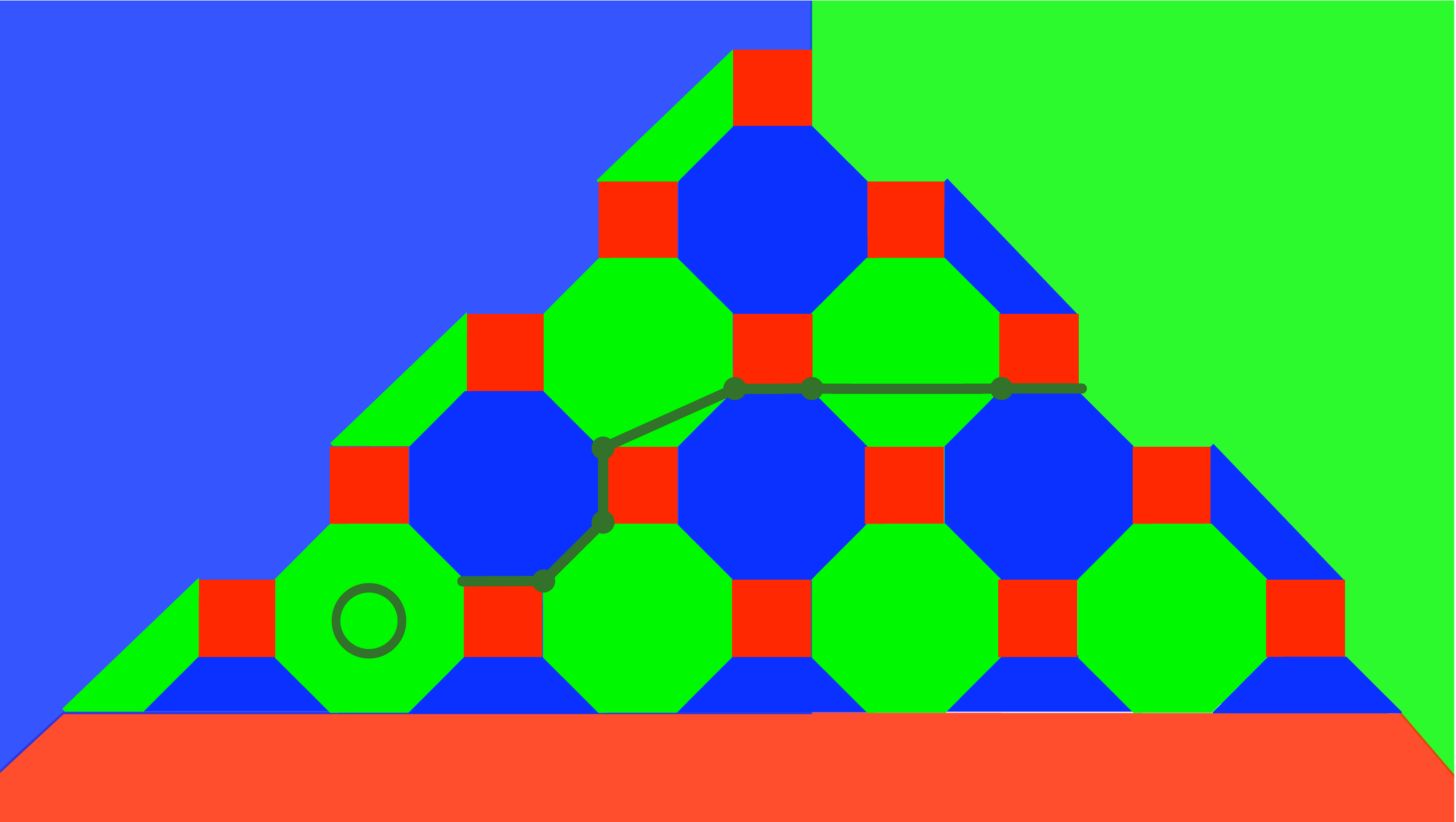}} 
\caption{\small{\label{fig:colored-chain}A green-colored chain in a
triangular code.  The chain connects a green-colored side of the 4.8.8
triangular code to a green octagonal face.  If qubits are flipped (are in
error) along this chain, it will only be detected by this terminal octagonal
check operator.}}
\end{figure}

Although the colors of the faces in a color code have no intrinsic meaning
for the algebraic structure of the code other than constraining the class of
graphs on which color codes are defined, it is useful to use the colors as
placeholders in discussions from time to time.  To that end, we will refer
to the colors of the faces as ``red,'' ``green,'' and ``blue.''  We will
further assign a color to each edge so that an edge's color is complementary
to the colors of the two faces upon which it is incident.  We will call a
set of vertices lying on a collection of edges of the same color connected
by faces also having that color a ``colored chain;'' an example of a colored
chain is depicted in Fig.~\ref{fig:colored-chain}.  We will assign colors to
each side of a triangular code so that the color of the side is 
complementary to the colors of the faces terminating on that side; for
example, in Figs.~\ref{fig:triangular-codes} and
\ref{fig:4.8.8-triangular-codes}, the left sides of the triangles are blue, the
right sides are green, and the bottoms are red.  These side colors are
indicated explicitly in Fig.~\ref{fig:colored-chain}.

%
\subsection{Syndrome extraction}

To record each error-syndrome bit, the relevant data qubits interact with
one or more ancilla qubits and the ancilla qubits are then measured.  Shor
\cite{Shor:1996a}, Steane \cite{Steane:1998a}, and Knill \cite{Knill:2004a}
have devised elaborate methods for extracting an error syndrome to minimize
the impact of ancilla-qubit errors spreading to the data qubits.  For
topological codes, however, such elaborate schemes are not necessary; a
single ancilla qubit per syndrome bit suffices.  This is because, by
choosing an appropriate order in which data qubits interact with the ancilla
qubit, the locality properties of the code will limit propagation of errors
to a constant-distance spread.  Using more elaborate ancillas is possible,
and in general there is a tradeoff in the resulting accuracy threshold one
must examine between the reduction in error propagation complexity 
offered versus the additional verification procedures required.  Here, we
examine the simplest case, with one ancilla qubit per syndrome bit.  By
placing two syndrome qubits at the center of each face $f$ (one for the
$X_f$ measurement and one for the $Z_f$ measurement), the syndrome
extraction process can be made spatially local, in keeping with the spirit
of the semiregular 2D geometry constraints we are imposing.

\begin{figure}[!h]
  \begin{tabular}{l}
    \Qcircuit @R=.2em @C=.52em {
& & \text{\small$\ket{+}$\hspace{1.3em}} & \ctrl{4} & \ctrl{3} & \ctrl{2} & \ctrl{1} & \measureD{M_X} \\
\push{\rule[-.29em]{0em}{1em}}& \qw & \qw & \qw & \qw & \qw & \targ & \qw & \qw \\
\push{\rule[-.29em]{0em}{1em}}& \qw & \qw & \qw & \qw & \targ & \qw & \qw & \qw \\
\push{\rule[-.29em]{0em}{1em}}& \qw & \qw & \qw & \targ & \qw & \qw & \qw & \qw \\
\push{\rule[-.29em]{0em}{1em}}& \qw & \qw & \targ & \qw & \qw & \qw & \qw & \qw 
    }
    \raisebox{-3.5em}{$\hspace{1em},\hspace{.7em}$}
    \Qcircuit @R=.2em @C=.52em {
& & \text{\small$\ket{0}$\hspace{1.3em}} & \targ & \targ & \targ & \targ & \measureD{M_Z} \\
\push{\rule[-.29em]{0em}{1em}}& \qw & \qw & \qw & \qw & \qw & \ctrl{-1} & \qw & \qw \\
\push{\rule[-.29em]{0em}{1em}}& \qw & \qw & \qw & \qw & \ctrl{-2}& \qw & \qw & \qw \\
\push{\rule[-.29em]{0em}{1em}}& \qw & \qw & \qw & \ctrl{-3} & \qw & \qw & \qw & \qw \\
\push{\rule[-.29em]{0em}{1em}}& \qw & \qw & \ctrl{-4} & \qw & \qw & \qw & \qw & \qw 
    }
  \end{tabular}
\caption{Six-step circuits for measuring $X^{\otimes 4}$ and $Z^{\otimes
4}$.\label{fig:synd-extract-circuit}}
\end{figure}
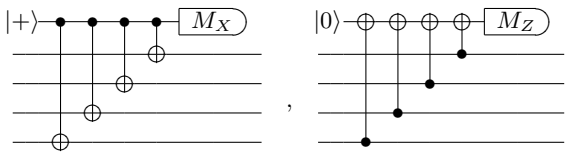

Because color codes are Calderbank-Shor-Steane (CSS) codes
\cite{Calderbank:1996a,Steane:1996b}, syndrome bits can be separated into
those which identify $Z$ errors (phase flips) and those which identify $X$
errors (bit flips).  These correspond to the bits coming from measuring the
$X_f$ and $Z_f$ operators respectively.  The circuit for measuring an
operator $X_f$ is identical to the one for measuring the operator $Z_f$, except
with the basis conjugated by a Hadamard gate; examples of bit-flip and
phase-flip extraction circuits for the square faces in the 4.8.8 color code
are depicted in Fig.~\ref{fig:synd-extract-circuit}.

\begin{figure}[!h]
\center{\includegraphics[width=0.8\columnwidth]{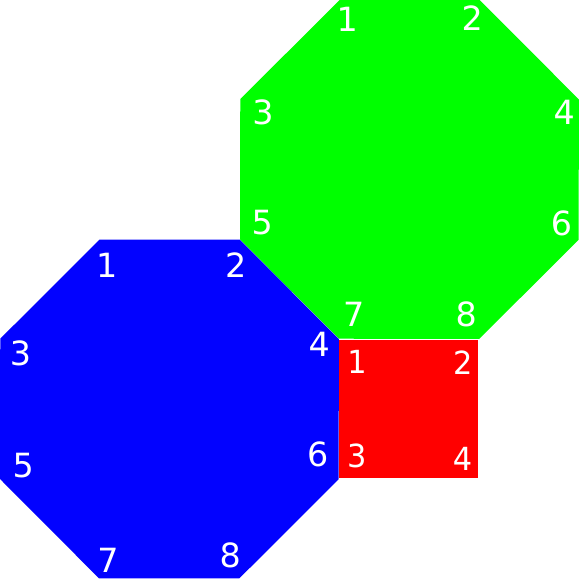}} 
\caption{Simple syndrome extraction circuit schedule. A round of $X$ checks
is followed by a round of $Z$ checks. The number at each vertex corresponds
to the discrete time step in which the physical qubit at that vertex
interacts with the syndrome qubit at the face's center via a $\CNOT$ gate.
The same schedule is used for both $X$ and $Z$ checks, but with the
direction of the $\CNOT$ gates reversed.\label{fig:naive-schedule}}
\end{figure}

In a full round of syndrome extraction, both $X_f$ and $Z_f$ must be
measured for each face $f$.  One way of scheduling this is to perform all
$X_f$ measurements in parallel followed by all $Z_f$ measurements in
parallel.  The minimal number of steps (ignoring preparation and
measurement) for parallel $X_f$ measurements is eight; an example of such a
schedule is depicted in Fig.~\ref{fig:naive-schedule}.  The $Z_f$
measurements can be carried out by the same schedule, but in the
Hadamard-conjugated basis as depicted in
Fig.~\ref{fig:synd-extract-circuit}.  A complete syndrome extraction round
using this schedule then takes 20 steps: 10 for the $X_f$ measurement and 10
for the $Z_f$ measurement.  For this schedule, one only needs to have one,
not two, syndrome qubits at the center of each face.

\begin{figure}[!h]
\center{\includegraphics[width=0.8\columnwidth]{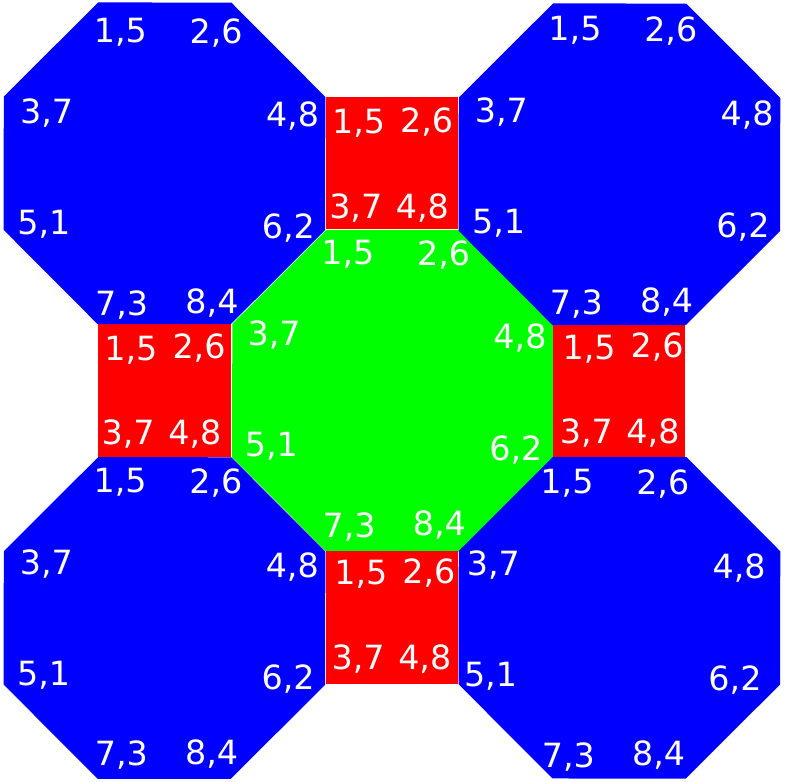}} 
\caption{Schedule with $X$ and $Z$ syndromes measured concurrently, in
``interleaved'' fashion.  This schedule takes $8$ steps, plus an extra step
for syndrome qubit preparations, plus an extra step for syndrome qubit
measurements. The label $m,n$ at a vertex indicates that at time step $m$
the qubit at that vertex interacts with the $X$-syndrome qubit via a $\CNOT$
gate and at time step $n$ the qubit at that vertex interacts with the
$Z$-syndrome qubit via a $\CNOT$ gate.\label{fig:interleaved-schedule}}
\end{figure}

The circuit for a full syndrome extraction round can be optimized to use
fewer time steps when both syndrome qubits in a face can be processed in
parallel.  An example of an ``interleaved'' schedule that uses ten steps is
depicted in Fig.~\ref{fig:interleaved-schedule}.

\begin{figure}[!h]
\center{\includegraphics[width=0.8\columnwidth]{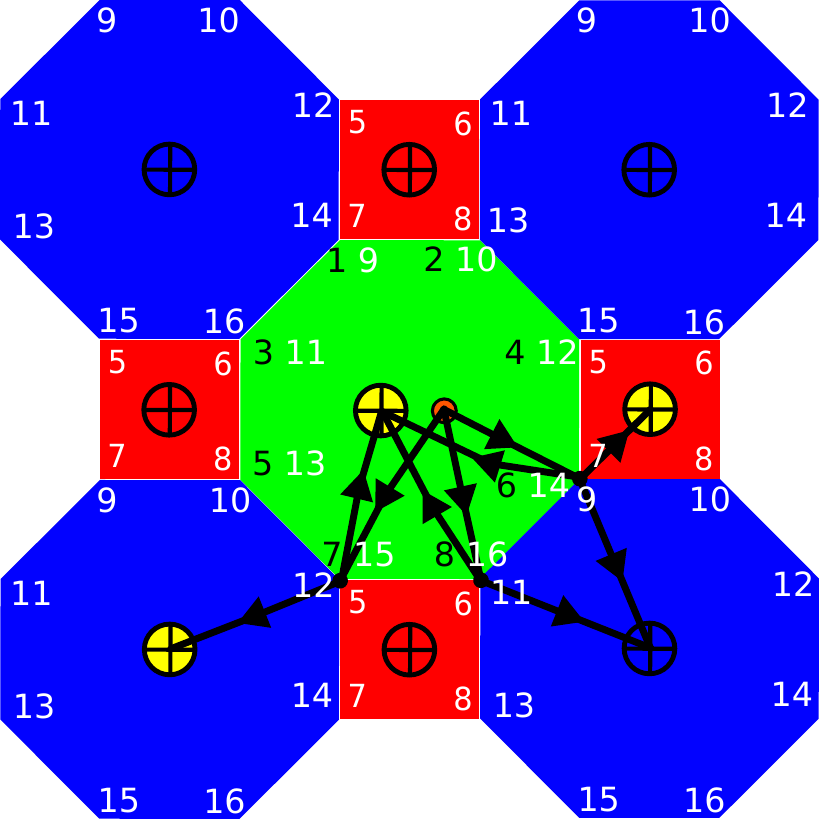}} 
\caption{A single $X$ error that occurs between time steps five and six on
the syndrome qubit for measuring $X^{\otimes 8}$, indicated by the small red
circle, will propagate to other $X$ errors according to the arrows.  Note
that an even number of $X$ flips is equivalent to no flip at all.  Errors
that propagate to other syndrome qubits will not propagate further because
the syndrome qubits are refreshed before each syndrome extraction round.
This particular error causes three data qubits to flip.  These flips are
correctly detected by the yellow-colored syndrome
bits.\label{fig:interleaved-schedule-error-propagation}}
\end{figure}

We calculate estimates for the accuracy threshold for both schedules, to
assess the impact of compressing the schedule.  Some authors who have
reported improved thresholds for concatenated-coding schemes using
Bacon-Shor codes attribute the improvement in large part to the simplicity
of the fault-tolerant Bacon-Shor-code syndrome-extraction circuit
\cite{Aliferis:2007a}.  For color codes, \textit{a priori}, it is not clear
that using a simpler syndrome-extraction circuit will yield an analogous
improvement.  This is because these circuits are not constructed using any
fault-tolerant design principles---catastrophic error propagation is halted
by the codes' structure, not by circuit-design principles.  It may be the
case, in fact, that a simpler circuit will allow errors to propagate to a
larger set of qubits than a less simple one.  The set of errors to which
individual errors are propagated by a syndrome-extraction circuit are called
``hooks'' in Ref.~\cite{Dennis:2002a}.  An example of how an error can
propagate to a ``hook'' using the schedule of
Fig.~\ref{fig:interleaved-schedule} is depicted in
Fig.~\ref{fig:interleaved-schedule-error-propagation}.

Neither the 20-step nor the 10-step schedule is necessarily optimal in the
sense of yielding the highest threshold for a fixed number of time steps; we
leave that optimization to others.  Indeed any schedule that satisfies two
constraints is valid:  (1) no qubit can be acted upon by two gates at the
same time and (2) any stabilizer generator for an error-free input state
(including ancilla syndrome qubits) must propagate to an element of the
stabilizer group for an error-free output state.  Satisfying this second
criterion is not trivial; for example, an ``obvious'' schedule that acts on
each face in a clockwise fashion in a manner obeying constraint (1) will not
satisfy constraint (2).

The number of steps in the syndrome extraction round can be reduced further
to eight steps if we prepare the ancillas for the octagon measurements not
in single-qubit states but in cat-states $(|0\>^{\otimes 8} + |1\>^{\otimes
8})/\sqrt{2}$ and use Shor's method of syndrome extraction \cite{Shor:1996a}.  (One
can also use four-qubit cat states and create an eight-step schedule, as
demonstrated in Ref.~\cite{Fowler:2008c}.)  Eight steps is the absolute
minimum possible for syndrome extraction, since each qubit must be checked
by six different syndrome bits, which must also be prepared and measured.
While using cat states reduces the circuit depth, the cat states need to be
verified.  We opted not to study this schedule because the verification is
stochastic, which would lead to a difficult synchronization problem for a
large-sized code.  That said, such a schedule has the potential to offer a
larger accuracy threshold.

Because there is an inherent asymmetry in the order in which we choose to
perform $X_f$ and $Z_f$ measurements, we will report two threshold results,
one for the $X_f$ measurements and one for the $Z_f$ measurements.  When we
only report one value, we are reporting the lower of the two threshold
values.  For the phenomenological noise model, we choose to model the $X_f$
and $Z_f$ syndrome extraction processes as occurring synchronously rather than
one followed by the other, since so many of the details of the circuit are
washed away in the model anyway.  This has the advantage of enabling the
accuracy threshold in the phenomenological model to be identified with a
phase transition in an associated random-bond Ising model, as described in
Ref.~\cite{Katzgraber:2009a}.  We will discuss this connection in more
detail in Secs.~\ref{sec:code-capacity-noise-model-threshold} and
\ref{sec:RBIM-conclusions}.

Finally, it is worth reminding that the entire syndrome extraction round is
repeated a number of times equal to the distance of the code when
measurements are allowed to be faulty, such as in the circuit-level and
phenomenological noise models that we study.  This ensures that errors in
the syndrome bit values can be suppressed as well as errors in the data
qubits can be suppressed.

%
\subsection{Decoding algorithm}

The process of \emph{decoding} refers to a classical algorithm for
identifying a recovery operation given an error syndrome, regardless of
whether the code from which the syndrome was derived is classical or
quantum.  Importantly, decoding does not refer to ``unencoding,'' or
performing the inverse of encoding.  For classical linear codes, the optimal
decoding algorithm is the Most Likely Error (MLE) algorithm, which
identifies the recovery operation to be the most likely pattern of bit-flip
errors given the syndrome.  In general, this algorithm is NP-hard
\cite{Berlekamp:1978a}, but there are many families of codes for which the
algorithm is known to be efficient.

For quantum stabilizer codes, MLE decoding identifies the recovery operation
to be the most likely $n$-qubit Pauli-group error given the syndrome.  (The
process of extracting the syndrome forces every error to ``collapse'' onto a
definite $n$-qubit Pauli-group operator, which is why it is sufficient to
restrict to this family of operators.) MLE decoding is not necessarily optimal
for quantum stabilizer codes.  This is because quantum error-correcting
codes can be \emph{degenerate}, meaning that two distinct correctable errors
can map to the same error syndrome.  Color codes are examples of highly
degenerate codes.  The optimal decoding algorithm for quantum stabilizer
codes instead identifies the recovery operation to be one that causes the
most likely logical operator to be applied after recovery.  This is akin to
a doctor prescribing medicine that is most likely to cure the ailment rather
than prescribing medicine that cures the most likely ailment.

Once a decoding algorithm has identified a recovery operation, which is some
$n$-qubit Pauli-group operator, it need not necessarily be applied.  Because
the process of applying the recovery operation is subject to faults, it is
wiser to wait until the end of the computation and apply the net recovery
operation rather than apply it after each decoding step.  One can even
propagate the correction past the final qubit measurements at the end of the
quantum computation, where the recovery operation becomes completely
classical and fault-free.  The catch is that one must (classically)
adaptively update one's ``Pauli frame'' after each decoding iteration by
permuting the interpretation of the Pauli operators $X$, $Y$, and $Z$ on
each qubit as suggested by the recovery operation.  (The Pauli operators get
conjugated by the Pauli error identified by the decoder.)

For fault-tolerant quantum error correction and a number of interesting
encoded quantum circuits, only Clifford gates are required.  Since Clifford
gates propagate Pauli operators to Pauli operators in the Heisenberg
picture, one can efficiently track the changing Pauli frame through these
gates, as guaranteed by the Gottesman-Knill theorem \cite{Gottesman:1997a}.
One can safely defer applying recovery operations until after final
measurement in each of these circuits.  However, for universal quantum
computation, at least one non-Clifford gate is required.  In our protocols,
the only such gate we use is the classically-controlled $S^\dagger$ gate,
depicted later in the circuit of Fig.~\ref{fig:T-circuit}.  Because this
gate propagates a Pauli error to a sum of Pauli errors, it is necessary to
actually apply the recovery operation before all but a constant number of
these gates in order to prevent the number of terms required to track one's
``Heisenberg frame'' from growing exponentially.

We develop MLE decoders for triangular 4.8.8 color codes for the three noise
model settings we study: code capacity, phenomenological, and circuit-based.
For the code-capacity and phenomenological settings, the only operations are
single-qubit measurements and identity gates.  This means that they involve
no circuitry that could map $X$ errors to $Z$ errors or vice-versa.  Because
of this, and because our noise model is one in which single-qubit operations
are subject to BP channel noise (which applies $X$ errors and $Z$ errors
independently), decoding can factor into bit-flip decoding and phase-flip
decoding separately.  Because color codes are also ``strong'' CSS codes
\cite{Preskill:1998b}, the MLE decoders for bit-flip and phase-flip errors
are in fact identical; for concreteness, we formulate the decoder for $Z_f$
syndrome bits here.

%
\subsubsection{Code capacity MLE decoder}

In the code-capacity setting, we have a single error-free $m$-bit syndrome $\mathbf{s}
= (s_1, \ldots, s_m)^T$ where $s_f = 0$ when $Z_f$ is measured to have
eigenvalue $+1$ and $s_f = 1$ when $Z_f$ is measured to have eigenvalue
$-1$.  (The value of $m$ is a function of the code size; for the triangular
$n$-qubit distance-$d$ 4.8.8 color code, $m = (d+1)^2/4-1$ and $n= (d+1)^2/2 -
1$.)  We assign a binary variable $x_v$ to each vertex $v$ indicating
whether or not the recovery operation calls for the qubit at vertex $v$ to
be bit-flipped (have Pauli $X$ applied).  The objective of MLE decoding is
to minimize the number of $x_v$ variables that are assigned the value $1$
subject to the constraint that the parity of the $x_v$ variables on each
face is consistent with the observed syndrome.  This can be expressed as the
following mathematical optimization problem:
\begin{align}
\label{eq:2D_optimization_problem}
\text{min}&\ \sum_v x_v \\
\text{sto}&\ \bigoplus_{v \in f} x_v = s_f \qquad \forall f \\
x_v &\in \BB := \{0, 1\}.
\end{align}
This optimization problem can be expressed as a linear binary integer
program (IP) over the finite field $GF(2)$
as follows:
\begin{align}
\label{eq:2D_GF2_LP}
\text{min}&\ \mathbf{1}^T \mathbf{x} \\
\label{eq:2D_GF2_LP_constraint}
\text{sto}&\ H\mathbf{x} = \mathbf{s} \bmod 2 \\
\mathbf{x} &\in \BB^n,
\end{align}
where $\mathbf{1}$ denotes the all-ones vector and $H$ is the parity check
matrix associated with the $Z_f$-checks.  (For color codes, this is the
face-vertex incidence matrix.)

To take advantage of well-developed numerical optimization software, it is
helpful to replace the linear algebra over $GF(2)$ in this mathematical
program with linear algebra over $\RR$.  One way to do this is to introduce
``slack variables'' into the optimization problem.  Because each check
operator in the code has Pauli weight four or Pauli weight eight, each row
of $H$ has Hamming weight four or Hamming weight eight.  This means that the
$f$th component of the vector on the left hand side of constraint
(\ref{eq:2D_GF2_LP_constraint}) is a sum of four or eight binary $x_v$
variables that must equal $s_f$ modulo 2.  The modulo 2 restriction can be
dropped by replacing $\mathbf{s}$ by $\mathbf{s} + 2\mathbf{z}_1 +
4\mathbf{z}_2 + 8\mathbf{z}_3$ in the constraint, where the $\mathbf{z}_i$
are binary ``slack variable'' vectors that allow the LHS to sum to any
integer from $0\ldots 15$.  While there can be many degenerate solutions to
this revised optimization problem having different $\mathbf{z}_i$ values,
any solution generates the same optimal $\mathbf{x}$ as before.  By
combining the $\mathbf{z}_i$ variables and the $\mathbf{x}$ variables into a
single vector $\mathbf{y} = (\mathbf{x}^T, \mathbf{z}_1^T, \mathbf{z}_2^T,
\mathbf{z}_3^T)^T$, the slack-variable version of the program becomes the
following linear binary integer program in which the variables are
restricted to be binary but in which the linear algebra is over $\RR$:
\begin{align}
\text{min}&\ \mathbf{c}^T \mathbf{y} \\
\text{sto}&\ A\mathbf{y} = \mathbf{s} \\
\mathbf{y} &\in \BB^n, \end{align}
where $c$ is a vector containing $n$ ones followed by $3m$ zeros and $A$ is
the matrix generated by adjoining matrices to $H$ as 
\begin{align}
\label{eq:A-matrix1}
A := \begin{pmatrix}
  H\, |\, {-}2I\, |\, {-}4I\, |\, {-}8I
     \end{pmatrix},
\end{align}
in which each $I$ denotes the $m \times m$ identity matrix.

There are a number of symmetries that color codes possess which allow one to
significantly reduce the complexity of this binary IP.  For example, if
$\mathbf{y}$ satisfies the constraints of the IP, then so does $\mathbf{y}$
with any number of faces complemented.  Since complementing the face of any
optimal solution will not reduce its weight, we know that each face's sum
will never be more than half the weight of that face.  This means that for
any particular instance of the IP specified by the syndrome vector
$\mathbf{s}$, the sums for the octagon and square faces can only take the
syndrome-dependent values listed in Table~\ref{tab:2D-IP}, thereby reducing
the number of slack variables required.  We take advantage of these kind of
symmetries in the software we developed code for estimating the code capacity of 4.8.8
triangular color codes.  For example, we never need to use three slack
variables and some times we need none at all.
\begin{table}[ht]
\centering
\begin{tabular}{c|c|c} \hline \hline
 & Octagon & Square \\ [0.5ex] \hline
$s=0$ & 0, 2, 4 & 0, 2 \\
$s=1$ & 1, 3 & 1 \\ [1ex] \hline
\end{tabular}
\caption{Possible values octagonal and square face check sums can take for an
optimal IP solution if the face check sum's parity $s$ is fixed.}
\label{tab:2D-IP}
\end{table}

Maximum likelihood decoding is generally an NP-hard problem, and the color
codes do not appear to fall into an ``easy'' subset of instances.  This is
unfortunate because their close cousins, the surface codes, do have
efficient MLE decoders that can be solved as a minimum-weight perfect
matching problem \cite{Dennis:2002a}.  Nevertheless, we can solve the
associated IP for reasonably small instance sizes.

%
\subsubsection{Phenomenological noise MLE decoder}

\begin{figure}[!h]
\center{
  \subfigure[\ Measurement Error.]{\includegraphics[width=0.4\columnwidth]{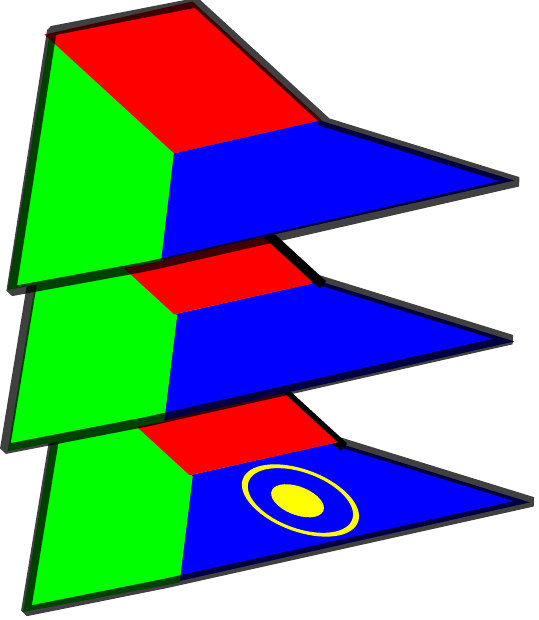}}
  \subfigure[\ Data Error.]{\includegraphics[width=0.4\columnwidth]{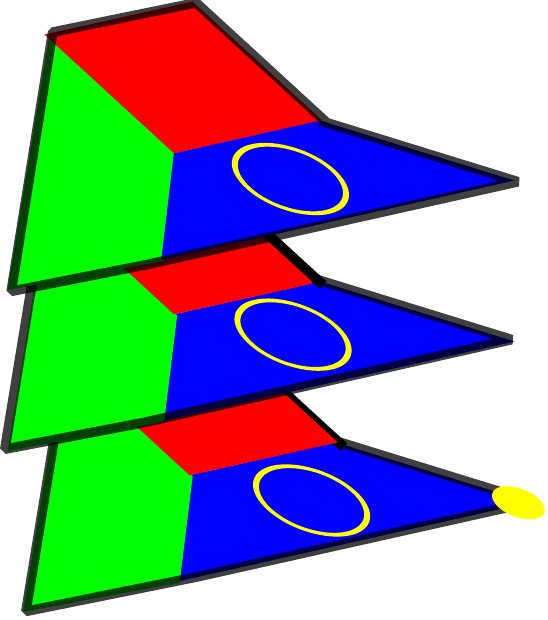}}
}
\caption{If syndrome qubits are also allowed to be in error, we repeat
syndrome measurements. Time advances from bottom to top.  Yellow circles
indicate syndrome bits with the value 1.  Solid yellow circles indicate
bit-flip errors.\label{fig:3D-errors}}
\end{figure}

In the phenomenological noise model, the syndrome values themselves can be
faulty so we repeat the syndrome extraction process a number of times equal
to the distance of the code.  In this setting, it is the \emph{difference}
in syndrome bit values from one time step to the next rather than the
absolute values at particular times step that indicate data errors.  This is
because a single data error at one time step will lead to flipped syndrome
bits for all future time steps (assuming that the syndrome extraction is not
faulty), and such a syndrome-bit history should not imply that data errors
occurred at each time step---it should imply that a data error occurred only
at the time step when the syndrome bit first changed its value.  The
difference in persistence between data and syndrome errors is depicted in
Fig.~\ref{fig:3D-errors}.  The input to a MLE decoder is therefore the
collection of syndrome \emph{difference} vectors for all time steps, namely
\begin{align}
\Delta \mathbf{s}_t = \mathbf{s}_t - \mathbf{s}_{t-1} = (\mathbf{s}_t +
\mathbf{s}_{t-1}) \bmod 2 \qquad \forall t,
\end{align}
where $\mathbf{s}_0 := \mathbf{0}$.

For a distance $d$ color code, the optimization problem to solve is again to
minimize the number of errors given the observed syndrome, except we now
have $d$ time steps' worth of data-error vectors, $\mathbf{x}_1, \ldots,
\mathbf{x}_d$, and $d$ time steps' worth of syndrome-error vectors,
$\mathbf{r}_1, \ldots, \mathbf{r}_d$, as variables in the optimization
problem.  Mathematically, we can write the optimization problem as
\begin{gather}
\text{min}\ \sum_t \mathbf{1}^T \mathbf{x}_t \\
\label{eq:3D-constraints}
\text{sto}\ (H\mathbf{x}_t + \mathbf{r}_t + \mathbf{r}_{t-1}) \bmod 2 =
\Delta \mathbf{s}_t \bmod 2 \quad \forall t \\
\mathbf{x} \in \BB^n.
\end{gather}

As we did for the code-capacity scenario, we can collect these constraints
into a single constraint and add slack variables to make the problem a
linear binary IP over the reals.  Because the left-hand side of the
constraints in Eq.~(\ref{eq:3D-constraints}) can sum to up to ten for octagon
constraints and up to six for square constraints, three slack variables
again suffice, allowing us to formulate the optimization problem as
\begin{align}
\text{min}&\ \mathbf{c}^T \mathbf{y} \\
\text{sto}&\ A\mathbf{y} = \Delta \mathbf{s} \\
\mathbf{y} &\in \BB^n, \end{align}
where $\mathbf{c}$ is a vector containing $(n+m)d$ ones followed by $3md$
zeros, $\Delta \mathbf{s}$ is the vector $(\Delta \mathbf{s}_1^T, \ldots,
\Delta \mathbf{s}_d^T)^T$,  $\mathbf{y}$ is the vector $(\mathbf{x}_1^T,
\ldots, \mathbf{x}_d^T, \mathbf{r}_1^T, \ldots, \mathbf{r}_d^T,
\mathbf{z}_1^T, \mathbf{z}_2^T, \mathbf{z}_3^T)^T$ and $A$ is the matrix
\begin{align}
\label{eq:A-matrix2}
A &= \left(\!\!\!\begin{array}{c|c|c|c|c}
  \begin{matrix}
    H &   &        &   \\
      & H &        &   \\
      &   & \ddots &   \\
      &   &        & H 
  \end{matrix}
  &
  \begin{matrix}
    I &   &        &   & \\
    I & I &        &   & \\
      &   & \ddots &   & \\
      &   &        & I & I 
  \end{matrix}
  &
  -2I & -4I & -8I 
\end{array}\!\!\! \right).
\end{align}

Finally, as we did for the code capacity setting, we can use symmetries to
reduce the complexity of solving this IP; Table~\ref{tab:3D-IP} summarizes
what the possible values are for the square-faced and octagonal-faced
constraints.
\begin{table}[ht]
\centering
\begin{tabular}{c|c|c} \hline \hline
 & Octagon & Square \\ [0.5ex] \hline
$s=0$ & 0, 2, 4, 6 & 0, 2, 4 \\
$s=1$ & 1, 3, 5 & 1, 3 \\ [1ex] \hline
\end{tabular}
\caption{Possible values octagonal and square face check sums can take for an
optimal IP solution if the face check sum's parity $s$ is fixed.}
\label{tab:3D-IP}
\end{table}

%
\subsubsection{Circuit-level decoder}

In the circuit-level noise model, each component of the syndrome extraction
circuit can fail with a probability that is a function of a parameter $p$,
so that the overall probability of a syndrome bit being in error, $p_s$, is
a complicated function of $p$.  Even more dauntingly, the circuits can
induce correlated errors between syndrome bits and between syndrome bits and
data qubits.  The phenomenological-noise model does not capture these noise
correlations.

We developed an MLE decoder for the circuit-level noise model that 
accounts for both these induced error correlations and the fact that in this
noise model, single-qubit operations are subject to BP-channel noise while
$\CNOT$ gates are subject to DP-channel noise.  However, this decoder uses
exponentially many more constraints than the phenomenological decoder as a
function of code size.  Because the IP decoder is already NP-hard, we opted
not to study this truly MLE decoder but rather use the
phenomenological-noise MLE decoder, which ignores these subtleties.  Taking
correlations into account will likely boost the accuracy threshold, but
probably not by large factors \cite{Fowler:2011a}.  By way of comparison,
the threshold for the square-lattice surface code in the circuit-level noise
model is $0.68\%$ when the phenomenological decoder is used
\cite{Harrington:2008a} ($0.75\%$ \cite{Raussendorf:2007a} when using a
non-MLE decoder that takes into account some entropic effects), a threshold
value that has recently been boosted to $1.1\%$ \cite{Wang:2011a} by
accounting for some of the correlations in the noise.  We leave the
refinement of true MLE decoding of this noise model to others.

%
\section{Numerical estimate of the accuracy threshold for fault-tolerant
quantum error correction}
\label{sec:numerical-estimate-of-the-accuracy-threshold}

%
\subsection{Code capacity noise
model\label{sec:code-capacity-noise-model-threshold}}

Because the $[\![n, 1, d]\!]$ triangular 4.8.8 color codes are CSS codes,
when they are subject to BP-channel noise of strength $p$, their code
capacity is the same as their bit-flip or phase-flip capacity; we focus on
the bit-flip capacity here for definiteness.  The number of distinct
bit-flip syndromes is $2^{(n-1)/2}$ and the number of distinct bit-flip
errors is $2^n$.  For small $n$, one can pre-solve the MLE decoding IP for each
of the $2^{(n-1)/2}$ distinct bit-flip syndromes.  One can then iterate
through each of the $2^n$ distinct error patterns, compute its syndrome, and
determine whether the combination of the error pattern plus the inferred
correction by the IP leads to a logical operator, indicating failure of the
decoding algorithm.  Since error-correction is assumed to be error-free in
this noise model, the corrected state is guaranteed to be in the codespace.
Because ($a$) the logical bit-flip operator is transversal, ($b$) all
stabilizer group elements have even weight, and ($c$) there are an odd
number of qubits in every triangular code, it follows that one can identify
a decoding failure quickly by computing whether the parity of the error
pattern equals the parity of its IP-inferred correction; this means that is
suffices to just store the parity of the inferred correction for each
pre-computed IP instance.  The probability of failure, $p_{\text{fail}}$ is
therefore
\begin{align}
p_{\text{fail}}
  &= \sum_{\text{failing patterns $E$}} p^{|E|}(1-p)^{n - |E|},
\end{align}
where $|E|$ denotes the Hamming weight of the bit-flip error pattern $E$.

We carried out this tabulation for the smallest triangular 4.8.8 color codes
of distances 1, 3, 5, and 7 (corresponding to 1, 7, 17, and 31 qubits
respectively) and computed the corresponding exact polynomials.  To speed up
the computation, we used several symmetries.  For example, it suffices to
examine only half of the error patterns because if the decoding algorithm
succeeds on an error pattern, it fails on its complement and vice versa.
Also, up to overall complementation, every error pattern can be uniquely
expressed as the modulo-2 sum of an IP-inferred minimal-weight error pattern
and a pattern where a bit-flip stabilizer group element has support.
Finally, the decoding algorithm is guaranteed to work on all errors whose
weight is less than the code's distance, so those error patterns do not need
to be examined.

The formulas we obtained for the smallest codes of distance $1$, $3$, and $5$
(code sizes 1, 7, and 17) are:
\begin{align}
p_{\text{fail}}^{(1)} &= p \\
p_{\text{fail}}^{(3)} &= p^7 + 7p^6(1-p) + 28p^4(1-p)^3 \\
  &\phantom{= }+ 7p^3(1-p)^4 + 21p^2(1-p)^5 \\
p_{\text{fail}}^{(5)} &=
 p^{17} +
 17p^{16}(1-p) +
 136p^{15}(1-p)^2  \\
 &\phantom{= }+
 348p^{14}(1-p)^3 +
 725p^{13}(1-p)^4  \\
 &\phantom{= }+
 3861p^{12}(1-p)^5 +
 4764p^{11}(1-p)^6 \\
 &\phantom{= }+
 12136p^{10}(1-p)^7 +
 9747p^9(1-p)^8  \\
 &\phantom{= }+
 14563p^8(1-p)^9 +
 7312p^7(1-p)^{10} \\
 &\phantom{= }+
 7612p^6(1-p)^{11} +
 2327p^5(1-p)^{12} \\
 &\phantom{= }+
 1655p^4(1-p)^{13} +
 332p^3(1-p)^{14}.
\end{align}

The formula we obtained for the distance-7 triangular 4.8.8 color code (31
qubits) is a bit more hefty:

\begin{widetext}
\begin{align}
p_{\text{fail}}^{(7)} &=
  p^{31} + 
  31       p^{30} (1-p) + 
  465      p^{29} (1-p)^2 +
  4495     p^{28} (1-p)^3 + 
  25658    p^{27} (1-p)^4 +
  96790    p^{26} (1-p)^5 \\ \nonumber
 &\phantom{= }+ 
  344858   p^{25} (1-p)^6 +
  1288630  p^{24} (1-p)^7 + 
  3742943  p^{23} (1-p)^8 +
  10488241 p^{22} (1-p)^9 \\ \nonumber
 &\phantom{= }+
  21436239 p^{21} (1-p)^{10} +
  44259329 p^{20} (1-p)^{11} +
  67781868 p^{19} (1-p)^{12} +
  106951476p^{18} (1-p)^{13} \\ \nonumber
 &\phantom{= }+
  127137964p^{17} (1-p)^{14} +
  155845748p^{16} (1-p)^{15} +
  144694447p^{15} (1-p)^{16} +
  138044561p^{14} (1-p)^{17} \\ \nonumber
 &\phantom{= }+
  99301599 p^{13} (1-p)^{18} +
  73338657 p^{12} (1-p)^{19} +
  40412986 p^{11} (1-p)^{20} +
  22915926 p^{10} (1-p)^{21} \\ \nonumber
 &\phantom{= }+
  9671834  p^9  (1-p)^{22} +
  4145782  p^8  (1-p)^{23} +
  1340945  p^7  (1-p)^{24} +
  391423   p^6  (1-p)^{25} +
  73121    p^5  (1-p)^{26} \\ \nonumber
 &\phantom{= }+
  5807     p^4  (1-p)^{27}.
\end{align}
\end{widetext}

Our computing resources did not allow us to compute the exact polynomial for
the next-sized code (distance 9 code on 49 qubits), so we resorted to a
Monte Carlo estimate for $p_{\text{fail}}(p)$.  We did this by first
selecting three values of $p$ near where we believed the threshold to be.
For each $p$, we generated $N$ trial error patterns drawn from the Bernoulli 
distribution, namely in which we applied a bit-flip on each of the $n$
qubits with probability $p$.  We then inferred the syndrome for each error
pattern and checked whether or not it led to a decoding failure for the MLE
decoder.  The optimal unbiased estimator for $p_{\text{fail}}$ that we used is
\begin{align}
\label{eq:p-fail-estimator-mean}
p_{\text{fail}}^{(\text{est})} = \frac{N_{\text{fail}}}{N}
\end{align}
with a variance of
\begin{align}
\label{eq:p-fail-estimator-variance}
({\sigma^2_{\text{fail}}})^{(\text{est})} =
\frac{p_{\text{fail}}^{(\text{est})}\left(1-p_{\text{fail}}^{(\text{est})}\right)}{N}.
\end{align}

\begin{figure}[!h]
\center{\includegraphics[width=\columnwidth]{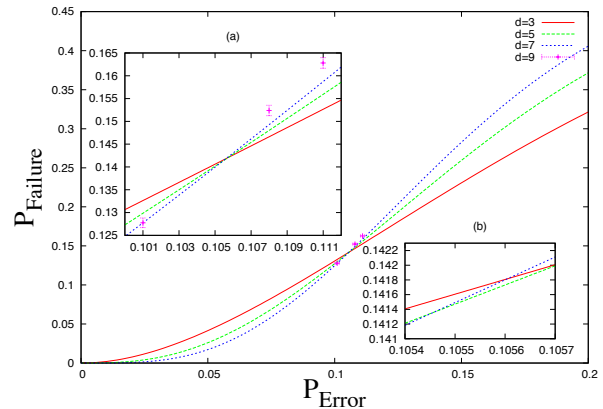}} 
\caption{Code capacity for the 4.8.8 triangular color codes.  $p_{th} =
10.56(1) \%$.  Error bars on Monte Carlo data reflect  $10^5$ instances
studied at each of the three corresponding values of $p$.  The inset figures
are zoom-ins near the crossing point to show greater resolution
there.\label{fig:code-capacity}}
\end{figure}

To get reasonably small error bars in these estimates, given where we
believed the threshold to be, we chose $N = 10^5$.  The polynomials for
$p_{\text{fail}}(p)$ are plotted in Fig.~\ref{fig:code-capacity}, including
our three points of Monte Carlo data.  From these plots, we estimate the
accuracy threshold for this noise model to be $10.56(1)\%$.  The error we
report in this value comes from the error analysis method we describe in
detail in the next section.

To put our result in context, we reference Table~\ref{tab:code-thresholds}.
The threshold value of 10.56(1)\% we find is is slightly higher than the
corresponding MLE threshold for the code capacity 10.31(1)\% of 4.4.4.4
surface codes.  Intuitively this makes sense, as the 4.8.8 color code has
both weight-8 and weight-4 stabilizer generators, both of which are modeled
as being measured instantaneously and ideally.  Being able to measure
high-weight generators quickly should improve the performance of a code,
which is the effect we observe.

\begin{figure}[htb]
\center{\includegraphics[width=0.9\columnwidth]{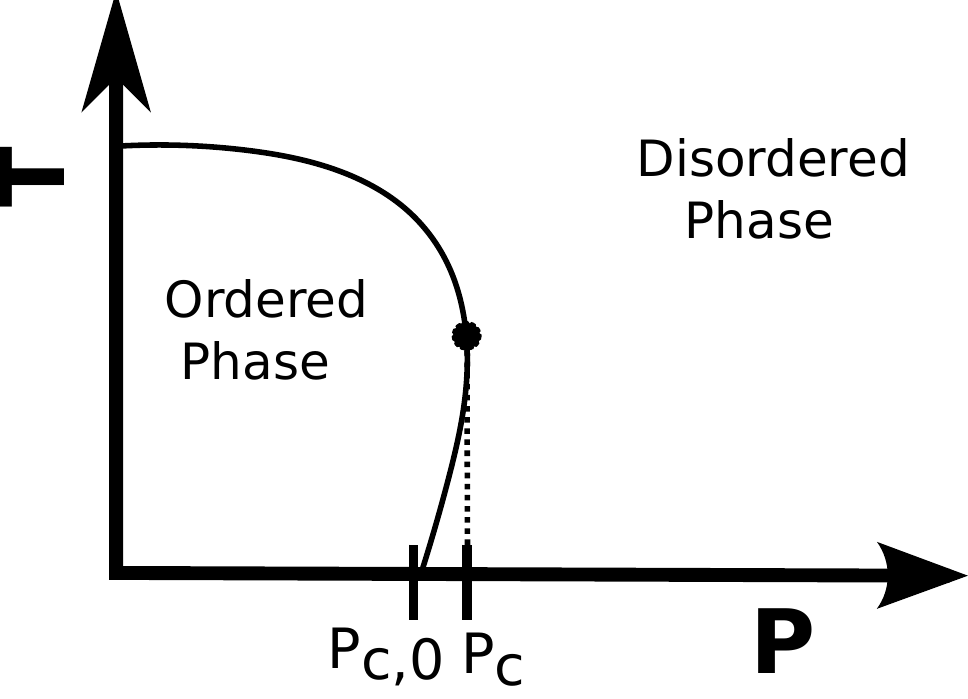}}
\caption{\label{fig:Nishimori} Phase diagram for 3-body random-bond Ising
model.  The dark circle is called the Nishimori point. The dotted line is
the expected phase boundary given by the Nishimori conjecture. Our value of
code capacity ($10.56(1)\%$) establishes that the $T=0$ intercept is
$P_{c,0}$, while results of Ohzeki \cite{Ohzeki:2009b} ($10.925(5)\%$)
establish that the Nishimori point occurs at $P_{c}$.  Because $P{c} \neq
P_{c,0}$, the Nishimori conjecture for this model is false.}
\end{figure}

Our threshold is also less than the threshold value of 10.925(5)\% for
optimal decoding, which is also not surprising.  As with the 4.4.4.4 surface
codes, the reduction in threshold is not very significant.  For both the
surface codes and the 4.8.8 color codes, the accuracy threshold in the code
capacity noise model corresponds to a phase transition in a random-bond
Ising model (RBIM) of classical spins \cite{Dennis:2002a, Katzgraber:2009a}.
For the color codes, the Ising model features 3-body interactions, whereas
for the surface codes, the Ising model features 2-body interactions.  The
MLE decoder in both settings corresponds to the order-disorder transition in
the spin model at zero temperature, whereas the optimal decoder corresponds
to the order-disorder transition at the temperature along the so-called
``Nishimori line,'' where the randomness in the bond couplings equals the
randomness in the state arising from finite temperature fluctuations.  In
both the surface-code and color-code settings, the small decrease in
accuracy threshold when going from optimal to MLE decoding reflects that the
phase-boundary in these models is re-entrant, but only by a small amount.
Our results therefore imply a violation of the so-called Nishimori
conjecture \cite{Nishimori:1981a, Nishimori:1986a}, which conjectures that
the spin model shouldn't become more ordered as the temperature increases.
The violation that our results imply is depicted in cartoon fashion in
Fig.~\ref{fig:Nishimori}.  To our knowledge, the violation of the Nishimori
conjecture for the 3-body RBIM is unknown before our work.  We expand more
on this connection in Sec.~\ref{sec:RBIM-conclusions}.

%

%
\subsection{Phenomenological noise
model\label{sec:phenomenological-noise-model-threshold}}

In the phenomenological noise model, our fault-tolerant quantum error
correction protocol repeats syndrome extraction multiple times to increase
the reliability of the syndrome bits.  This causes the number of possible
error patterns for a given code size to grow so rapidly that obtaining exact
curves for $p_{\text{fail}}(p)$ even for small code sizes is intractable.
We therefore resorted to Monte Carlo estimates for these curves for even the
smallest code sizes.  The specific Monte Carlo algorithm we used for
computing $p_{\text{fail}}$ at a fixed value of $p$ is listed in Algorithm
\ref{alg:MC-estimator}.

In words, Algorithm \ref{alg:MC-estimator} creates an estimator for
$p_{\text{fail}}$ by assessing the performance of many simulated trials of
faulty quantum error correction.  In each trial, errors are laid down,
giving rise to an observed syndrome history.  From the syndrome history, a
correction is inferred.  The actual error history and the inferred error
history are XORed onto a single effective time slice, but the state in this effective
time slice is not necessarily in the codespace.  To achieve this, a
fictional ideal (error-free) round of error correction is simulated.  If
this succeeds (\ie, if it does not generate a logical bit-flip operation),
then the trial is deemed a success; otherwise it is deemed a failure.  By
repeating many trials, one obtains an optimal unbiased estimator for the
failure probability $p_{\text{fail}}$, with mean and variance given by
Eqs.~(\ref{eq:p-fail-estimator-mean}--\ref{eq:p-fail-estimator-variance}),
identical to the formulas relevant in the code capacity noise model setting.

\renewcommand{\algorithmiccomment}[1]{// \textsl{#1}}

\algsetup{indent=1em}
\begin{algorithm}[h!]

  \caption{: $p_{\text{fail}}(p)$ by Monte Carlo\label{alg:MC-estimator}}

  \begin{algorithmic}[1]
  
    \medskip

    \STATE $n_{\text{faces}} \leftarrow \tfrac{1}{4}(d+1)^2 - 1$.

    \FOR{ $i = 1$ \TO $N$}

      \medskip

      \STATE \COMMENT{Generate data and syndrome errors for $d$ time slices.}

      \FOR{$t = 1$ \TO $d$}

        \FOR{$j = 1$ \TO $n$}
 
          \STATE $E[t,j] \leftarrow 1$ with probability $p$. \COMMENT{Data errors.}

        \ENDFOR

        \FOR{$j = n+1$ \TO $n + 1 + n_{\text{faces}}$}

          \STATE $E[t,j] \leftarrow 1$ with probability $p$.  \COMMENT{Synd. errors.}

        \ENDFOR

      \ENDFOR

      \medskip

      \STATE $E_{\text{min}} \leftarrow \text{Decode}(\text{Syndrome}(E))$.
\COMMENT{3D error volume.}

      \STATE $E' \leftarrow \bigoplus_{t} E[t] \oplus E_{\text{min}}[t]$. \COMMENT{2D error plane.}
      
      \STATE $E'_{\text{min}} \leftarrow \text{Decode}(\text{Syndrome}(E'))$.
\COMMENT{Ideal decoding}.

      \IF{ $(\bigoplus_i E'[i] \oplus E'_{\text{min}}[i] = 1)$ }

        \STATE $N_{\text{fail}} \leftarrow N_{\text{fail}} + 1$.

      \ENDIF

    \medskip

    \ENDFOR

    \RETURN $p_{\text{fail}}^{(\text{est})} = N_{\text{fail}}/N$.

  \end{algorithmic}

\end{algorithm}

\begin{figure}[!h]
\center{\includegraphics[width=\columnwidth]{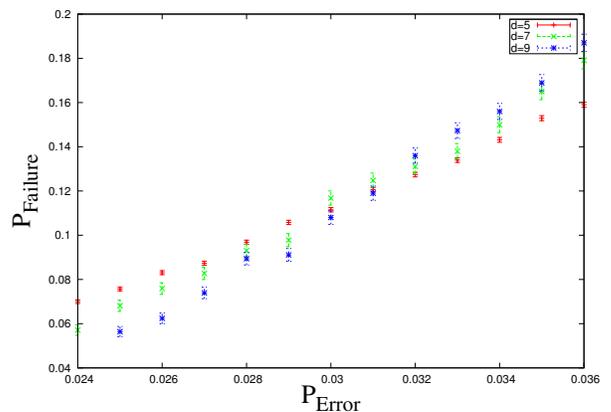}} 

\caption{Monte Carlo data used to estimate the accuracy threshold in the
phenomenological noise model.\label{fig:phenom-plots}}
\end{figure}

Our plots of $p_{\text{fail}}$ versus $p$ for small-distance color codes are
depicted in Fig.~\ref{fig:phenom-plots}.  Just as for surface codes, the
phenomenological noise MLE decoder can be mapped to a random-plaquette gauge
model (RPGM) on classical spins such that the zero-temperature
order-disorder phase transition in the spin model corresponds to the
accuracy threshold of the color codes.  Because of this, as argued in
Ref.~\cite{Wang:2003a}, the mutual intersection of the curves in
Fig.~\ref{fig:phenom-plots} at the threshold $p_{c}$ corresponds to critical
behavior in the spin model such that the spin correlation length $\xi$
scales as
\begin{align}
\xi \sim |p - p_{c}|^{-\nu_0},
\end{align}
where $\nu_0$ is a critical exponent set by the universality class of the
spin model.

For a sufficiently large code distance $d$, then, the failure probability
should scale as
\begin{align}
p_{\text{fail}} = (p - p_{c})d^{1/\nu_0}.
\end{align}

We use our Monte Carlo data to fit to this form, but as in
Ref.~\cite{Wang:2003a}, we allow for systematic corrections coming from
finite-size effects that create a constant offset.  Specifically, we use the
method of differential corrections \cite{Pezzullo:2011a} to fit the curves
to the form
\begin{align}
\label{eq:pf-pc-curve-fit}
p_{\text{fail}} = A + B(p - p_c)d^{1/\nu_0}.
\end{align}

\begin{figure}[!h]
\center{\includegraphics[width=\columnwidth]{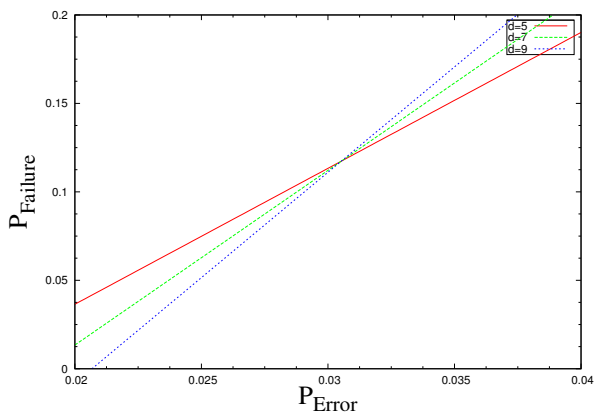}}
\caption{Linear fit near curve crossings of phenomenological-noise-model
Monte Carlo data.  Estimated accuracy threshold is $p_{\textit{th}} =
3.05(4)\%$.\label{fig:phenom-line-fits}} \end{figure}

The linear fits to our data are plotted in Fig.~\ref{fig:phenom-line-fits}.
Using the software of Ref.~\cite{Pezzullo:2011a}, we found the following
values for $p_c$ and $\nu_0$:
\begin{align}
p_c &= 0.030\,534 \pm 0.000\,385 \\
\nu_0 &= 1.486\,681 \pm 0.166\,837.
\end{align}

To put our results in context, as we did in the code capacity setting, we
reference Table~\ref{tab:code-thresholds}.  For the same reasons as in the
code capacity noise model setting, the threshold we compute is larger than
the MLE decoder's threshold for the 4.4.4.4 surface codes.  We conjecture
that is it also measurably less than the threshold for the optimal
color-code decoder, as is the case for optimal vs. MLE decoding for surface
codes.  So far, the threshold for optimal decoding of 4.8.8 color codes has
not been estimated, but the analysis for optimal decoding of 6.6.6 color
codes suggests that the threshold will be near $4.5\%$.  If true, our data
would signal a violation of the Nishimori conjecture for the RPGM associated
with the 4.8.8 color code, something we are not aware of being reported
elsewhere.

Finally, we note that while the value of $\nu_0$ is consistent with value of
$\nu_0 = 1.463(6)$ obtained for the 4.4.4.4 surface code \cite{Wang:2003a}
and the 6.6.6 color code, the uncertainty in the value we obtained is too
high to draw any meaningful conclusions.

%
\subsection{Circuit-level noise model}

As with the phenomenological noise model, computing $p_{\text{fail}}(p)$
exactly even for small code sizes is intractable, so we again appeal to
Monte Carlo estimation.  Our Monte Carlo simulation algorithm is similar to
Algorithm \ref{alg:MC-estimator}, except the manner in which the error
pattern $E$ is generated is different.  To generate $E$, we simulate BP and
DP channel noise as described by the noise model on the explicit circuit
given for syndrome extraction.  This results in a correlated error model for
syndrome and data qubits.  We then use the phenomenological noise MLE
decoder and assess success or failure as we did for that noise model.

\begin{figure}[!h]
\center{\includegraphics[width=\columnwidth]{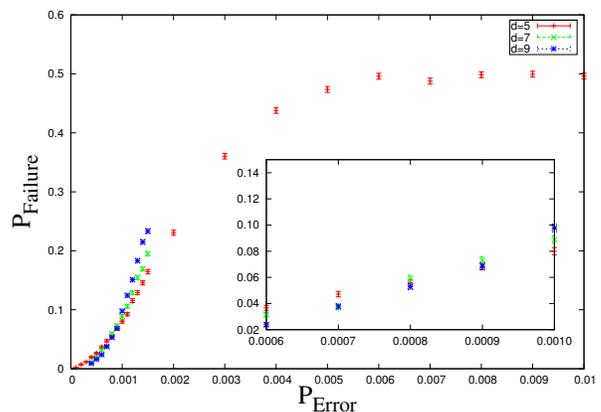}} 
\caption{Monte Carlo data used to estimate accuracy threshold in the
circuit-based noise model in which the noninterleaved syndrome extraction
circuit is used.\label{fig:XZ-sched-plot}}
\end{figure}

\begin{figure}[!h]
\center{\includegraphics[width=\columnwidth]{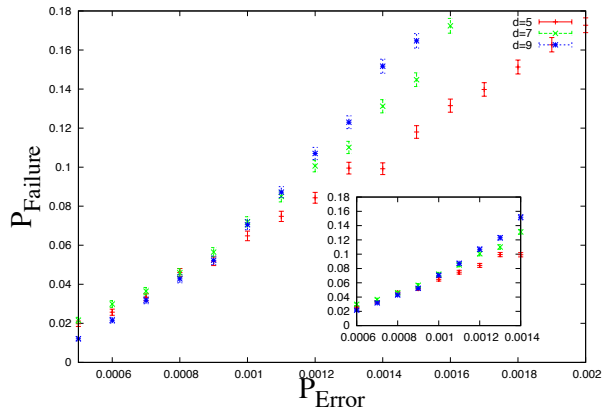}} 
\caption{Monte Carlo data used to estimate the accuracy threshold in the
circuit-based noise model in which the interleaved syndrome extraction
circuit is used.\label{fig:XZ-interleaved-plot}}
\end{figure}

We estimated the $p_{\text{fail}}(p)$ curves for several small 4.8.8
triangular color codes for both the $X$-then-$Z$ schedule of
Fig.~\ref{fig:naive-schedule} and the interleaved $X$-$Z$ schedule of
Fig.~\ref{fig:interleaved-schedule}.  Our results are plotted in
Figs.~\ref{fig:XZ-sched-plot} and \ref{fig:XZ-interleaved-plot}.

To compute the accuracy thresholds from our data, we again fit our data near
the crossings to an equations whose form is similar to that of by
Eq.~(\ref{eq:pf-pc-curve-fit}).  However, the motivation for such a fit is a
bit more tenuous in this case because while the MLE decoder we are using
maps to a RPGM, the noise model which generates it is correlated.  For this
reason, as also found in Ref.~\cite{Wang:2003a}, we found it necessary to
include a quadratic term, unlike the case for the pure phenomenological
noise model.  In other words, we fit our data to an equation of the form
\begin{align}
p_{\text{fail}} = A + B(p - p_c)d^{1/\nu_0} + C(p-p_c)^2 d^{2/\nu_0}.
\end{align}

\begin{figure}[!h]
\center{\includegraphics[width=\columnwidth]{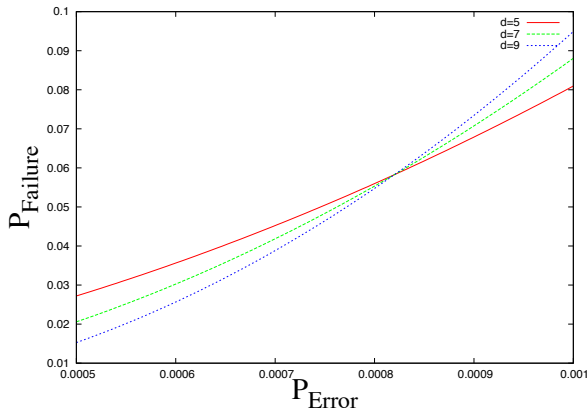}}
\caption{Quadratic fit near curve crossings of noninterleaved-circuit
circuit-based-noise-model Monte Carlo data.  Estimated accuracy threshold is
$p_{\textit{th}} = 0.082(3)\%$.\label{fig:quadratic-fits}}
\end{figure}

The quadratic fits to our data for the $X$-then-$Z$ schedule are plotted in
Fig.~\ref{fig:quadratic-fits}.  Again using the software of
Ref.~\cite{Pezzullo:2011a}, we found the following values for $p_c$ and
$\nu_0$ for the $X$-then-$Z$ schedule:
\begin{align}
p_c &= 0.000\,820 \pm 0.000\,022 \\
\nu_0 &= 1.350\,954 \pm 0.079\,188.
\end{align}

To be clear, there is both a $Z$-error and an $X$-error accuracy threshold;
we report the smaller of the two here.

Similarly, for the $XZ$-interleaved schedule we found
\begin{align}
p_c &= 0.000\,800 \pm 0.000\,037 \\
\nu_0 &= 1.509\,871 \pm 0.151\,690.
\end{align}

To remind, our results are for the smaller of the $X$-error and $Z$-error
thresholds.

Our results show that despite our efforts to shorten the schedule of the
syndrome extraction circuit, the impact on the resulting accuracy threshold
is essentially indistinguishable.  The value of $0.082(3)\%$ for the
accuracy threshold for MLE decoding of the 4.8.8 color codes in the
circuit-level noise model is about a factor of ten less than the the
corresponding $0.68\%$ accuracy threshold for MLE decoding of 4.4.4.4
surface codes in the circuit-level noise model.  We believe that the
difference comes from the fact that the 4.8.8 codes have some weight-8
stabilizer generators while the 4.4.4.4 codes only have weight-4 stabilizer
generators.  This causes the circuits for extracting the syndrome for the
weight-8 generators in the 4.8.8 codes to be larger, inviting more avenues
for failure.  Indeed, we have investigated the finite-sized
error-propagation patterns for the 4.8.8 codes such as the one depicted in
Fig.~\ref{fig:interleaved-schedule-error-propagation}, and they are
significantly larger and more complex than the corresponding patterns for
the 4.4.4.4 surface codes.  Expanding this line of reasoning, we predict
that the 6.6.6 color codes will have an MLE-decoded accuracy threshold in
the circuit-based noise model that is somewhere between the 4.8.8 and
4.4.4.4 accuracy thresholds in this noise model.

%
\section{Analytic bound on the accuracy threshold for fault-tolerant quantum
error correction}
\label{sec:analytic-bound}

While numerical estimates of the accuracy threshold are valuable, equally
valuable are analytic proofs that the accuracy threshold is no smaller than
a given value.  One method of obtaining such a lower bound is to use the
self-avoiding walk (SAW) method, first proposed in Ref.~\cite{Dennis:2002a}.
The idea behind this method begins with the observation that our goal is to
lower-bound the failure probability of decoding, which is the probability
that the actual errors plus the inferred correction (modulo 2) lead to an
error chain that corresponds to a logical operator.  For color codes,
logical operators can be not only string-like but also string-net like, as
described in the original paper on color codes \cite{Bombin:2006b}.  They
must also have a Pauli-weight at least as large as the distance of the code.
The probability that a logical operator is present in the post-corrected
state is therefore at least as large as the probability that an error-chain
string of Pauli-weight equal to the code distance is present.  Certainly
this is a very pessimistic bound; there are many error chain strings and
string-nets of this Pauli weight that do not result in failure!

The SAW lower-bound method can be applied relatively straightforwardly to
the code-capacity and phenomenological noise models with MLE decoding.  The
method begins to break down when applied to the circuit-level noise model
with phenomenological MLE decoding.  One reason for this is that the circuit
introduces correlated errors, called ``hooks'' in Ref.~\cite{Dennis:2002a},
which suggest that the SAW bounding the failure probability should be
allowed to sometimes take more than one step in a single iteration.  With
some finesse, this can be accounted for and bounded as in
Ref.~\cite{Dennis:2002a}.  However, for the color codes, the steps need not
be path-connected either.  For example, the circuit may create three
separated errors on a single octagon plaquette.  Calling such a process a
``walk'' or attempting to bound the behavior of the process by a true SAW
method is dubious at best.  For this reason, we have chosen to omit bounding
the accuracy threshold in the circuit-level noise model and instead have
bounded the accuracy threshold only for the other two noise models, as
described below.

%
\subsection{Code capacity noise model}

As argued by Dennis \etal\ in Ref.~\cite{Dennis:2002a}, the probability that
an $[\![n, k, d]\!]$ topological code decoded by an error-free MLE decoder
fails is upper-bounded by the probability that a self-avoiding walk creates
a closed path (\ie, a self-avoiding polygon or SAP) of length $d$ or
greater:
\begin{align}
\label{eq:SAP-sum}
p_{\text{fail}} &\leq \sum_{L \geq d} \text{Prob}_{\text{SAP}}(d) \\
 &\leq n\sum_{L \geq d} n_{\text{SAP}}(L)\,(4p(1-p))^{L/2}.
\end{align}

Self-avoiding walks on the 4.8.8 lattice have been studied, and it is known
that the number of self-avoiding polygons of length $L$ on the lattice
scales asymptotically as \cite{Jensen:1998a}
\begin{align}
n_{\text{SAP}}(L) \leq P(L)\mu_{4.8.8}^L, \quad \mu_{4.8.8} \approx
1.808\,830\,01(6),
\end{align}
where $P$ is a polynomial and $\mu_{4.8.8}$ is the so-called
\emph{connective constant} for the $4.8.8$ lattice.  (The value
$\mu_{4.8.8}$ has been rigorously bounded to be $1.804\,596 \leq \mu_{4.8.8}
\leq 1.829\,254$ \cite{Jensen:2004a, Alm:2005a}.)  For small $p$, each
summand in Eq.~(\ref{eq:SAP-sum}) is upper-bounded by the term with $L = d$,
and the number of summands is at most a polynomial in $d$, so that
$p_{\text{fail}} \to 0$ as $d \to \infty$ as long as
\begin{align}
\label{eq:p-transcendental}
p(1-p) \leq \frac{1}{4\mu_{4.8.8}^2}.
\end{align}

Solving this equation for $p$, we find that the code capacity threshold is
at least
\begin{align}
p_{c} \geq 8.335\,745(1)\%.
\end{align}

Despite the crudeness of the SAW bound, it comes surprisingly close to the
numerical value of $10.56(1)$ that we estimate in
Sec.~\ref{sec:code-capacity-noise-model-threshold}.

%
\subsection{Phenomenological noise model}

\begin{figure}[htb]
\center{\includegraphics[width=0.9\columnwidth]{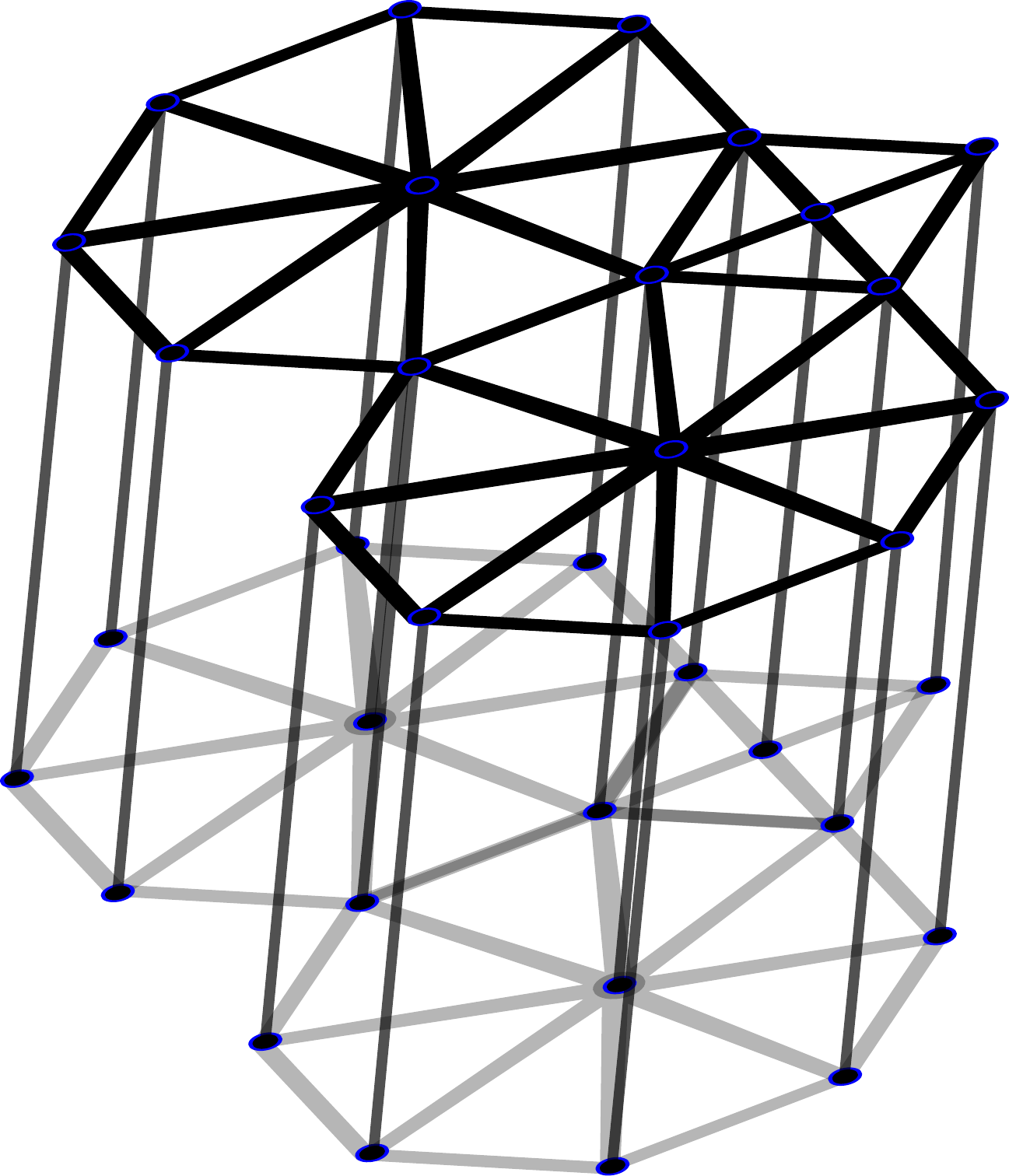}}
\caption{\small{\label{fig:4.8.8-pizza-prism}Prismatic lattice on which a
self-avoiding walk occurs in the analysis of the accuracy threshold for
fault-tolerant quantum error correction using color codes in the
phenomenological noise model.}}
\end{figure}

The SAW bound method is essentially the same as for the code capacity noise
model, except now errors can happen on syndrome qubits as well as data
qubits and the set of all relevant qubits forms a three-dimensional volume.
The relevant SAW traverses a 3D lattice that connects syndrome qubits and
data qubits both with themselves and each other as dictated by the color
code; the corresponding nonregular prismatic lattice is depicted in
Fig.~\ref{fig:4.8.8-pizza-prism}.  To our knowledge, the connective constant
for this lattice is not known, but it could be computed in principle using
standard methods, \eg, those outlined in Refs.~\cite{Jensen:1998a,
Jensen:2004a, Alm:2005a}.  We opted to bypass this analysis and instead
compute a coarser bound on the failure probability.

Because the lattice in Fig.~\ref{fig:4.8.8-pizza-prism} has vertices of
degree $\Delta$ equal to $6$, $8$, and $10$, we can bound the number of SAPs
of length $L$ by 
\begin{align}
n_{\text{SAP}}(L) \leq 2\Delta_{\text{max}}(2\Delta_{\text{max}} - 1)^{L -
1}.
\end{align}
Using $\Delta_{\text{max}} = 10$, we obtain a formula similar to that of
Eq.~(\ref{eq:p-transcendental}), namely
\begin{align}
p(1-p) \leq \frac{1}{4(9)^2} = \frac{1}{324}.
\end{align}

Solving this equation for $p$, we find that the phenomenological noise threshold is
at least
\begin{align}
p_{c} \geq \frac{9 - 4\sqrt{5}}{18} \approx 0.3096\%.
\end{align}

This bound is nearly a factor of ten less than the value of $p_c =
3.05(4)\%$ that we estimate in
Sec.~\ref{sec:phenomenological-noise-model-threshold}.  With further
computational effort in determining the connective constant of the governing
lattice, we suspect that the SAW bound will still be below our numerical
estimate, but significantly closer, in analogy with the relationship between
our SAW bound for the code capacity and the value we estimate numerically.
We leave this analysis to others wishing to tighten this bound.

%
\section{Fault-tolerant computation with color codes}
\label{sec:ftqc-with-color-codes}

To establish a threshold for fault-tolerant quantum computation, it is
sufficient to establish three things:  1) a threshold for fault-tolerant
quantum error correction, 2) a procedure for performing a universal set of
gates in encoded form, and 3) that a failure in an encoded gate that occurs
with probability $p$ leads to failures in each output codeword with
probability at most $p$.  These three ingredients establish that each gate
in a quantum circuit can be simulated fault-tolerantly by performing it in
encoded form followed by fault-tolerant quantum error correction.  We
previously established the first criterion in
Sec.~\ref{sec:fault-tolerant-error-correction}.  We establish the second two
criteria here for two possible computer architectures.

In the first, which we call the ``pancake architecture,'' each logical qubit
is stored in its own triangular 4.8.8 color code and the logical qubits are
stacked atop one another.  This architecture is essentially the same as the
one proposed in Ref.~\cite{Dennis:2002a}.  Almost all encoded operations are
implemented transversally in this model, acting on single ``logical qubit
pancakes'' or between two such ``pancakes.''  In the second, which we call
the ``defect architecture,'' each logical qubit is stored as a connected
collection of missing check operators, which we call a ``defect,'' in a
single 2D 4.8.8 substrate.  This architecture is essentially the same as the
one proposed in Ref.~\cite{Raussendorf:2007a}.  Almost all encoded
operations are performed in one of two ways: encoded single-qubit gates are
performed by disconnecting a region containing the defect, operating
transversally on the region, and reconnecting the region, while the encoded
CNOT gate is performed by a sequence of local measurements that cause one
defect to circulate around another.

%
\subsection{Fault-tolerance by transversal gates}
\label{sec:fault-tolerance-by-transversal-gates}

In this section, we compute the threshold for fault-tolerant quantum
computation with triangular 4.8.8 color codes when (almost) all encoded
gates are implemented \emph{transversally}.  To remind, by calling a gate
``transversal,'' we mean that it acts identically on all physical qubits in
a code block.  For example a two-qubit transversal gate between two
triangular codes acts as the same two-qubit physical gate between
corresponding physical qubits in each code block.  Some authors refer to
this notion of transversality as \emph{strong} transversality
\cite{Eastin:2007a}.

%
\subsubsection{Identity gate}

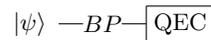
\begin{figure}[htb]
\centerline{
\Qcircuit @C=1em @R=1em {
 \lstick{\ket{\psi}} & \push{BP} \qw & \gate{\text{QEC}}
} 
} 
\caption{\small{\label{fig:noisy-I}Noisy identity gate.  $BP$ indicates the
action of the BP channel.}}
\end{figure}

The accuracy threshold for the identity gate is exactly the same as the
accuracy threshold for fault-tolerant quantum error correction, by
definition.  Schematically, Fig.~\ref{fig:noisy-I} depicts the noisy
identity gate circuit.

Formally, we can express the equivalence between the accuracy threshold for
the identity gate and the accuracy threshold for fault-tolerant quantum
error correction as
\begin{align}
p_{th}^{(I)} = p_{th}^{(\text{QEC})}.
\end{align}

%
\subsubsection{$\CNOT$ gate}

\begin{figure}[htb]
%
%
\begin{align}
\raisebox{ 1.5em}{
\Qcircuit @C=1em @R=1em {
 & \push{BP} \qw & \ctrl{1} & \push{D} \qw & \gate{\text{QEC}} \\
 & \push{BP} \qw & \targ & \push{D} \qw & \gate{\text{QEC}}
} 
} 
&\hspace{0.5em}=
\raisebox{ 1.5em}{
\Qcircuit @C=1em @R=1em {
 & \ctrl{1} & \push{BPD} \qw & \gate{\text{QEC}} \\
 & \targ & \push{BPD} \qw & \gate{\text{QEC}}
} 
} 
\end{align}
%
%
\caption{\small{\label{fig:noisy-CNOT}Noisy $\CNOT$.  $BP$ indicates the
action of the BP channel; $D$ indicates the action of the DP channel.}}
\end{figure}
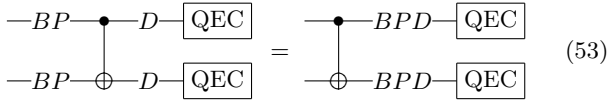

The color codes are Calderbank-Shor-Steane (CSS) codes
\cite{Calderbank:1996a, Steane:1996b}, and for all such codes, the encoded
controlled-NOT ($\CNOT$) gate can be implemented \emph{transversally},
namely by applying $\CNOT$ gates between corresponding pairs of physical
qubits in two color codes.  (For color codes, fewer $\CNOT$ gates than a
fully transversal set also suffice.) Schematically,
Fig.~\ref{fig:noisy-CNOT} depicts a noisy $\CNOT$ gate.  Each physical
$\CNOT$ gate propagates the BP channel on its control to the BP channel on
its target and vice versa, so that the effective noise model seen by the
fault-tolerant quantum error correction procedure on each code block after
the encoded $\CNOT$ gate is the BP channel followed by the projection of the
two-qubit DP channel onto a single qubit.  Although the DP channel can
create correlated errors between output code blocks, it will never cause a
correlated error within a code block.  Since our decoder treats the noise
model phenomenologically, it does not account for DP-channel features such
as the fact that in the DP channel a $Y$ error is more probable than the
combination of separate $X$ and $Z$ errors.  For this reason, since half of
the DP-channel errors act as a bit-flip on a given code block and half of
them act as a phase-flip on a given code block, our decoder interprets the
post-$\CNOT$ noise model as a BP channel with an effective error rate of $p
+ p/2$ for bit flips and $p + p/2$ for phase flips.  This means that the
accuracy threshold for the $\CNOT$ gate is actually $2/3$ of the value for
the identity gate.  The $\CNOT$ gates used in an encoded $\CNOT$ gate must
therefore meet a more stringent requirement than the identity gate to be
implemented transversally fault-tolerantly.  (However, the $\CNOT$ gates
used in fault-tolerant quantum error correction still only need to meet the
threshold for the encoded identity gate.)
\begin{align}
p_{th}^{(\CNOT)} = \frac{2}{3}p_{th}^{(I)}.
\end{align}

%
\subsubsection{Hadamard gate}

The color codes are \emph{strong} CSS codes, meaning that the $X$-type and
$Z$-type stabilizer generators have the same structure.  As with all strong
CSS codes, the encoded Hadamard gate ($H$) can be implemented transversally.

\begin{figure}[htb]
%
%
\begin{align}
%
%
\Qcircuit @C=1em @R=1em {
 & \push{BP} \qw & \gate{H} & \push{BP} \qw & \gate{\text{QEC}}
} 
%
%
&\hspace{0.5em}=
%
%
\Qcircuit @C=1em @R=1em {
 & \gate{H} & \push{BPBP} \qw & \gate{\text{QEC}}
} 
%
%
\end{align}
%
%
\caption{\small{\label{fig:noisy-H}Noisy Hadamard  $BP$ indicates the
action of the BP channel; $D$ indicates action of the DP channel.}}
\end{figure}
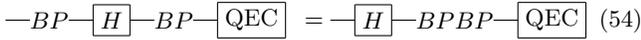

Like the $\CNOT$ gate, the Hadamard gate propagates the BP channel to the BP
channel.  However, since faults in the Hadamard gate are modeled as
an ideal Hadamard gate followed by the BP channel, the effective noise model
is not one but \emph{two} actions of the BP channel, as depicted in
Fig.~\ref{fig:noisy-H}.

It is straightforward to show that two successive applications of the BP
channel with probability $p$ is equivalent to one application of BP channel
with probability $2p(1-p)$.  This is therefore the effective post-Hadamard
noise channel, so that the threshold for the Hadamard gate is about half of
that for fault-tolerant quantum error correction:
\begin{align}
p_{th}^{(H)} = \frac{1}{2} - \frac{1}{2}\sqrt{1 - 2 p_{th}^{(I)}} 
 &\approx \frac{1}{2}{p_{th}^{(I)}}.
\end{align}

%
\subsubsection{Phase gate}

The color codes have the feature that each stabilizer generator for the code
has a Pauli weight equal to $0 \bmod 4$ and each pair of generators are
incident on $0 \bmod 2$ qubits.  One can show that because of this, the
encoded phase gate ($S$) has a transversal implementation \cite{Knill:1996a,
Bombin:2006b}.  (Technically, it is the transversal $S^\dagger$ operation
that acts as an encoded $S$.) 

While a faulty phase gate acts as an ideal phase gate followed by a BP
channel, the phase gate itself does not propagate the BP channel preceding
it symmetrically for bit flips and phase flips.  This follows from the
conjugation actions
\begin{align}
SXS^{\dagger} &= Y = iXZ &  SZS^{\dagger} &= Z.
\end{align}

The phase gate therefore propagates a phase flip to a phase flip and a bit flip to both a
bit-flip and a phase flip, as depicted in Fig.~\ref{fig:noisy-S}.
\begin{figure}[htb]
\begin{align}
\Qcircuit @C=1em @R=1em {
 & \push{B} \qw & \gate{S} & \push{B} \qw & \gate{\text{QEC}}
} 
&\hspace{0.5em}=
\Qcircuit @C=1em @R=1em {
 & \gate{S} & \push{BPB} \qw & \gate{\text{QEC}}
} 
\\
\Qcircuit @C=1em @R=1em {
 & \push{P} \qw & \gate{S} & \push{P} \qw & \gate{\text{QEC}}
} 
&\hspace{0.5em}=
\Qcircuit @C=1em @R=1em {
 & \gate{S} & \push{PP} \qw & \gate{\text{QEC}}
} 
\end{align}
\caption{\small{\label{fig:noisy-S}Noisy phase gate. $B$ indicates the
action of the bit-flip channel; $P$ indicates the action of the phase-flip
channel.}}
\end{figure}

The phase gate is correspondingly more sensitive to phase-flip noise because
the effective phase-flip strength is $p^3 + 3p(1-p)^2$.  The phase gate thus
has separate thresholds for bit-flip and phase-flip noise.  For bit-flip
noise, the threshold is
\begin{align}
p_{th}^{(S, \text{bit-flip})} &= \frac{1}{2} - \frac{1}{2}\sqrt{1 - 2
p_{th}^{(I)}} 
 \approx \frac{1}{2}{p_{th}^{(I)}}.
\end{align}
For phase-flip noise, one must solve a cubic equation to get a closed-form
solution for the threshold as a function of the threshold for the identity
gate.  While this is possible in principle, to save space we simply state
the cubic equation in the variable $x = p_{th}^{(S, \text{phase-flip})}$
that must be solved and its approximate solution, which we can estimate because
we know that the accuracy threshold is very close to 0:
\begin{align}
x^3 + 3x(1-x)^2 &= {p_{th}^{(I)}}, \\
 x \approx \frac{1}{3}{p_{th}^{(I)}}.
\end{align}

%
\subsubsection{Single-qubit measurements}

To \emph{destructively} apply the encoded single-qubit measurements $M_X$
and $M_Z$, we transversally measure $X$ or $Z$ on each of the qubits in the
code block.  We then perform classical error correction on the measurement
outcomes (because they may be faulty) to infer the outcome of the encoded
measurement, as depicted schematically in Fig.~\ref{fig:noisy-measurements}.
\begin{figure}[htb]
\begin{align}
\Qcircuit @C=1em @R=1em {
& \push{B} \qw & \measureD{M_Z} & \push{\framebox{CEC}} \cw
} 
&\hspace{0.5em}=
\Qcircuit @C=1em @R=1em {
 & \measureD{M_Z} & \push{B} \cw & \push{\framebox{\text{CEC}}} \cw
} 
\\
\Qcircuit @C=1em @R=1em {
 & \push{P} \qw & \measureD{M_X} & \push{\framebox{\text{CEC}}} \cw
} 
&\hspace{0.5em}=
\Qcircuit @C=1em @R=1em {
 & \measureD{M_X} & \push{B} \cw & \push{\framebox{\text{CEC}}} \cw
} 
\end{align}
\caption{\small{\label{fig:noisy-measurements}Noisy measurements.  $B$
denotes the bit-flip channel, $P$ denotes the phase-flip channel, and CEC
denotes classical error correction of the measurement outcomes.
Post-measured states are drawn with double lines to indicate that they are
``classical.''}}
\end{figure}

The correctness of this procedure follows from the fact that $X$ and $Z$
operators can be expressed as $Z = S^2$ and $X = HZH$, and the encoded
operations $H$ and $S$ have previously been demonstrated to have transversal
encoded implementations.  Bit or phase errors (as relevant) before a
measurement then map to bit errors on the observed classical bit pattern.

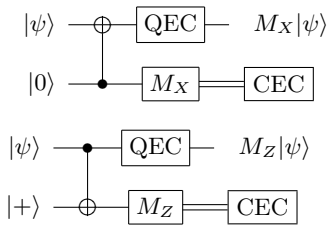
\begin{figure}[!htb]
\center{
  \subfigure{
    \Qcircuit @C=1em @R=1em {
     \lstick{{\ket{\psi}}} & \targ     & \gate{\text{QEC}}   &
\push{\quad{M}_X {\ket{\psi}}} \qw \\
     \lstick{{\ket{0}}}    & \ctrl{-1} & \gate{{M}_X} &
\push{\framebox{\text{CEC}}} \cw  \\
    } 
  } 
\hspace{4.5em}
%
  \subfigure{
    \Qcircuit @C=1em @R=1em {
     \lstick{{\ket{\psi}}} & \ctrl{1} & \gate{\text{QEC}} & \push{\quad{M}_Z
{\ket{\psi}}} \qw \\
     \lstick{{\ket{+}}}    & \targ    & \gate{{M}_Z}      &
\push{\framebox{\text{CEC}}} \cw & \\
    } 
  } 
} 
\caption{\label{fig:MZ-MX-transversal}Circuits for nondestructive encoded
${M}_X$ and ${M}_Z$, using the states $|0\>$ and $|+\>$ as ``magic
states.''}
\end{figure}

The reason the measurement is destructive is that after the measurement, the
qubits are no longer in the codespace of the color code; the post-measured
state is not projected onto an $X$ or $Z$ eigenstate in the codespace.
However, as pointed out by Steane \cite{Steane:1998a}, given the ability to
prepare encoded $|+\>$ states, a circuit composed of transversal ${\CNOT}$
and transversal destructive ${M}_X$ measurements can implement
\emph{nondestructive} $M_X$ measurements transversally.  A similar story
holds for encoded $|0\>$ states and $M_Z$ measurements.  The circuits for
generating these nondestructive measurements transversally are depicted in
Fig.~\ref{fig:MZ-MX-transversal}.  Because the encoded $|0\>$ and $|+\>$
states are being used to enable gates, namely nondestructive encoded
measurements, these states are called ``magic states'' for the gates
\cite{Bravyi:2005a}.  Ordinarily, quantum error correction would follow not
just one, but both of the outputs of the encoded $\CNOT$ gate in these
circuits, but because one of the encoded qubits is destructively measured
immediately after the $\CNOT$ gate, that encoded qubit does not require
quantum error correction; it will be effectively performed by the classical
error correction process occurring after the destructive measurement.

The threshold for destructive $M_Z$ and $M_X$ measurements is the same as
the code capacity threshold for the code, regardless of which noise model we
are considering.  This is because the physical measurements are made only
once, as repetition cannot improve their effective error rate.  The
(flawless) classical error correction performed in post-processing has a
threshold equal to the code capacity threshold.  Hence, we have the result
that
\begin{align}
p_{th}^{(M_X, \text{destructive})} &= p_{th}^{(I, \text{code capacity})}, \\
p_{th}^{(M_Z, \text{destructive})} &= p_{th}^{(I, \text{code capacity})}.
\end{align}

Although these measurements need only be smaller than the code capacity
threshold to implement the encoded measurement, when these measurements are
used in the fault-tolerant quantum error correction protocol, they must be
smaller than the threshold set by the prevailing noise model---a threshold
that may be significantly lower.

To compute the threshold for nondestructive $M_Z$ and $M_X$ measurements, we
examine how errors propagate through the circuits in
Fig.~\ref{fig:MZ-MX-transversal}.  As with the analysis of
Fig.~\ref{fig:noisy-CNOT}, the effective noise channel we need to consider
after the $\CNOT$ gate is the BP channel followed by the DP channel on each
output.  One of these enters a destructive measurement, which, as we found
in the analysis of Fig.~\ref{fig:noisy-measurements}, has a rather high
threshold equal to the code capacity even in the circuit-level noise model.
However, it is lowered slightly by the fact the effective error rate is
$\frac{2}{3}p$, as discussed in the analysis of the encoded $\CNOT$ gate.
The other output enters a standard quantum error correction circuit, also
subject to noise of strength $\frac{2}{3}p$.  Since the lowest threshold of
these two thresholds is this one, the overall threshold for an encoded
nondestructive measurement is the same as the threshold for the encoded
$\CNOT$ gate.  Namely, we have the result that
\begin{align}
p_{th}^{(M_X, \text{nondestructive})} &= p_{th}^{(\CNOT)} =
\frac{2}{3}p_{th}^{(I)}, \\
p_{th}^{(M_Z, \text{nondestructive})} &= p_{th}^{(\CNOT)} =
\frac{2}{3}p_{th}^{(I)}.
\end{align}
%

%
\subsubsection{$|0\>$ and $|+\>$ preparation}

It is tempting to assert that the way to fault-tolerantly prepare the
encoded $|0\>$ state is to perform an encoded nondestructive $M_Z$
measurement.  The flaw with this reasoning is that the nondestructive $M_Z$
measurement requires the encoded $|+\>$ state as a magic state, and the
analogous way of preparing a $|+\>$ state requires a $|0\>$ state.

To get out of this chicken-and-egg cycle, one must use an independent
process.  We describe a two-step process that works for preparation of an
encoded $|0\>$ state; the process for preparing an encoded $|+\>$ state is
similar.

The first step is to prepare the product state $|0\>^{\otimes n}$ by
transversally measuring $M_Z$ on each physical qubit.  This state is a
stabilizer state, having $n$ check operators, with check operator $i$ being
$Z$ on qubit $i$ for $i = 1, \ldots n$.  The second step is to
fault-tolerantly measure the $X$ checks for the color code.  Because the
only $Z$-type operators consistent with all the $X$ checks are the color
codes' $Z$ checks for the color code and the logical $Z$ operator, these
measurements will transform the state into the logical $|0\>$ state.

It turns out that it is not necessary to also fault-tolerantly measure the
$Z$ checks for the color code.  The state is already in an eigenstate of
these operators at this point, so all the measurements can do is yield
syndrome bits.  Had one obtained these bits and processed them, the
post-corrected state would still have been subject to $X$ errors drawn from
the same distribution as the $X$ errors afflicting the initial
$|0\>^{\otimes n}$ preparation---fault-tolerant error correction doesn't
suppress the final error rate to zero, it only keeps it at the same rate one
started with.

The threshold for preparation of encoded $|0\>$ and $|+\>$ states is
therefore the same as the threshold for fault-tolerant quantum error
correction, namely,
\begin{align}
p_{th}^{(|0\>)} &= p_{th}^{(|+\>)} = p_{th}^{(I)}.
\end{align}

It is worth noting that while the process for fault-tolerantly preparing
$|0\>$ and $|+\>$ states is not strictly transversal, the only
nontransversal operation is fault-tolerant quantum error correction, a
process that is required in addition to transversal operations in any event
in order to achieve fault-tolerant quantum computation.

%
\subsubsection{$T$ gate}
\label{sec:T-gate}

\begin{figure}[htb]
\centerline{
\Qcircuit @C=1em @R=1em {
 \lstick{\ket{\psi}} & \ctrl{1} & \gate{\text{QEC}} & \qw & \gate{S} &
\gate{\text{QEC}} \\
 \lstick{\ket{\pi/4}} & \targ & \gate{M_Z} & 
\push{\framebox{\text{CEC}}} \cw & \control \cw \cwx  \\
} 
} 
\caption{\small{\label{fig:T-circuit}Magic-state circuit for the $T$ gate.}}
\end{figure}
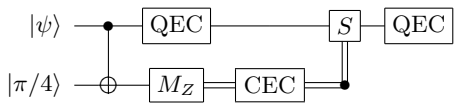

Another gate that admits a transversal implementation with a magic state is
the $T$ gate, also called the $\pi/8$ gate, defined as
\begin{align}
T := \begin{bmatrix}1&0\\0&e^{-i\pi/4}\end{bmatrix} 
   = e^{-i\pi/8} \begin{bmatrix}e^{i\pi/8}&0\\0&e^{-i\pi/8}\end{bmatrix}.
\end{align}
If we we have an encoded version of the state
\begin{align}
|\pi/4\> &:= TH|0\> \\
 &=  \frac{1}{\sqrt{2}}\left(|0\> + e^{i\pi/4}|1\>\right),
\end{align}
also called $|A\>$ and $|A_{\pi/4}\>$ in the literature, we can implement
the $T$ gate transversally using the circuit of Fig.~\ref{fig:T-circuit}.
This circuit is not a Clifford circuit, because the classically-controlled
$S$ gate is not a Clifford gate.  Nevertheless, it only uses gates that we
have previously shown how to implement in encoded form by purely transversal
operations.

To compute the $T$ gate threshold, we again study error propagation through
its defining circuit, \viz the circuit in Fig.~\ref{fig:T-circuit}.  As
shown previously, the $\CNOT$ gate creates an input to the first QEC cycle
that has a threshold of $2/3$ of the standard QEC threshold.  The $S$ gate
creates an input to the second QEC cycle which splits the threshold into
bit-flip and phase-flip thresholds approximately equal to $1/2$ and $1/3$ of
the standard QEC threshold.  The threshold for the $T$ gate is set by the
smallest of these, namely the $S$ gate threshold, which is
\begin{align}
p_{th}^{(T, \text{bit-flip})} &= \frac{1}{2} - \frac{1}{2}\sqrt{1 - 2
p_{th}^{(I)}} 
 \approx \frac{1}{2}{p_{th}^{(I)}}, \\
 p_{th}^{(T, \text{phase-flip})} &= x  \approx \frac{1}{3}{p_{th}^{(I)}}.
\end{align}

%
\subsubsection{$|\pi/4\>$ preparation}
\label{sec:T-state}

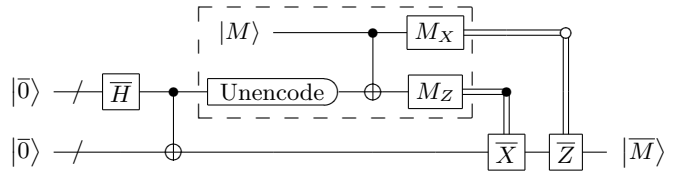
\begin{figure}[htb]
\centerline{
\Qcircuit @C=1em @R=1em {
         &         & & & \lstick{\ket{M}}        & \ctrl{1} & \gate{M_X} &
\cw           & \controlo \cw & &  \\
\lstick{\ket{\overline{0}}}  & {/} \qw & \gate{\overline{H}} & \ctrl{1} &
\measureD{\text{Unencode}} & \targ    & \gate{M_Z} & \control \cw  & \cwx & &
\\
\lstick{\ket{\overline{0}}} & {/} \qw & \qw & \targ & \qw & \qw & \qw &
\gate{\overline{X}} \cwx & \gate{\overline{Z}} \cwx & \qw & \quad
{\ket{\overline{M}}} \gategroup{1}{5}{2}{7}{0.7em}{--} \\  
} 
} 
\caption{\small{\label{fig:M-injection}Circuit for injecting a single-qubit
magic state $M$.  The circuit for multi-qubit magic states is similar.}}
\end{figure}

There are two alternatives for preparing encoded $|\pi/4\>$ states
fault-tolerantly described in the literature.  In the first, low-fidelity
$|\pi/4\>$ states are ``injected'' into the code by teleportation, using the
circuit in Fig.~\ref{fig:M-injection} \cite{Knill:2004a}, and then
``distilled'' using encoded gates until the resultant $|\pi/4\>$ states have
an error below the accuracy threshold.  In the second, high-quality
$|\pi/4\>$ states are first distilled and then injected into the code.
The circuit depicted in Fig.~\ref{fig:M-injection} is not fault-tolerant,
but faults are already suppressed by the code on the encoded qubits; only
operations from the latter-half of the decoding circuit onwards are
unprotected.

Unlike all of the previous encoded gates, this method for implementing an
encoded $|\pi/4\>$ preparation requires an operation which is neither
transversal nor fault-tolerant quantum error correction.  The ``unencoding''
portion of the circuit is the time-reversed coherent circuit for encoding a
state in the color code, derivable via standard stabilizer codes as shown in
Ref.~\cite{Gottesman:1997a}.  This unencoding circuit does not appear to
have a transversal implementation.  While the Eastin-Knill theorem
\cite{Eastin:2008a} asserts that at least one nontransversal operation is
required to generate a universal set of encoded gates, it does not guarantee
that no transversal implementation of this circuit exists.  That is because
the process of fault-tolerant quantum error correction used to prepare
$|0\>$ and $|+\>$ states is not transversal.  For 3D color codes
\cite{Bombin:2007b}, in which $T$ is intrinsically transversal and in which
encoded $|0\>$ and $|+\>$ states still require fault-tolerant quantum error
correction for preparation, only transversal and FTQEC operations are
needed, for example.  It would be interesting to develop a variant of the
circuit in Fig.~\ref{fig:M-injection} which only uses transversal operations
and possibly fault-tolerant quantum error correction to inject a $|\pi/4\>$
state into 2D color codes.  We leave that for others to explore.

While the portion of the circuit in Fig.~\ref{fig:M-injection} in which the
physical $|M\>$ state interacts with the unencoded qubit via a $\CNOT$
appears to also not be transversal, it can be made so with slight
modification.  In principle, one could prepare $n$ states of the form
$|M\>$ and transversally apply the $\CNOT$  gate between these and the
code block, but only the one qubit corresponding to the unencoded state will
be used to classically control the $\overline{X}$ and $\overline{Z}$ gates
that are used to inject the correct state.  As usual, these corrections do
not need to be actually implemented, only used to update the Pauli frame.

\begin{figure}[htb]
\centerline{
\Qcircuit @C=1em @R=1em {
 \lstick{\ket{\widetilde{\pi/4}}} & \ctrl{14} & \qw       & \qw       & \qw
& \qw       & \gate{M_X} & \cw \\
 \lstick{\ket{\widetilde{\pi/4}}} & \qw       & \ctrl{13} & \qw       & \qw
& \qw       & \gate{M_X} & \cw \\
 \lstick{\ket{\widetilde{\pi/4}}} & \targ     & \targ     & \qw       & \qw
& \ctrl{12} & \qw        & \rstick{\ket{\pi/4}} \qw \\
 \lstick{\ket{\widetilde{\pi/4}}} & \qw       & \qw       & \ctrl{11} & \qw
& \qw       & \gate{M_X} & \cw \\
 \lstick{\ket{\widetilde{\pi/4}}} & \targ     & \qw       & \targ     & \qw
& \targ     & \gate{M_Z} & \cw \\
 \lstick{\ket{\widetilde{\pi/4}}} & \qw       & \targ     & \targ     & \qw
& \targ     & \gate{M_Z} & \cw \\
 \lstick{\ket{\widetilde{\pi/4}}} & \targ     & \targ     & \targ     & \qw
& \qw       & \gate{M_Z} & \cw \\
 \lstick{\ket{\widetilde{\pi/4}}} & \qw       & \qw       & \qw       &
\ctrl{7}  & \qw       & \gate{M_X} & \cw \\
 \lstick{\ket{\widetilde{\pi/4}}} & \targ     & \qw       & \qw       &
\targ     & \targ     & \gate{M_Z} & \cw \\
 \lstick{\ket{\widetilde{\pi/4}}} & \qw       & \targ     & \qw       &
\targ     & \targ     & \gate{M_Z} & \cw \\
 \lstick{\ket{\widetilde{\pi/4}}} & \targ     & \targ     & \qw       &
\targ     & \qw       & \gate{M_Z} & \cw \\
 \lstick{\ket{\widetilde{\pi/4}}} & \qw       & \qw       & \targ     &
\targ     & \targ     & \gate{M_Z} & \cw \\
 \lstick{\ket{\widetilde{\pi/4}}} & \targ     & \qw       & \targ     &
\targ     & \qw       & \gate{M_Z} & \cw \\
 \lstick{\ket{\widetilde{\pi/4}}} & \qw       & \targ     & \targ     &
\targ     & \qw       & \gate{M_Z} & \cw \\
 \lstick{\ket{\widetilde{\pi/4}}} & \targ     & \targ     & \targ     &
\targ     & \targ     & \gate{M_Z} & \cw
} 
} 
\caption{\small{\label{fig:state-distillation}Distillation circuit for
$|\pi/4\>$ states; it is the 15-qubit Reed-Muller code's encoding circuit in
reverse.}}
\end{figure}
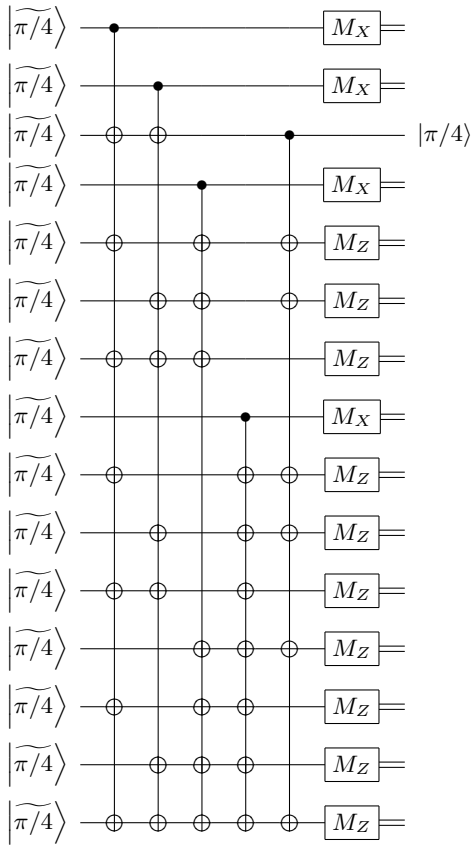

Both alternatives for preparing high-quality encoded $|\pi/4\>$ states
require a procedure for magic-state distillation.  One option is to use the
encoding circuit for the 15-qubit Reed-Muller code \cite{Knill:1998a} (also
the smallest 3D color code \cite{Bombin:2007b}) run in reverse, as depicted
in Fig.~\ref{fig:state-distillation}.  For it to work, the initial states
must have an error less than the $|\pi/4\>$ distillation threshold.  For the
circuit depicted in Fig.~\ref{fig:state-distillation}, the distillation
threshold for independent, identically distributed (iid) depolarizing noise
is $(6-2\sqrt{2})/7 \approx 45.3\%$ \cite{Reichardt:2006a, Buhrman:2004a},
for dephasing iid noise is $(\sqrt{2}-1)/\sqrt{2} \approx 29.3\%$
\cite{Reichardt:2006a, Virmani:2004a}, and for worst-case iid noise is
$(\sqrt{2}-1)/2\sqrt{2} \approx 14.6\%$ \cite{Reichardt:2006a,
Virmani:2004a}.  The entire circuit must be run
$\bigO(\poly(\varepsilon^{-1}))$ times to achieve an output error less than
$\varepsilon$; convergence should be quite rapid in practice given the
actual polynomial \cite{Reichardt:2006a}.  Various tricks can be used to
boost the distillation threshold and reduce the resources required to
achieve high-fidelity states; any of these can be readily adapted to this
setting.

%
\subsubsection{Synthesis}

It is well-known result the gate basis $\{H, S, \CNOT, M_X, M_Z, |0\>, |+\>,
|\pi/4\>\}$ is universal for quantum computation \cite{Nielsen:2000a} (in
fact, it is even overcomplete).  We have presented transversal methods for
performing color-code encoded versions of each of these except for the state
preparations.  By the Eastin-Knill theorem \cite{Eastin:2008a}, it is
impossible to generate a complete universal encoded gate basis in
transversal form.  However, color codes offer a particularly gentle way
around this theorem.  There are only two nontransversal operations used.
The first is fault-tolerant quantum error correction, a process that is
required in addition to encoded computations in any event for the entire
protocol to be fault tolerant.  The second is the time-reversed coherent
encoding circuit for color codes.  Such a circuit is useful for encoding
unknown quantum states, but in an actual quantum computation, the input state
is known so it is not needed for this purpose.  Whether this ``unencoding
circuit'' can be replaced with another operation which uses only transversal
operations and fault-tolerant quantum error correction is an interesting
open question.  For 3D color codes, we know that the answer is ``yes.''

The ``pancake architecture,'' described in Ref.~\cite{Dennis:2002a} for the
Kitaev surface-codes, realizes the encoded gate set we described using only
gates between spatially neighboring qubits.  One difference in our analysis
from that performed in Ref.~\cite{Dennis:2002a} is that we have analyzed the
accuracy threshold not only for fault-tolerant quantum memory but also for
fault-tolerant quantum computation, a feat made tractable by the strong CSS
nature of the color codes.

%
\subsection{Fault-tolerance by code deformation}

The method of fault-tolerance described in
Sec.~\ref{sec:fault-tolerance-by-transversal-gates} requires a
three-dimensional architecture to allow the transversal $\CNOT$ gates to
remain spatially local.  This violates the spirit of using two-dimensional
codes in the first place.  Fortunately, it is possible to use \emph{code
deformation} to achieve fault-tolerance in a strictly two-dimensional
architecture.  Our construction here mirrors that of Raussendorf \etal's
construction for surface codes \cite{Raussendorf:2007a, Raussendorf:2007b}.
Fowler has independently constructed a method for using code deformation in
4.8.8 color codes that is similar to ours \cite{Fowler:2008c}.  Some salient
differences between our method and Fowler's are that ($i$) Fowler's logical
qubits are always encoded in a triple of defects whereas ours are encoded in
single defects except during certain logical gates, and ($ii$) Fowler's
scheme disallows different defect types from occupying the same plaquette
location while ours does not.  Each of these differences allows our scheme to
encode a higher density of information.  Specifically, our scheme
allows a six-fold increase in logical qubit density over the Fowler scheme.

To begin, we generate a sufficiently large 4.8.8 triangular color code by
performing fault-tolerant quantum error correction on a collection of
qubits.  We are not interested in what state the triangular code
encodes---all we require is that the state is in the codespace with
arbitrarily high fidelity.  We consider any logical qubits associated with
the entire surface to be ``gauge'' qubits in the language of subsystem
stabilizer code theory \cite{Poulin:2005a}.  We will use this state as a
substrate for generating and manipulating encoded qubits.

Each element of the standard set of stabilizer generators for a color code
can be labeled by a face of a definite color (red, green, or blue) and an
operator of a definite Pauli type ($X$ or $Z$).  Notationally, we will refer
to a generator as a $(c, P)$ generator if it is of color $c$ and Pauli type
$P$.  To prepare an encoded qubit in our color code substrate, we remove a
connected product of stabilizer generators of the same color and type.
(Generally removal of any element of the stabilizer group will yield a
logical qubit; we restrict attention to this class for simplicity.)  We call
this removed region a \emph{defect} in analogy with the language used by
Raussendorf \etal\ in Ref.~\cite{Raussendorf:2007a}.  This removal is
entirely passive---we simply cease measuring this product of stabilizer
generators in future quantum error correction rounds.  For this reason, it
is manifestly a fault-tolerant process.

In the following sections, we describe how to perform a universal repertoire
of encoded logic gates on defect-based logical qubits, with arbitrarily high
fidelity.  This is therefore a prescription for fault-tolerant quantum
computation using code deformation.

%
\subsubsection{Preparing a defect in $|0\>$ or $|+\>$}

In principle, the generator removed to form a defect qubit can be identified
with any element of the encoded Pauli group for that encoded qubit.  For
concreteness, we make the choice of calling the removed generator a logical
$Z$ when it is $Z$-type defect (also called a `primal' or `smooth' defect in
the language of Ref.~\cite{Raussendorf:2007a}) and a logical $X$ when it is
$X$-type defect (also called a `dual' or `rough' defect in the language of
Ref.~\cite{Raussendorf:2007a}).  Thus removing a $c$-colored $X$- or
$Z$-type generator corresponds to preparing a logical $|+\>_{(c,X)}$ or
$|0\>_{(c,Z)}$ state respectively.

The logical $Z$ operator for a $(c, X)$ defect acts as $Z$ on a $c$-colored
chain of qubits connecting the defect to another $c$-colored boundary, which
may itself be another defect.  If no such other boundary exists, then the
defect fails to encode a logical qubit.  To avoid this complication, we have
chosen our substrate to be a triangular code, having boundaries of each of
the three colors.  Similarly, the logical $X$ operator for a $(c, Z)$ defect
acts as $X$ on a $c$-colored chain of qubits connecting the defect to a
$c$-colored boundary.

Preparing a $|+\>_{(c, Z)}$ or $|0\>_{(c, X)}$ state requires more care.  To
do this, we measure $M_X$ or $M_Z$ respectively along a $c$-colored chain of
qubits from the plaquette we wish to store the logical qubit in and the
nearest $c$-colored boundary.  This projects each qubit along the chain into
either $|+\>$ or $|-\>$ (resp. $|0\>$ or $|1\>$), which we can interpret as
$|+\>$ (resp. $|0\>$) for each qubit by changing local Pauli bases.  We then
measure the $Z$-checks (resp. $X$-checks) incident on this chain except the
one at the defect location and correct any errors, which places the defect
back into the substrate in the desired state.

An arbitrarily large $c$-colored defect can be prepared in a single step by
ceasing to measure a collection of $c$-connected defects by a similar
process, enabling the preparation process to be made arbitrarily reliable.
This introduces a number of ``gauge'' qubits in the interior of the defect
that can be ignored; the details of this are described in the next section.

%
\subsubsection{Growing, shrinking, and moving defects}

\begin{figure}[!ht]
\centerline{
  \subfigure[\ Octagonal green $Z$
defect.]{\includegraphics[width=0.45\columnwidth]{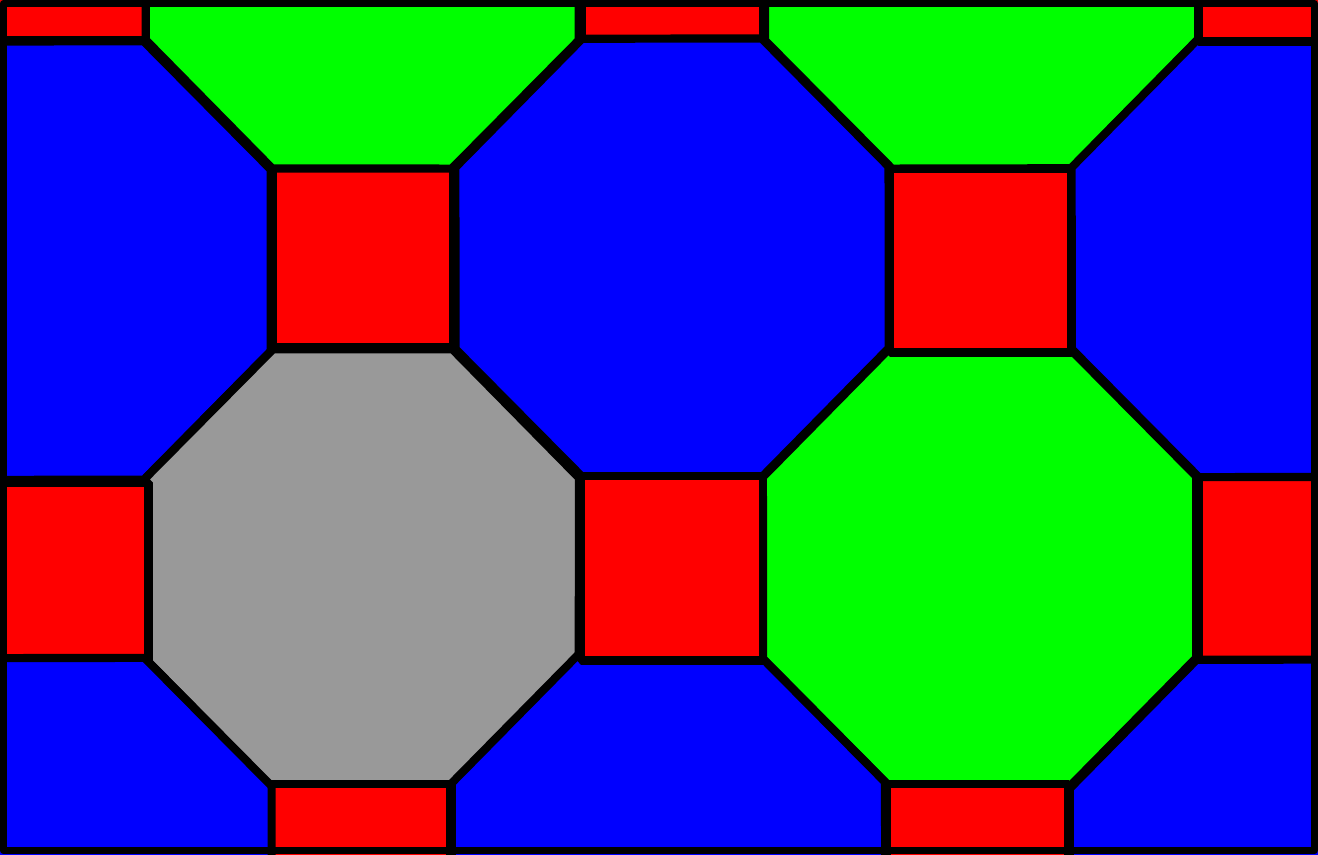}}\hspace{1em}
  \subfigure[\ Growth to two
defects.]{\includegraphics[width=0.45\columnwidth]{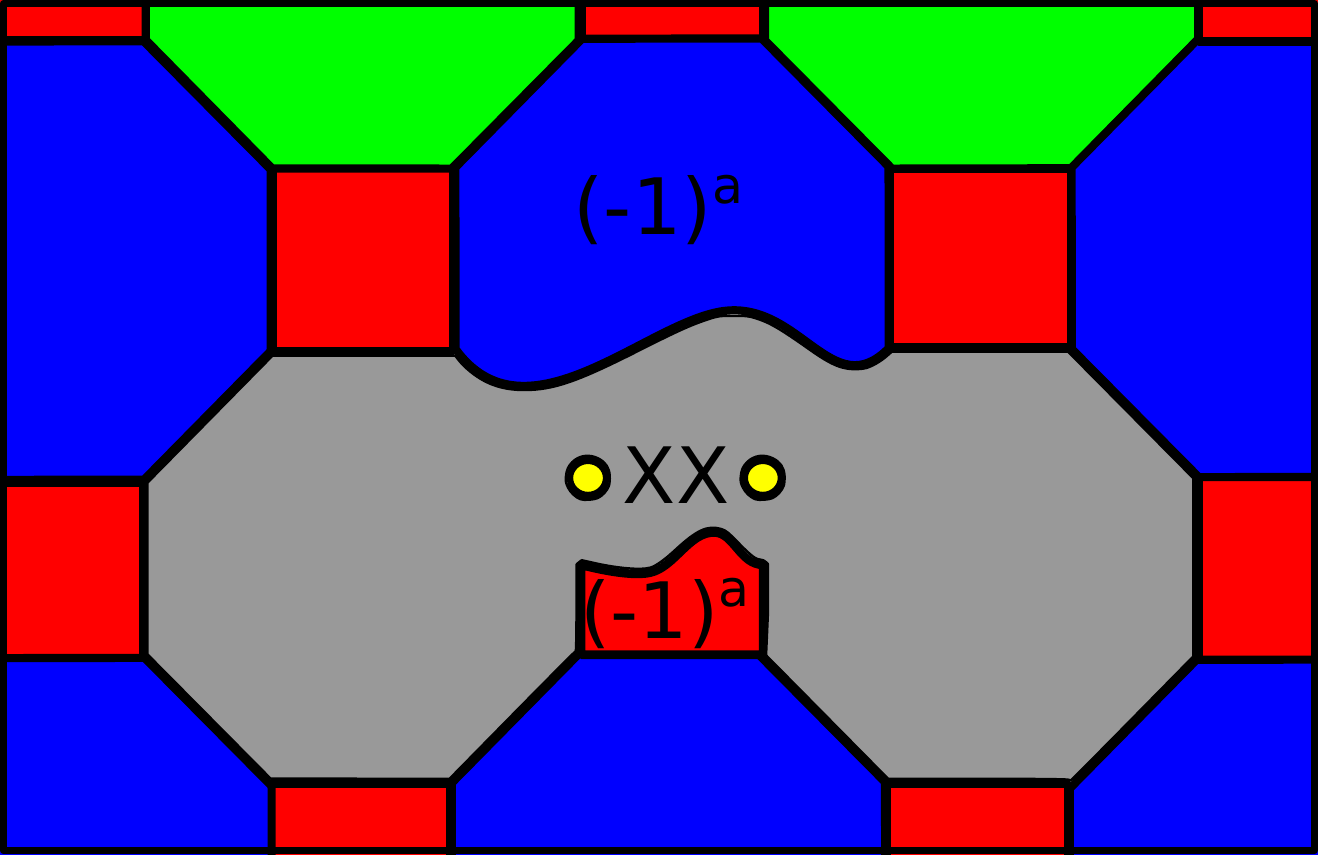}}\hspace{1em}
}
\caption{\label{fig:defect-growth1}Growth of an octagonal green $Z$ defect
by one site.}
\end{figure}

\begin{figure}[!ht]
\centerline{
  \subfigure[\ Square red $Z$
defect.]{\includegraphics[width=0.45\columnwidth]{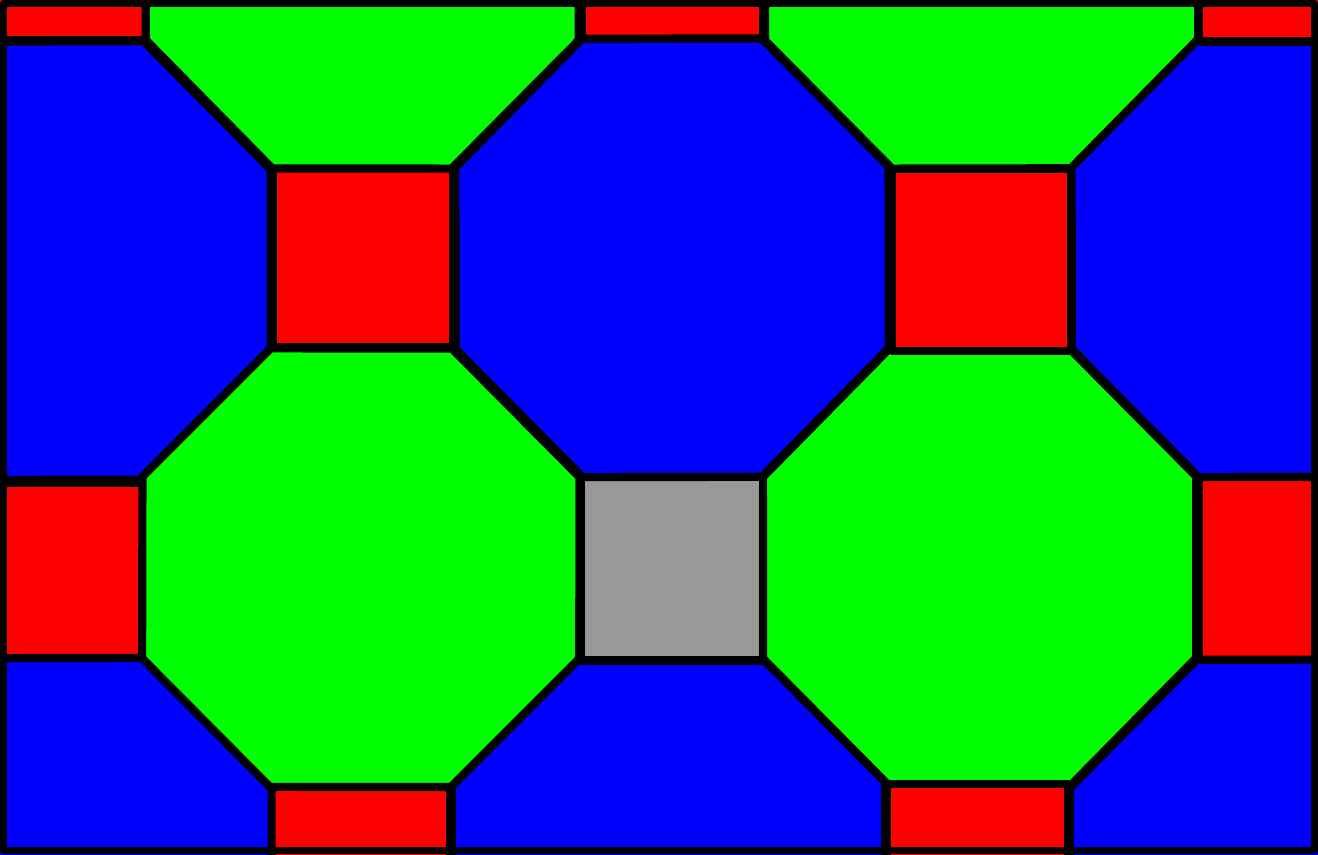}}\hspace{1em}
  \subfigure[\ Growth to two
defects.]{\includegraphics[width=0.45\columnwidth]{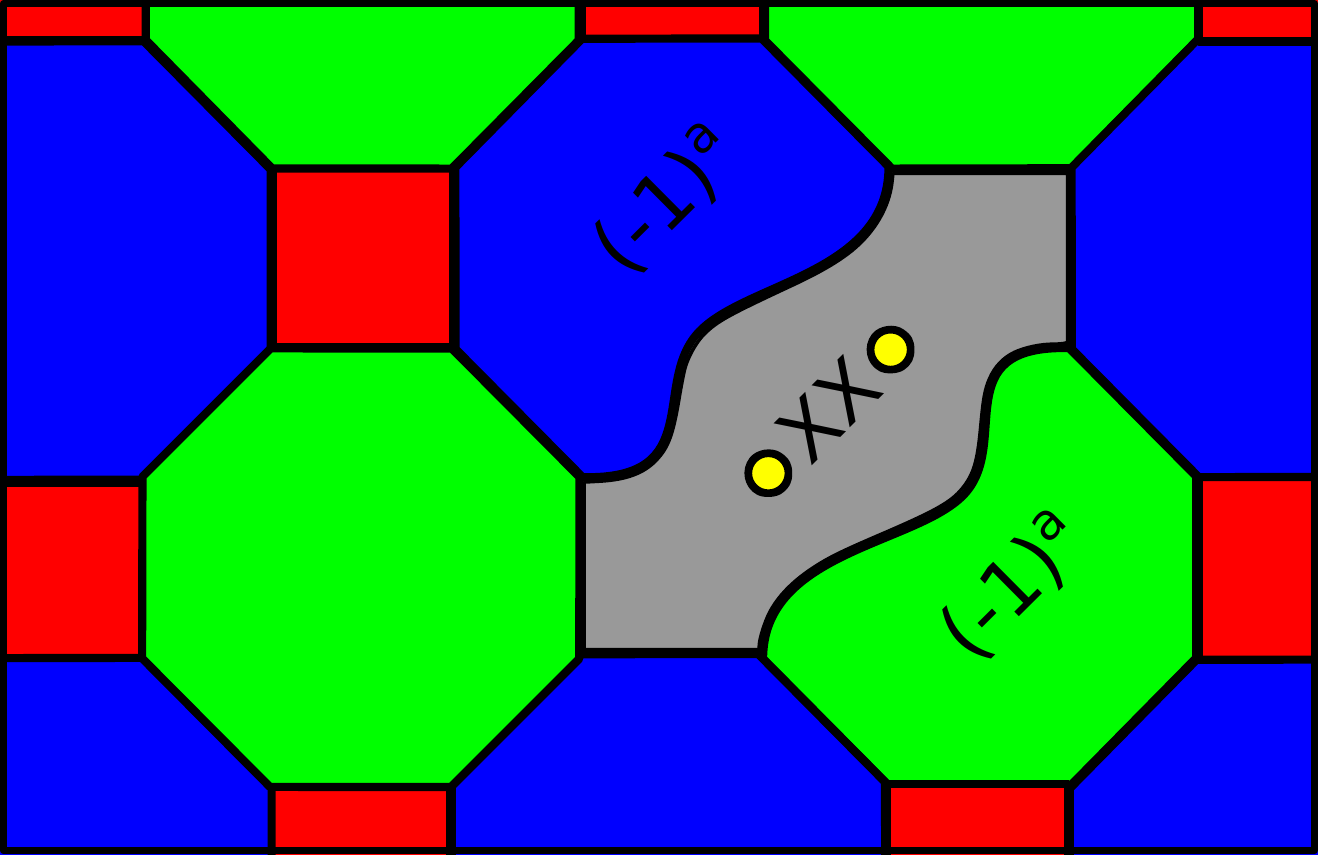}}\hspace{1em}
}
\caption{\label{fig:defect-growth2}Growth of a square red $Z$ defect
by one site.}
\end{figure}

\begin{figure}[htb]
\centerline{
\Qcircuit @C=1em @R=1em {
 \lstick{\ket{0}} & \ctrl{1} & \ctrl{2} & \qw & \gate{M_Z} \\
 \lstick{\ket{\psi}} & \targ & \qw & \qw \\
\lstick{\ket{\psi}} & \qw & \targ & \qw \\
} 
} 
\caption{\label{fig:growth-circuit}Measurement of $XX$ to grow a $Z$-type
defect. The measurement can be performed with existing circuitry already in
place for syndrome extraction. }
\end{figure}
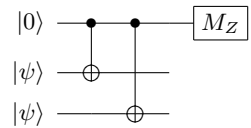

We grow a $(c, P)$ defect qubit on region $q$ in the following way.
Suppose we would like to extend the defect so that it includes an adjacent
region $q'$ of the same color and type. (By adjacent, we mean that the
regions can be connected by a single two-qubit $c$-colored link.)  To do
this, we first perform the following conditional operation.  If $P = X$,
then we measure $ZZ$ on a $c$-colored link connecting the regions, while if
$P = Z$, then we measure $XX$ on a $c$-colored link connecting the regions.
Examples of how this works for octagonal and square defects are depicted in
Figs.~\ref{fig:defect-growth1} and \ref{fig:defect-growth2}; the circuit in
Fig.~\ref{fig:growth-circuit} implements this transformation.  A $YY$
operator can be used to grow a $X$ and $Z$-type defect at the same time. 

After this measurement, the new collective defect operator is the product of the $q$
and $q'$ defect operators. The $\pm XX$ or $\pm ZZ$ operator has also been
added to the list of stabilizer generators.  As usual, we do not need to
actually correct the result to a $+1$ outcome: it suffices to update the Pauli
frames of the stabilizer generators incident on these two interior qubits.

Because we will no longer use the weight-two operator, we may consider it to
also be a ``gauge'' operator in the language of subsystem stabilizer codes
\cite{Poulin:2005a}.  This also makes its anticommuting partner a gauge
operator, which we may interpret to be either of the original defect
operators (on $q$ or $q'$).  By introducing these two new gauge operators,
we may reinterpret the defect logical operator on the collective $q$ and
$q'$ region as acting solely on its boundary.  In particular, the interior
of the collective $q$ and $q'$ region need never be involved in future
syndrome extractions.

An important question is whether the defect growth process is
fault-tolerant.  The simplest circuit for measuring $XX$ or $ZZ$ would
perform $\CNOT$ gates into or out of an ancilla qubit to each of the two
relevant qubits, as depicted in Fig.~\ref{fig:growth-circuit}.  Although a
single error in this ancilla qubit could propagate to two errors on the two
interior qubits, because we subsequently treat these qubits as encoding a
gauge qubit, we do not worry about errors on these.  It could still be the
case that the value of the measurement obtained is incorrect, which impacts
the update of the Pauli frame of the two adjacent stabilizer generators in a
correlated way.  Thus a single syndrome measurement error would propagate to
two syndrome-bit errors.  To prevent this happening to first order in the error
probability, we repeat the $XX$ or $ZZ$ measurement twice and use the
majority vote of the three outcomes to update the Pauli frame.

Compared to the process of defect growth, defect contraction is much
simpler: to shrink a defect by a single-plaquette, one simply measures that
plaquette operator in the next round of fault-tolerant quantum error
correction.

By a combination of local growth and shrinking processes, one can deform the
code with a $(c, P)$ defect at one plaquette to a code with a $(c, P)$
defect anywhere else.  In other words, the \emph{move} operation for a
defect can be decomposed into a sequence of more elementary \emph{grow} and
\emph{shrink} operations.

%
\subsubsection{Measuring a defect}

To destructively measure the logical operator encircling a defect, one first
shrinks the defect to size of a single plaquette. Then one measures the
defect with the existing circuitry at that plaquette as though it were a
local stabilizer generator. The shrunken defect will have a significantly
lower tolerance to one type of Pauli error but that error type is in the
basis being measured in and will not disturb the measurement outcome. To
destructively measure the string-like logical operator connecting two
defects, one brings the two operators as close together as possible. One
then measures the weight-two operator connecting the defects using the
circuitry used to grow a defect from one site to encompass the other. Again,
the tolerance to errors of one Pauli type will be significantly lower, but
this will not be of the type that disturbs the measurement.

To nondestructively measure a defect, one uses the circuit of
Fig.~\ref{fig:MZ-MX-transversal}, which uses destructive measurement of
$M_Z$ or $M_X$, preparation of $|0\>$ or $|+\>$, and the $\CNOT$ gate
described in the next section.

%
\subsubsection{$\CNOT$ gate between defects}

\begin{figure}[htb]
\centerline{
\Qcircuit @C=1em @R=1em {
 \lstick{\ket{\text{control}}_{(c,Z)}}   & \qw      & \ctrl{1} & \qw & \qw &
\rstick{\ket{\text{control}}_{(c,Z)}} \qw \\
 \lstick{\ket{0}_{(c'',X)}} & \targ    & \targ    & \targ & \qw & \gate{M_Z}
\\
 \lstick{\ket{+}_{(c', Z)}} & \qw    & \qw    & \ctrl{-1} & \qw &
\rstick{\ket{\text{target}}_{(c',Z)}} \qw \\
 \lstick{\ket{\text{target}}_{(c',Z)}}        & \ctrl{-2} & \qw & \qw & \qw
& \gate{M_X} \\
} 
} 
\caption{\label{fig:CNOT-same-type}Circuit for braiding a $\CNOT$ gate
between $Z$-type defects.  The colors $c$ and $c'$ may be the same or
different, but the color $c''$ is a color different from these.  The circuit
for braiding a $\CNOT$ gate between $X$-type defects is similar: the $\CNOT$
gate directions are reversed, the types of the defects and the types of the
measurements have their Pauli types swapped from $X$ to $Z$ and vice-versa,
and the $|0\>$ state becomes a $|+\>$ state and vice-versa.}
\end{figure}
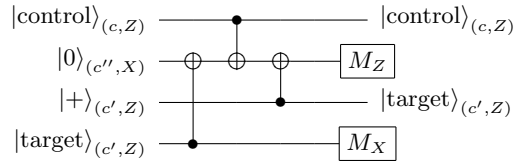

It is straightforward to show that moving a $(c, Z)$ defect qubit around a
$(c', X)$ defect qubit (or vice-versa) generates an encoded $\CNOT$ gate
controlled by the $(c, Z)$ defect when $c$ and $c'$ are different colors;
the construction is essentially the same as that in
Refs.~\cite{Raussendorf:2007a, Raussendorf:2007b, Fowler:2008c}.  Since this process
traces out a braid in spacetime, we call this process ``braiding defects.''
Also drawing upon Refs.~\cite{Raussendorf:2007a, Raussendorf:2007b}, one can
generate a $\CNOT$ gate between two $Z$-type defects or two $X$-type
defects, whether they are the same color or not.  The circuit for doing this
between two $Z$-type defects is depicted in Fig.~\ref{fig:CNOT-same-type};
the circuit for doing this between two $X$-type defects is similar.

\begin{figure}[htb]
\begin{tabular}{l}
\Qcircuit @C=1em @R=1em {
 \lstick{\ket{\psi}_{(c,X)}} & \targ & \gate{M_Z} \\
 \lstick{\ket{+}_{(c', Z)}} & \ctrl{-1} & \rstick{\ket{\psi}_{(c', Z)}} \qw 
} 
\\[10ex]
%
\Qcircuit @C=1em @R=1em {
 \lstick{\ket{\psi}_{(c,Z)}} & \ctrl{1} & \gate{M_X} \\
 \lstick{\ket{0}_{(c', X)}} & \targ & \rstick{\ket{\psi}_{(c', X)}} \qw 
} 
\end{tabular}
\caption{\label{fig:X-to-Z}Circuits for converting a $Z$-type defect into an
$X$-type defect and vice-versa.}
\end{figure}
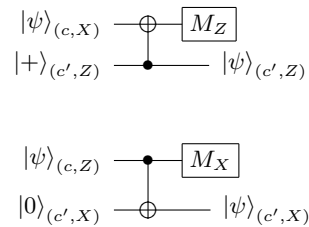

One can convert an $X$-type defect into a $Z$-type defect, or vice-versa,
(changing its color as a side effect) using one of the circuits in
Fig.~\ref{fig:X-to-Z}.  In conjunction with the other type of $\CNOT$ gates
mentioned, this allows $\CNOT$ gates between two defects regardless of the
colors or Pauli types they have.

%
\subsubsection{Phase gate on a defect}

To perform an $S$ (phase) gate on a $(c, P)$ defect, we prepare two more
qubits of $(c', P)$ and $(c'', P)$ type, each in the state $|0\>$ and use
$\CNOT$ gates to put the defect into a three-defect repetition code.  This
maps our single-defect logical qubits into the three-defect logical qubits
Fowler uses in his construction \cite{Fowler:2008c}.  We then grow the
defects and connect them so that they separate an interior triangular region
from an exterior region, just as described in Fowler's construction.  If $P
= Z$, then as Fowler noted, it suffices to apply $S$ transversally
(actually, $S^\dagger$ must be applied transversally) to generate a logical
$S$ on the triple-defect qubit.  However, if $P = X$, then Fowler's
construction fails, because the ``exterior trees'' in his language fail to
undergo the action $SXS^\dagger = Y$.  ``Pruning'' the exterior tree as
Fowler suggests for his implementation of the Hadamard gate fails as well,
because such an operation yields only the ``byproduct operator'' for logical
$X$ or $Z$ on the triple-defect qubit, but not both.  To perform the $S$
gate on $X$-type defects, we propose the following two-step procedure.
First, we arrange the defects to separate a triangular interior from the
exterior and apply $S^\dagger$ transversally on the interior.  Second, we
rearrange the defects so that part of what was the exterior becomes the new
interior, and perform $S^\dagger$ transversally on this new triangular
interior.  In this way, both the interior and exterior trees experience the
$S$ gate.  Once the gate is complete, we run the three-defect encoding
circuit in reverse and absorb the two ancilla qubit regions back into the
substrate in subsequent quantum error correction rounds.

Of course, the $S$ gate can also be achieved via magic states of the form
$|\pi/2\> := \frac{1}{\sqrt{2}}\left(|0\> + e^{i\pi/2}|1\>\right)$ (also
called $|Y\>$ or $|+i\>$ in the literature) in a manner similar to what is
done for surface codes.  But this is one of the great benefits of 4.8.8
color codes---no magic state distillation and usage is required to realize
this gate in encoded form.  It may well be worth the lower accuracy
threshold of color codes relative to surface codes in order to achieve the
resource reduction for performing encoded $S$ gates.

%
\subsubsection{Hadamard gate on a defect}

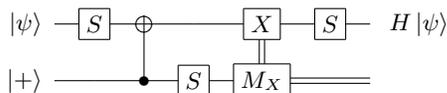
\begin{figure}[htb]
\centerline{
\Qcircuit @C=1em @R=1em {
 \lstick{\ket{\psi}} & \gate{S} & \targ & \qw & \gate{X} & \gate{S} & \qw &
{\qquad H\ket{\psi}} \\
 \lstick{\ket{+}} & \qw & \ctrl{-1} & \gate{S} & \gate{M_X} \cwx[-1]& \cw &
\cw \\
} 
} 
\caption{\small{\label{fig:H-circuit}Circuit for simulating $H$ with
previously-described gates.}}
\end{figure}

From one point of view, a logical Hadamard gate is unnecessary because it
can be implemented using the gates we have previously described, for example
by the circuit of Fig.~\ref{fig:H-circuit}.
However, we have developed a more resource-efficient way to perform this
gate that we describe here.

If we want to perform a Hadamard gate on a $(c, P)$ defect, we first prepare
an ancilliary $(c,P)$ defect in the state $|0\>$ and perform a $\CNOT$ gate
from the defect qubit to this defect ancilla using the circuit of
Fig.~\ref{fig:CNOT-same-type}.  This encodes the original defect qubit into
the two-qubit bit-flip repetition code across the two defects.  The $ZZ$
operator for the two defect qubits is in the stabilizer group of this
repetition code, so we can measure $ZZ$ without disturbing the encoded
qubit.  The logical $Z$ operator is a $c'$-colored chain of $Z$ operators
around either of the defects and the logical $X$ operator is a $c$-colored
chain of $X$ operators connecting the defects, where $c' \neq c$.  This
encoding is the one used at all times in the Raussendorf \etal\ scheme
\cite{Raussendorf:2007a, Raussendorf:2007b}, but here we only use it to
perform the Hadamard gate and go back to our original single-defect encoding
once the gate is completed.

After we've encoded the defect qubit into two, we then perform individual
$M_Z$ measurements on a $c'$-colored chain of qubits surrounding both
defects, where $c' \neq c$.  This separates the region of the two qubits
from the substrate, so we can then apply $H$ transversally on the cut-out
region without influencing the external substrate.  This operation applies a
logical Hadamard gate to the two defect qubits in the interior, but also
turns them into $P'$-type defect qubits in the process, where $P'$ is
conjugate to $P$ (\ie, $P' = Z$ if $P = X$, and $P' = X$ if $P = Z$).  We
then stitch the cut out region back into the code by measuring the $P$-check
operators incident on the cut.  The encoding circuit for the repetition code
is run in reverse, and the resulting defect can be converted back to its
original type and color using circuits of the form depicted in
Fig.~\ref{fig:X-to-Z}.

%
\subsubsection{Injecting $|\pi/4\>$ into a defect}

To perform universal encoded quantum computation with defects, our approach
requires defects encoded into the state $|\pi/4\>$ with an error below its
distillation threshold, as discussed in the previous ``pancake''
architecture.  We therefore need a method for injecting magic states into
defects such that the injection process introduces errors at a rate below
the distillation threshold.  The single-qubit preparation threshold for a
magic state is therefore the difference between its distillation threshold
and the error introduced by its injection process.

It is worth remarking that this kind of injection process is used in
defect-based surface code schemes as well \cite{Raussendorf:2006a,
Raussendorf:2007a, Fowler:2008c}.  In these schemes, one must not only
inject $|\pi/4\>$ states, but also inject $|\pi/2\>$ states as well.
However, the impact of errors introduced by errant injection has not been
studied to our knowledge.  It is unclear whether considering it will
significantly alter the high threshold values numerically estimated for
surface codes---the difference between a $1\%$ accuracy threshold and a
$14\%$ distillation threshold is not that great, so it is reasonable to
expect that it may be quite important, especially because injection
generates small-sized defects that are not arbitrarily well-protected from
noise at first.  We do not investigate the impact of the injection process on the
threshold for color codes here either, but we expect that it will be less
consequential because the value of the color-code accuracy threshold is much
lower than that for the surface codes.

To inject into a $(c, Z)$ defect, we identify the corner of the triangular
substrate containing the $c$-colored plaquette and measure $M_Z$ on the
qubit in the corner, isolating it from the code.  We then apply $TH$ to the
corner qubit and then measure the weight-four $X$ check in the corner,
bringing the corner qubit back into the code.  We then cease measuring the
$Z$ check in the standard way, creating a single-plaquette $Z$-type defect
in the corner.  This defect is not well protected from noise, so we move it
from the corner and grow it as fast as we can, so that the ambient noise
doesn't degrade the fidelity of the encoded state.

%
\subsubsection{$T$ gate on a defect}

Given $|\pi/4\>$ defect qubits, we can distill them and use them to perform
the $T$ gate in the same way as described in Secs.~\ref{sec:T-gate} and
\ref{sec:T-state} for the ``pancake'' architecture.

%
\section{Conclusions}
\label{sec:conclusions}

%
\subsection{Fault-tolerant quantum computation}

We studied fault-tolerant quantum computation using color codes, inspired by
($a$) the need to minimize qubit transport in real technologies having 2D
layouts and ($b$) the high accuracy thresholds reported for similar
topological codes.  We framed our study with a well-defined quantum control
model and three physically-motivated noise models of increasing realism
which we call the code-capacity noise model, the phenomenological noise
model, and the circuit-based noise model.

The strategy behind our study was to first understand how to
fault-tolerantly simulate the identity gate via fault-tolerant quantum error
correction and then extend this understanding to how to fault-tolerantly
simulate a universal set of quantum gates capable of general-purpose quantum
computation.
  
In the course of studying fault-tolerant quantum error correction, we
formulated most-likely-error decoding for color codes as a mathematical
optimization problem known as an integer program.  We also developed
feasible schedules for parallelized syndrome extraction for the most
efficient family of color codes, the 4.8.8 color codes.  To better
understand the performance of our decoder, we elaborated a
previously-established connection between the performance of our decoder and
some statistical-mechanical classical spin models.

Our numerically-estimated value for most-likely-error fault-tolerant quantum
error correction for 4.8.8 color codes in the code-capacity noise model is
10.56(1)\%.  This is not significantly different from what had previously
been estimated for optimal decoding of these and the 6.6.6 color codes, or
most-likely-error or optimal decoding of Kitaev's 4.4.4.4 surface codes.
Indeed, the upper bound for any CSS code is slightly more than 11\%, so all
of these codes perform close to optimally in this noise model.  To support
our numerical estimate, we proved that the threshold is at least
$8.335\,745\,(1)\%$ using a self-avoiding walk technique.

Our numerically-estimated value for the accuracy threshold of
most-likely-error fault-tolerant quantum error correction for 4.8.8 color
codes in the phenomenological noise model is 3.05(4)\%.  Again, this is not
significantly different from what had previously been estimated for optimal
decoding of the 6.6.6 color codes, or most-likely-error or optimal decoding
of Kitaev's 4.4.4.4 surface codes.  We attribute the nominal improvement we
find relative to Kitaev's surface codes for both this and the previous noise
model to the fact that the color codes have higher-weight stabilizer
generators, which should be modeled as more errant, but which aren't in
these noise models.  To support our numerical estimate, we proved that the
threshold is at least $0.3096\%$ using a self-avoiding walk technique.

Our numerically-estimated value for most-likely-error fault-tolerant quantum
error correction for 4.8.8 color codes in the circuit-based noise model is
0.082(3)\%.  By attempting to optimize the syndrome extraction circuit by
hand, we ended up surprisingly \emph{decreasing} our threshold estimate to
0.080(3)\%, suggesting that optimizing the syndrome extraction circuit to
find the highest threshold is a nontrivial task.  Unlike our findings for
the previous two noise models, our accuracy-threshold estimate is in fact
significantly different from what had previously been estimated for
most-likely-error decoding of Kitaev's 4.4.4.4 surface codes---it is nearly
a tenth the comparable value of $0.68\%$.  That said, it is consistent with
the value of ``about 0.1\%'' estimated using a different suboptimal decoder
for these codes considered in Ref.~\cite{Wang:2009b}.  However, the estimate
in Ref.~\cite{Wang:2009b} lacked any error analysis, so it is hard to
determine how consistent these results truly are.  We believe that the
reduction in threshold relative to the surface code threshold comes from the
increased weight of the stabilizer generators for the 4.8.8 color code.
Based on this, we predict that the 6.6.6 color codes will have a quantum
error-correction accuracy threshold for this noise model somewhere between
0.082(3)\% and 0.68\% without any additional optimizations.  We did not
prove a lower bound on the threshold in this noise model, as the
self-avoiding walk technique breaks down for this noise model.

To extend our results to general-purpose fault-tolerant quantum computing,
we considered two different approaches.  In the first, the architecture
consisted of 2D surfaces stacked like pancakes in which each surface
corresponded to a logical qubit and almost all operations were either global
transversal operations or local syndrome extraction operations.  In the
second, the architecture consisted of an extended 2D surface in which logical
qubits were associated with ``defects'' and almost all operations were either
defect braiding by local measurements or local syndrome extraction
operations.

In the ``pancake'' architecture, we showed that encoded universal quantum
computation was possible using only local stabilizer measurements, global
transversal operations, and the time-reversed coherent encoding circuit for
the color code, which was used to inject magic states.  Each gate in this
architecture has its own accuracy threshold that is a significant fraction
of the quantum error correction (memory) threshold.

In the ``defect'' architecture, we showed that encoded universal quantum
computation was possible using only local stabilizer measurements, code
deformation, and transversal operations on isolated regions.  These
deformations came in different forms, including growing small defects into
large ones, braiding defects around each other for encoded $\CNOT$ gates,
and isolating defects from the rest of the code.  Each gate has the same
accuracy threshold as the quantum error correction (memory) threshold,
although errors afflicting injected magic state defects before they are
grown to full size may dominate the threshold for the less realistic
code-capacity and phenomenological noise models.

Because the defect architecture has a higher threshold and is more
consistent with the original motivation for our study---namely that many
technologies are restricted to a single 2D layout---we believe the
defect-based approach to be the most practical.  To that end, we extended
some of the defect-based approach for color codes presented in
Ref.~\cite{Fowler:2008c} so that a significantly higher density of defects
can be stored and processed in the surface.

%
\subsection{Relation to statistical-mechanical phase transitions}
\label{sec:RBIM-conclusions}

It has been previously established that there is a mapping between quantum
color codes and a classical statistical-mechanical model known as the
three-body random-bond Ising model (3BRBIM).  In this mapping, each check
maps to a classical $\pm 1$ spin and each qubit maps to a three-body
interaction, with the interaction being ferromagnetic if the qubit is not in
error and antiferromagnetic if it is.  Specifically, the Hamiltonian
constructed by this mapping is
\begin{align}
H &= \sum_{\text{qubits } q} J_q \prod_{\text{checks }c \ni q} S_c,
\end{align}
where $J_q \in \pm 1$ indicates a flip on qubit $q$ and $S_c \in \pm 1$
indicates the eigenvalue of the check $c$.

A feature of the mapping is that the code capacity for any particular
decoding algorithm represents a point on the boundary of the order-disorder
transition of the associated 3BRBIM.  Our integer-programming decoder
is an ``energy-minimizing'' decoder in this paradigm, corresponding to the
phase boundary at zero temperature.  Because our code-capacity value of
10.56(1)\% is lower than the code capacity of 10.925(5)\% of a
``free-energy-minimizing'' decoder implicitly explored by Ohzeki
\cite{Ohzeki:2009b}, this demonstrates that the phase boundary of the 3BRBIM
is ``re-entrant'' as depicted in Fig.~\ref{fig:Nishimori}, violating the
so-called Nishimori conjecture for this system.  This result is
counterintuitive because it states that the 3BRBIM can become \emph{more}
ordered by increasing the temperature, depending on the system's quenched
disorder parameter.  It would be exciting to see experimental confirmation
of this effect.

%
\subsection{Future directions}

While we have been able to answer many questions about fault-tolerant
quantum computing using color codes, practicalities have necessarily
limited the focus of our analysis, leaving other related questions open.
Our results also also raise new questions that we believe are worthy of
study.

One future direction we mentioned is optimizing the syndrome extraction
circuit.  One could also examine using more elaborate ancilla states in the
circuit, such as those used in the schemes proposed by Shor
\cite{Shor:1996a}, Steane \cite{Steane:1998a}, and Knill \cite{Knill:2004a}.
In any scheme one chooses, further improvement may still be possible by
transforming the circuit used in an implementation.

Another future direction we alluded to is optimizing the decoding algorithm.
One could examine the performance of the truly optimal decoder for the
circuit model which accounts for the correlations in the noise induced by
the syndrome extraction circuit.  This will yield an upper bound on the
accuracy threshold for the noise model(s) studied.  On the other end of the
spectrum, it would be useful to explore the performance of faster decoders
which don't yield as high a threshold as the MLE decoder but which may be
more valuable in practice.  The renormalization group decoder
\cite{Duclos-Cianci:2009a} and minimum-weight perfect matching decoder
\cite{Dennis:2002a} (using a mapping of one color code to two Kitaev surface
codes \cite{Duclos-Cianci:2011a}) are examples of this.  Another alternative
is to generalize the results of Feldman \etal, who developed an efficient
linear-program decoder for binary codes based on an  integer-program-based
decoder similar to the one we developed here \cite{Feldman:2005a}.

The lower bound technique of self-avoiding walks that we used is certainly
not the tightest, and it may be of interest to establish tighter lower
bounds.  For tighter bounds, it may be possible to use different techniques.
In the case of the circuit-based noise model, the self-avoiding walk bound
technique breaks down dramatically, and it would be worth exploring other
lower-bound techniques in this setting.

While we believe the noise and control model that we studied is reasonable,
it is certainly not unique and can be improved upon with more experimental
input.  As shown by Levy \etal, \cite{Levy:2009a, Levy:2011a}, when more
realistic models are included, conclusions regarding fault tolerance can
change dramatically.  Even at an abstract level, one could modify our
depolarizing noise model for $\CNOT$ gates so that it acted ideally with
probability $1-p$ and applied one of the fifteen nontrivial Pauli operators
with probability $p/15$ each rather than acting ideally with probability
$1-p$ and applying one of the sixteen Pauli operators with probability
$p/16$.

While we gave a prescription for injecting magic states into the color code
for both the pancake and 2D defect-based architectures, we did not carefully
study the threshold of the circuits used for injection.  To our knowledge,
this type of study has not been performed for Kitaev's surface codes either.
Such studies would be valuable, as it could be the case that the magic state
preparation threshold is actually less than the accuracy threshold reported
for all of the other gates, even though the distillation threshold for the
magic states is higher than the accuracy threshold for the other gates.

Finally, the connection between color codes and the three-body random-bond
Ising model allowed us to explore the structure of order-disorder transition
in the latter model by studying the former.  This is one of the rare
examples where a purely quantum information theoretic result has led to
greater understanding of a classical system.  Kitaev's surface codes and the
two-body random-bond Ising model have a similar connection and have admitted
a similar study \cite{Dennis:2002a, Wang:2003a}.  It is clear that it is the
CSS structure of these codes that admits these studies; one could argue that
every CSS code is a topological code for some topology, having an associated
classical statistical-mechanical model for a given quantum noise model.  It
might be interesting to use the fault-tolerant decoding of CSS codes
generally as a tool to explore related statistical-mechanical systems with
quenched disorder.

%
\begin{acknowledgments}

We would like to thank the following individuals for helpful discussions:
Hector Bombin, Bob Carr, Chris Cesare, Guillaume Duclos-Cianci, Bryan
Eastin, Austin Fowler, Anand Ganti, Peter Groszkowski, Jim Harrington,
Charles Hill, Lloyd Hollenberg, Uzoma Onunkwo, Cindy Phillips, David Poulin,
Robert Raussendorf, and David Wang.  We would also like to thank Dave Gay
for use of AMPL mathematical programming language.  PRR was supported by the
Quantum Institute at Los Alamos National Laboratories.  AJL and JTA were
supported in part by the National Science Foundation through Grant 0829944.
JTA and AJL were supported in part by the Laboratory Directed Research and
Development program at Sandia National Laboratories.  Sandia National
Laboratories is a multi-program laboratory managed and operated by Sandia
Corporation, a wholly owned subsidiary of Lockheed Martin Corporation, for
the U.S.  Department of Energy's  National Nuclear Security Administration
under contract DE-AC04-94AL85000.

\end{acknowledgments}

%

\bibliographystyle{landahl}
\bibliography{landahl.JabRef}

\end{document}